\documentclass[12pt]{article}

\usepackage[english]{babel}
\usepackage[T1]{fontenc}
\usepackage[latin1]{inputenc}
\usepackage{stmaryrd,amsthm}
\usepackage{upgreek}
\usepackage{amsmath}
\usepackage{bm}
\usepackage{mathrsfs}
\usepackage{tabularx}
\usepackage{amssymb}
\usepackage[mathscr]{eucal}
\usepackage{caption}
\usepackage{dsfont}
\usepackage{graphicx}
\usepackage{color}

\newcommand{\ket}[1]{\left|\,{#1}\,\right>}

\newcommand{\beq}{\begin{equation}}
\newcommand{\eeq}{\end{equation}}

\newcommand{\m}[1]{$\mathop{#1}$}

\begin{document}

\begin{titlepage}
\title{\vskip -60pt
{\textbf{Symmetries and currents of
\\the ideal and unitary Fermi gases}}
\author{Xavier Bekaert$^a$, Elisa Meunier$^a$ and Sergej Moroz$^{b,c}$}}
\date{}
\maketitle \vspace{-1.0cm}

\begin{center}
~~~\\
$^a$ Laboratoire de Math\'ematiques et Physique Th\'eorique\\
Unit\'e Mixte de Recherche $7350$ du CNRS\\ 
F\'ed\'eration de Recherche $2964$ Denis Poisson\\
Universit\'e Fran\c{c}ois Rabelais, Parc de Grandmont\\
37200 Tours, France\\
{\small{{{{xavier.bekaert@lmpt.univ-tours.fr}},~{{elisa.meunier@lmpt.univ-tours.fr}}}}}

~~~\\
$^b$ Institut f\"ur Theoretische Physik\\
Philosophenweg 16\\
D-69120 Heidelberg, Germany\\

~~~\\
$^c$ Department of Physics \\ 
University of Washington \\
Seattle, WA 98195-1560, USA \\
{\small{{{{morozs@uw.edu}}}}}

\end{center}

\vspace{1cm}

\begin{abstract}
\noindent

The maximal algebra of symmetries of the free single-particle Schr\"odinger equation is determined and its relevance for the holographic duality in non-relativistic Fermi systems is investigated. This algebra of symmetries is an infinite dimensional extension of the Schr\"odinger algebra,
it is isomorphic to the Weyl algebra of quantum observables, and it may be interpreted as a non-relativistic higher-spin algebra. The associated infinite collection of Noether currents bilinear in the fermions are derived from their relativistic counterparts via a light-like dimensional reduction. The minimal coupling of these currents to background sources is rewritten in a compact way by making use of Weyl quantisation. Pushing forward the similarities with the holographic correspondence between the minimal higher-spin gravity and the critical $O(N)$ model, a putative bulk dual of the unitary and the ideal Fermi gases is briefly discussed.
\end{abstract}




\thispagestyle{empty}
\end{titlepage}

\tableofcontents

\section{Introduction} \label{intro}

The quantum many-body problem of a non-relativistic two-component Fermi gas with short-range attractive interactions is a longstanding problem in condensed matter physics. At low temperature, the system is known to be superfluid and undergoes a smooth crossover from the Bardeen-Cooper-Schrieffer (BCS) to the Bose-Einstein-Condensate (BEC) regime as the two-body attraction is increased (see \cite{coldatomsreview} for recent reviews). Considerable progress in atomic physics in the two last decades allowed to study the BCS to BEC crossover with unprecedented accuracy. Of special theoretical interest is the regime in between BCS and BEC known as the unitary Fermi gas.\footnote{In this regime the two-body low-energy cross section saturates the maximal bound originating from the unitarity of the S-matrix. This property gives rise to the term ``unitary'' Fermi gas.} In three spatial dimensions, the unitary Fermi gas is intrinsically  strongly coupled and no obvious small parameter is available, precluding the reliable application of a perturbative expansion. In this way, the unitary Fermi gas provides a great theoretical challenge and requires the development and subsequent applications of advanced non-perturbative many-body methods. 

A special property of the unitary Fermi gas in vacuum (describing few-body physics) is the invariance of the action under the scale transformations and more generally under the Schr\"odinger group of Niederer and Hagen \cite{Niederer, Hagen}. This group of space-time symmetries provides a direct non-relativistic analogue of the conformal group. Although the general proof is still lacking, it is believed that for the unitary Fermi gas there is no conformal anomaly and thus that the Schr\"odinger symmetry survives quantisation \cite{Mehen, Werner}. Motivated by this, Nishida and Son extended the general methods of conformal field theory (CFT) to the realm of non-relativistic physics and applied them to the unitary Fermi gas \cite{Nishida, Nishida2}.

Due to the non-relativistic conformal symmetry of the unitary Fermi gas in vacuum, Son, Balasubramanian and McGreevy \cite{Son:2008ye,Balasubramanian:2008dm} recently have initiated an attempt to apply the methods of the gauge-gravity duality to this system.  While their seminal papers revived the interest of mathematical and high-energy physicists toward non-relativistic symmetries, they mostly triggered an intensive body of research for the putative holographic duals of various non-relativistic systems originating from condensed matter theory. However the initial target, \textit{i.e.} a holographic description of the unitary Fermi gas, remains tantalising despite several steps forward \cite{coldatomstests}. As mentioned by Son in \cite{Son:2008ye}, a possible direction of investigation is the unitary Fermi gas with $U(1)\times Sp\,(2N)$ symmetry introduced in \cite{Nikolic:2007zz,Veillette:2007zz} (see also \cite{Abuki}) whose gravity dual might be a classical theory in the large-$N$ limit. Notably, this gravity theory would have an infinite tower of fields with unbounded spin, similar to the conjectured anti de Sitter (AdS) dual of the critical $O(N)$ model \cite{Klebanov:2002ja}. Interestingly, an impressive check of this latter conjecture has recently been performed for three-point correlation functions \cite{Giombi:2009wh}. These encouraging results strengthen the natural expectation that this AdS/O(N) model correspondence provides a proper source of inspiration for an AdS/unitary-fermions correspondence.
Our recent letter \cite{Bekaert:2011cu} aims to represent a further step towards a precise conjecture along these lines. The goal of the present work is to present in detail some findings about the free and unitary fermions that support our proposal. Some of these results were already announced in \cite{Bekaert:2011cu} without a proof. Although the body of our paper focuses on the CFT (boundary) side, some comments on the gravity (bulk) side and the holographic correspondence are in order.

The AdS/O(N) correspondence proposed by Klebanov and Polyakov \cite{Klebanov:2002ja} pursuing earlier insights of Sezgin and Sundell \cite{Sezgin:2002rt} involves, on the boundary, a multiplet of $N$ massless scalar fields in the fundamental representation of $O(N)$ with a quartic $O(N)$-invariant interaction and, in the bulk, an infinite tower of symmetric tensor gauge fields with interactions governed by Vasiliev equations \cite{Vasiliev:2003ev} (see \cite{Vasiliev:2004qz} for introduction). 
The crucial point in this correspondence is that there is an infinite collection of $O(N)$-singlet symmetric currents of all even ranks, bilinear in the boundary scalar field, that precisely matches the spectrum of the higher-spin gauge theory. These boundary currents are conformal primary fields and are exactly conserved for the free theory (while only at leading order in $1/N$ for the interacting theory) so their bulk duals should indeed be gauge fields.
They are actually the Noether currents of the maximal symmetry algebra of the massless Klein-Gordon equation \cite{Eastwood:2002su}, that is the infinite-dimensional symmetry algebra of a free conformal scalar field. This algebra of rigid symmetries is isomorphic to the algebra which is gauged in the bulk higher-spin theory \cite{Vasiliev:2003ev}. 
A precise statement of the correspondence is that the generating functional of the connected correlators of the boundary currents is given, in
the large-$N$ limit, by the on-shell classical action of the bulk fields expressed in terms of the boundary data.
In the large-$N$ limit, the generating functionals of the critical theory and of the free theory are related by a Legendre transformation, which should be dual to a mere change of boundary conditions for the same bulk theory, as follows from the general analysis of \cite{Klebanov:1999tb,Gubser:2002vv}.

So what could be an educated guess for a gravity dual of unitary fermions? We will turn back to this cardinal issue in the conclusion but, before, let us start by looking for the non-relativistic analogue of the above-mentioned construction. As was found in \cite{Nikolic:2007zz,Veillette:2007zz}, a sensible large-$N$ extension of the unitary Fermi gas has $U(1)\times Sp\,(2N)$ symmetry\footnote{Note that we are following the physicist convention here and define the compact symplectic group as $Sp\,(2N):=U(2N)\cap Sp\,(2N,\mathbb C)$. Alternatively, physicists also frequently use the notation $USp\,(2N)$ while mathematicians usually refer to this group as $Sp\,(N)$.} and involves a multiplet of $2N$ non-relativistic massive fermions transforming in the fundamental representation of $Sp\,(2N)$.
The general arguments of \cite{Gubser:2002vv} imply that, in the large-$N$ limit, the Helmholtz free energies of unitary fermions and of non-interacting fermions are related by a Legendre transformation. Thus, in this limit the results obtained from the free theory are of direct interest for the theoretically more challenging critical regime at the unitarity point. This important observation motivated us to focus in this paper on a collection of free non-relativistic massive fermions in the fundamental representation of $Sp\,(2N)$ and to study its symmetries and currents.

The summary of our main results and the plan of the paper are as follows:
In Section \ref{Legendre}, we start with an introduction to the unitary Fermi gas and its large-$N$ extension.
We also present the general arguments of \cite{Gubser:2002vv} and demonstrate that, in the large-$N$ limit, the generating functionals of the unitary Fermi gas and of the ideal Fermi gas are related by a Legendre transformation.
In Section \ref{Symmetry}, the maximal symmetry algebra of the free Schr\"odinger equation is identified and shown to be isomorphic to the Weyl algebra of quantum observables in the time-reversed Heisenberg picture. It provides an infinite-dimensional extension of the Schr\"odinger algebra, as was recognised in \cite{Valenzuela:2009gu}.
In Section \ref{Currents}, 
an infinite collection of $Sp\,(2N)$ or $O(N)$ singlet symmetric tensors of all ranks, bilinear in the fermionic field is obtained from the corresponding relativistic currents through a dimensional reduction along a light-like direction. 
In Section \ref{Sources}, the coupling of these bilinears to external sources is considered and written in a compact form
by making use of the Weyl quantisation. This allows us to identify the algebra of gauge symmetries with the algebra of quantum observables with arbitrary time dependence. These symmetries can be thought as the higher-spin generalisations of the non-relativistic general coordinate and Weyl symmetries discussed in \cite{Son:2005rv}. 
In Section \ref{Conclusion}, we summarise our results and review our proposal \cite{Bekaert:2011cu} of a possible gravity dual of the unitary and the ideal Fermi gases: the
$O(N)$-singlet bilinear sector of the large-$N$ extension of the free or unitary fermions in $d$ space dimension should be dual to the null-reduction of classical Vasiliev theory on $AdS_{d+3}$ with $\mathfrak{u}(2)$-valued tensor gauge fields of all integer ranks and suitable boundary conditions. In particular, the bulk dual of the ``physical'' (\textit{i.e.} $N=1$, $d=3$) unitary UV-stable Fermi gas would be the null dimensional reduction of the $\mathfrak{u}(2)$ higher-spin gauge theory on $AdS_6$ with the exotic boundary condition for the complex scalar field dual to the Cooper-pair field.

Wherever possible, we will stick to the notations and conventions of \cite{Son:2008ye}. Except in Appendix \ref{Weylquantization}, we set $\hbar=1$.

\section{Unitary Fermi gas and its large-$N$ extension}\label{Legendre}

\subsection{Action and symmetries} 

Nowadays a dilute two-component Fermi gas near a broad Feshbach resonance can be cooled with the help of lasers to ultra-low temperatures $\sim 10^{-9}\text{K}$, and is studied extensively in experiments with ultracold atoms. In three spatial dimensions ($d=3$) at low densities it can be very well described by the microscopic action
\beq \label{ac1}
S[\,\psi\,;c_0]=\int dt \int d {\bf x} \left[ \sum_{\alpha=\uparrow,\downarrow}\psi_\alpha^*\left(i\partial_t+\frac{\Delta}{2m}+\mu \right)\psi_\alpha\,-\,c_0\, \psi^*_\downarrow \psi^*_\uparrow\psi_\uparrow\psi_\downarrow \right],
\eeq
where the two species of fermionic atoms of mass $m$ are represented by the Grassmann-odd fields $\psi_\uparrow$ and $\psi_{\downarrow}$, while $\mu$ stands for the chemical potential, and $c_0$ measures the microscopic interaction strength. In actual experiments with ultracold gases the two different components denote different hyperfine eigenstates which we denote here for simplicity by $\uparrow$ and $\downarrow$ but which have nothing to do with genuine spins ``up'' and ``down''.\footnote{Note that due to the lack of the spin-statistics theorem for non-relativistic quantum field theories, the spin of fermions (and thus the number of components) is not constrained \cite{LevyLeblond}. For example, we can have spinless one-component fermions.} This action has an internal $U(2)$ symmetry. Due to the contact nature of the interaction term, the non-relativistic quantum field theory defined by the action \eqref{ac1} must be regularised. This can be achieved, for example, by introducing a sharp ultraviolet cutoff. Subsequently, the bare interaction parameter $c_0$ is related via renormalisation to a low-energy observable: the s-wave scattering length $a$. The concrete functional relation between $c_0$ and $a$ depends on the regularisation scheme and will not be presented here. In this paper we will be mainly interested in excitations above the vacuum state, \textit{i.e.} a system at zero temperature and zero density. For $a\leqslant 0$ there are no bound states in the two-component Fermi gas and in this range the vacuum corresponds to $\mu=0$ (see e.g. \cite{Nikolic:2007zz} for a detailed explanation).  Due to the presence of a universal two-body dimer bound state for $a>0$, the chemical potential in vacuum is negative and is related to the scattering length via $\mu=-\frac{1}{2m a^2}$. In any case, the only length scale in the renormalised theory in vacuum is given by the scattering length $a$.

The non-interacting Fermi gas is obtained for $a=0$ which translates  into $c_0=0$. In vacuum it is obviously scale invariant. Another theoretically interesting regime is the strongly interacting unitary Fermi gas, where $a^{-1}=0$. The only length scale defined by the scattering length drops out in this regime. Hence the classical theory for the unitary Fermi gas is scale invariant. Although there is no general proof yet, there are numerous theoretical and experimental evidences collected by now that the quantum unitary Fermi gas in vacuum is also scale invariant.\footnote{In other words there is no quantum anomaly associated with the scale transformation. Interestingly, the unitary Bose gas suffers from a quantum scale anomaly, known in the nuclear and atomic physics as the Efimov effect. Presence of this anomaly hinders an experimental realisation of a stable unitary Bose gas in experiments with ultracold quantum gases.} More precisely, the action of the unitary Fermi gas is invariant under the Schr\"odinger symmetry, which will be introduced in Section \ref{Symmetry}, and the theory is believed to be an example of a strongly interacting non-relativistic CFT \cite{Mehen, Nishida}.

A sensible large-$N$ extension of the unitary Fermi gas that preserves the pairing structure of the interaction term was found in \cite{Nikolic:2007zz,Veillette:2007zz}. The model with $N$ ``flavors'' is defined by the action
\beq \label{ac2}
S[\,\psi\,;c_0,N]=\int dt \int d {\bf x} \left[ \psi^\dagger\left(i\partial_t+\frac{\Delta}{2m}+\mu \right)\psi-\frac{c_0}{4N}\, \big|\psi^T \mathbb{J}\, \psi\big|^2 \right],
\eeq
where $\psi$ denotes a multiplet of $2N$ massive fermions with components $\psi^A=\psi^{\alpha,a}$ with $\alpha=\,\uparrow,\downarrow$ and $a=1,\dots, N$. The symbol $\mathbb{J}$ represents the symplectic $2N\times2N$ matrix $\mathbb{J}_{AB}=\epsilon_{\alpha\beta}\otimes\delta_{ab}$ which has the block form
$\mathbb{J}=\left( \begin{array}{cc}
0 & 1 \\
-1 & 0 \end{array} \right)$.
For $N=1$, one recovers the original model \eqref{ac1}, \textit{i.e.} $S[\,\psi\,;c_0,N=1]\,=\,S[\,\psi\,;c_0]$. As far as internal symmetries are concerned, the kinetic term in Eq. \eqref{ac2} is invariant under $U(2N)$, while the quartic interaction is invariant under $U(1)\times Sp\,(2N,\mathbb C)$. As a result, the full interacting theory is invariant under the intersection of $U(2N)$ with $U(1)\times Sp\,(2N,\mathbb C)$, which happens to be $U(1)\times Sp\,(2N)$ (see the footnote in Section \ref{intro}).
For $N=1$, one finds as mentioned above $U(1)\times Sp\,(2)\cong U(2)$ as the internal symmetry group, since $Sp\,(2)\cong SU(2)$.

The preceding construction introduces a new integer parameter into the theory and resembles in various respects the structure of the relativistic linear $O(N)$ models. While the analogy is suggestive, there is an important difference which is worth to be emphasised already here. On the one hand, in the relativistic $O(N)$ model the internal symmetries of the free and of the critical theory happen to be the same. On the other hand, the internal symmetry of the kinetic part of the action \eqref{ac2} is larger than the internal symmetry of the full action. Thus, the $N>1$ extensions of the ideal and of the unitary Fermi gas have different internal symmetries. This makes the relation between these two theories more subtle than in the relativistic $O(N)$ case.

Let us finally note that for general $N$ flavors, $U(2)\times O(N)$ is a subgroup of $U(1)\times Sp\,(2N)$. Mathematically, the subgroups $U(2)$ and $O(N)$ are centralisers\footnote{Let $G_1$ and $G_2$ be two subgroups of $G$. The subgroup $G_1$ is the centraliser of $G_2\subseteq G$ if and only if $G_1$ is the largest subgroup of $G$ such that all its elements commute with all elements of $G_2$. Usually, the centraliser of $G_2\subseteq G$ is denoted by $C(G_2)$ ($=G_1$ here). Such a pair of subgroups $G_1$ and $G_2$ is sometimes called a Howe dual pair by mathematicians.} of each other inside $U(1)\times Sp\,(2N)$,\footnote{This follows from the fact that $Sp\,(2)\times O(N)$ is a subgroup of $Sp\,(2N)$ where the subgroups $Sp\,(2)$ and $O(N)$ are centralisers of each other. This property plays an important role (though for a different reason) in the construction of higher-spin algebras \cite{Vasiliev:2004qz}.} as they transform independently spin and flavor indices. In the following, $U(2)\times O(N)$ symmetry subgroup will play a central role in the suggestion of the putative holographic dual of the unitary Fermi gas.

\subsection{Ideal and unitary gases as Legendre conjugates}\label{Legendreconj} 

The celebrated BCS theory has taught us that the physical phenomena of superfluidity and superconductivity have their origin in the condensation of particle-particle Cooper pairs at low temperature. From this insight, it becomes clear that a proper understanding of physics of these Cooper pairs is of a central importance for quantum Fermi systems. By applying the general observation of Gubser and Klebanov on the double trace deformations of conformal field theories \cite{Gubser:2002vv} to the large-$N$ extension of the unitary Fermi gas, we show here that the generating functionals of Cooper pair connected correlators in the non-interacting and in the unitary Fermi gases are related by a Legendre transformation in the large $N$ limit (or, similarly, in the mean field approximation).

The following discussion will closely parallel the derivation of Gubser and Klebanov that was introduced for an infrared relevant double trace deformation of a conformal field theory like, for example, the relativistic linear $O(N)$ model in three space-time
dimensions. There is one important difference between the relativistic and the non-relativistic problems of interest that we would like to emphasize here. In the $O(N)$ model , the quartic interaction term is an infrared relevant perturbation of a free CFT triggering the renormalization group flow to approach the infrared stable Wilson-Fisher fixed point. Due to a distinct power counting in the non-relativistic physics, the four-fermion contact interaction in \eqref{ac1} is infrared irrelevant in the most physically interesting case of three spatial dimensions. This implies that the Gaussian fixed point is infrared stable and the unitarity fixed point (in vacuum) is in fact approached in the ultraviolet of the renormalization group flow.

With a slight abuse of terminology, by ``Cooper pair'' we mean here the $Sp\,(2N)$-singlet bilinear 
\begin{eqnarray}\label{kCooper}
k(t,{\bf x})\,:=\,\frac12\,\psi^T \mathbb{J}\, \psi\,=\,\frac12\, \psi^A\,\mathbb{J}_{AB} \, \psi^B\,=\,\sum\limits_{a=1}^N \psi_{\uparrow,\,a} \,\psi_{\downarrow,\,a} \,,
\end{eqnarray}
which reproduces the genuine Cooper pair $\psi_{\uparrow}\psi_{\downarrow}$ when $N=1$. 
The generating functional $W[\,\upvarphi\,;c_0,N]$ of Cooper-pair connected correlators in the Fermi gas
described by \eqref{ac2} is defined by the path integral
\beq\label{W1}
\exp i\,W[\,\upvarphi\,;c_0,N]\,=\,\int{\cal D}\psi{\cal D}\psi^\dagger\,\,\exp i \,S[\,\psi\,,\upvarphi\,;c_0,N]\,,
\eeq
where
\beq\label{S1}
S[\,\psi\,,\upvarphi\,;c_0,N]\,:=\,S[\,\psi\,;c_0,N]\,-\,\int dt \,d {\bf x}\, \big(\,k\,\upvarphi^*+k^*\upvarphi\,\big)
\eeq
is the action in the presence of an external charged scalar field
$\upvarphi$ coupled to the Cooper pair $k$.

In particular, the free ($c_0=0$, {infrared fixed point in $d=3$}) action in the presence of the source $\upvarphi$ reads
\beq\label{free1}
S_{\mbox{free}}[\,\psi\,,\upvarphi]
\,:=\,
S[\,\psi\,,\upvarphi\,;0,N]\,=\,\int dt \,d {\bf x} \left[ \psi^\dagger\left(i\partial_t+\frac{\Delta}{2m}+\mu \right)\psi\,-\big(\,k\,\upvarphi^*+k^*\upvarphi\,\big)\,\right]
\,,
\eeq
and is quadratic in the dynamical field $\psi$ (since the kinetic term and the Cooper pair $k$ are).
This quadratic functional is usually rewritten in a more elegant form by making use of the Nambu-Gor'kov field
\begin{eqnarray}\label{Nambu}
\Uppsi\,=\,\begin{pmatrix}\psi_\uparrow\, \\ \psi^{*}_\downarrow\end{pmatrix}\,.
\end{eqnarray}
Notice that $\psi$ and $\Uppsi$ are not related by a unitary transformation (not even by a linear or anti-linear transformation) but the canonical anti-commutation relations are preserved.
In terms of the Nambu-Gor'kov field, the quadratic action \eqref{free1} takes the form
\beq\label{free2a}
S_{\mbox{free}}[\,\Uppsi\,,\upvarphi]
\,=\,\int dt\, d {\bf x}\, \Uppsi^\dagger\left(
\begin{array}{cc} 
i\partial_t+(\frac{\Delta}{2m}+\mu) & \upvarphi\\
\upvarphi^* & i\partial_t-(\frac{\Delta}{2m}+\mu)
\end{array} 
 \right)\Uppsi\,.
\eeq
The generating functional of connected correlators of Cooper pairs in the ideal Fermi gas is $W_{\mbox{free}}[\,\upvarphi\,;N]
\,:=\,W[\,\upvarphi\,;0,N]$. It can easily be evaluated formally 
since the path integral \eqref{W1} is Gaussian in such case:
\beq\label{free2}
W_{\mbox{free}}[\,\upvarphi\,;N]
\,=\, -i
N\,\mbox{Tr}\log\left(
\begin{array}{cc} 
i\partial_t+(\frac{\Delta}{2m}+\mu) & \upvarphi\\
\upvarphi^* & i\partial_t-(\frac{\Delta}{2m}+\mu)
\end{array} 
 \right)\,=:\,N\,W_{\mbox{free}}[\,\upvarphi\,]
\,
\eeq
providing an explicit solution of the infrared stable conformal field theory in $d=3$.
In order to prepare the ground for the later discussion, let us already here introduce the field
\beq\label{mean2}
\pi\,:=\,
\frac{\delta W_{\mbox{free}}[\upvarphi]}{\delta \upvarphi^*}\,.
\eeq
conjugate to the free field $\upvarphi$, and the Legendre transformation
\beq\label{Leg}
\Gamma_{\mbox{free}}[\,\pi\,]\,:=\,
W_{\mbox{free}}[\,\upvarphi]\,-\,\int dt \,d {\bf x}\, \big(\,\upvarphi\,\pi^*+\upvarphi^*\,\pi\,\big)\,,
\eeq
of the free connected correlator generating functional $W_{\mbox{free}}[\upvarphi]$.

In order to relate this to the interacting theory, we use a standard trick: the Hubbard-Stratonovich transformation which reformulates any system of particles with a two-body interaction equivalently as a system of particles interacting only via a fluctuating auxiliary field. 
More precisely, here one transforms the path integral over the fundamental fermionic field $\Uppsi$ with quartic vertex into a Gaussian path integral via the introduction of an auxiliary complex scalar field $\upphi$ mediating the interaction in the particle-particle channel. This auxiliary field is called ``dimer'' in the literature on the unitary Fermi gas. More concretely, on the right-hand-side of \eqref{W1} one can introduce a Gaussian path integral over the auxiliary field $\upphi$ to get
\beq\label{HSpathint}
\exp i\,W[\,\upvarphi\,;c_0,N]\,\propto\,\int{\cal D}\Uppsi{\cal D}\Uppsi^\dagger\,{\cal D}\upphi{\cal D}\upphi^*\,\,\exp i\, S_{\mbox{HS}}[\,\Uppsi\,,\upphi\,,\upvarphi\,;c_0,N]\,,
\eeq
where $S_{\mbox{HS}}[\,\Uppsi\,,\upphi\,,\upvarphi\,;c_0,N]$ is the Hubbard-Stratonovich transformation of the action \eqref{S1}. It is equal to the sum of a chemical-potential like term for the dimer plus
the free action in presence of the source $\upvarphi$ shifted by the dimer $\upphi$,
\beq
S_{\mbox{HS}}[\,\Uppsi\,,\upphi\,,\upvarphi\,;c_0,N]\,:=\,\frac{4N}{c_0}\int dt \,d {\bf x}\,|\upphi|^2\,+\,S_{\mbox{free}}[\,\Uppsi\,,\upvarphi+\upphi]\,.
\eeq
In the following, it is convenient to work directly with the shifted dimer field $\phi=\upphi+\upvarphi$.
The integral over the dynamical field $\Uppsi$ in \eqref{HSpathint} can now be evaluated and gives as a result
\beq\label{Wpint}
\exp i\, W[\,\upvarphi\,;c_0,N]\,\propto\,\int{\cal D}\phi{\cal D}\phi^*\,\,\exp i\, S_{\mbox{eff}}[\phi\,,\upvarphi\,;c_0,N]\,,
\eeq
where the effective action for the dimer field depends linearly on the number $N$ of flavors: 
$S_{\mbox{eff}}[\,\phi\,,\upvarphi\,;c_0,N]\,=\,N\,S_{\mbox{eff}}[\,\phi\,,\upvarphi\,;c_0]$ 
and is the sum of the chemical-potential like term plus
the free effective action for the auxiliary field
\eqref{free2}
\beq\label{eff1}
S_{\mbox{eff}}[\phi\,,\upvarphi\,;c_0]\,:=\,\frac{4}{c_0}\int dt \,d {\bf x}\,|\,\phi-\upvarphi|^2\,+\,W_{\mbox{free}}[\phi]\,.
\eeq
The linear dependence of the effective action on the parameter $N$ means that $1/N$ controls the loop expansion of the dimer effective theory.
The large-$N$ limit allows for a saddle point approximation of the integral \eqref{Wpint} over the dimer field: 
\beq
W[\,\upvarphi\,;c_0,N]\,=\,N\,W_{\mbox{mean}}[\,\upvarphi\,;c_0]\,+\,{\cal O}(1)\,,
\eeq
where
\beq\label{mean}
W_{\mbox{mean}}[\,\upvarphi\,;c_0]\,:=\,S_{\mbox{eff}}[\,\phi(\upvarphi)\,,\upvarphi\,;c_0]
\eeq
is the celebrated ``mean field'' approximation  of the generating functional of connected correlators. Notice that in the physically relevant $N=1$ case, this term is \textit{a priori} of the same order as the $1/N$ corrections. Nevertheless, the mean field approximation becomes exact at $N=\infty$, providing an explicit relation between the generating functionals of the free and interacting theory:
\beq \label{exp}
W_{\mbox{mean}}[\,\upvarphi\,;c_0]\,=\,\frac{4}{c_0}\int dt \,d {\bf x}\,|\,\phi(\upvarphi)-\upvarphi|^2\,+\,W_{\mbox{free}}[\phi(\upvarphi)]\,.
\eeq
On the right-hand-side of \eqref{mean} and \eqref{exp}, the field $\phi$ depends on the source $\upvarphi$ because it should be understood as the solution of the classical equation of motion
\beq\label{mean2a}
\frac{\delta S_{\mbox{eff}}[\phi\,,\upvarphi\,;c_0]}{\delta \phi^*}=0\quad \Longrightarrow\quad
\frac{\delta W_{\mbox{free}}[\phi]}{\delta \phi^*}\,=\,\frac{4}{c_0}\,(\upvarphi-\phi)\,.
\eeq
Sometimes in this paper, the large-$N$ limit and the mean field approximation will be loosely said to be equivalent. By this, we mean that the equations \eqref{exp}-\eqref{mean2a} provide an approximation for the interacting generating functional which can either be understood as the leading-order approximation in the large-$N$ limit analogous to the 't Hooft limit in gauge theories (that is $N\to \infty$ at fixed $c_0$) or as the mean field approximation at fixed $N$ (say $N=1$). 

Now, two distinct limits of the approximated generating functional \eqref{exp}-\eqref{mean2a} can be considered: either a large-$c_0$ limit in which case the coefficient $c_0/N$ of the quartic term in the bare action \eqref{ac2} might be kept finite (though possibly small, e.g. in the ultraviolet) or instead a small-$c_0$ limit in which case the coefficient $c_0/N$ goes to zero, even if $N$ is kept finite (though possibly large for the validity of the saddle point approximation). In both cases, one finds that the generating functionals are Legendre conjugates, but expressed in terms of different rescaled fields in the distinct limits.
First, let us consider the limit $c_0\to\infty$\,. If one rescales the source $\varphi:=\frac{4}{c_0}\upvarphi$, then the equation of motion \eqref{mean2a} becomes
\beq
\frac{\delta W_{\mbox{free}}[\phi]}{\delta \phi^*}\,=\,\varphi +{\cal O}\Big(\frac1{c_0}\Big)\,,
\eeq
which means that the shifted dimer field and the rescaled source are exact Legendre conjugates at $c_0=\infty$.
Moreover, the approximated generating functional \eqref{exp} takes the suggestive form
\beq \label{exp2}
W_{\mbox{mean}}\Big[\,\frac{c_0\varphi}4\,;c_0\Big]
\,=\,-\int dt \,d {\bf x}\,\big(\,\varphi\,\phi^*+\varphi^*\phi\,\big)
\,+\, \frac{c_0}4\int dt \,d {\bf x}\,|\varphi|^2\,+\,W_{\mbox{free}}[\phi(\varphi)]
\,+\,{\cal O}\Big(\frac1{c_0}\Big)\,.
\eeq
Comparing with the definitions \eqref{mean2}-\eqref{Leg}, one is lead to the relation
\beq\label{Legendrerel}
\lim\limits_{c_0\to\infty}\Big\{\,W_{\mbox{mean}}\left[\,\frac{c_0\varphi}4\,\,;c_0\right]-\frac{c_0}4\int dt \,d {\bf x}\,|\varphi\,|^2\Big\}
\,\,=\,\,\Gamma_{\mbox{free}}[\varphi]\,.
\eeq
This result is very similar to the calculation performed in \cite{Gubser:2002vv}, the interpretation of which is very natural in the $O(N)$ model where the infrared stable Wilson-Fisher fixed point corresponds to an infinitely large dimensionful coupling. In the non-relativistic Fermi gas the above derivation is applicable to the spatial dimension $d<2$, where the unitary fixed point is infrared stable.

In $d>2$ the unitarity fixed point is ultraviolet stable which in regularisation with a sharp cutoff corresponds to the limit $c_0\to 0$. It appears therefore that the limit $c_0\to 0$ is necessary in $d>2$ for the unitary Fermi gas.\footnote{ Note, however, that $c_0\to\infty$ in any spatial dimension for the unitary Fermi gas in dimensional regularisation \cite{Nishida2}.} 
So let us now consider the limit $c_0\to 0$ and rescale the shifted dimer field $\tilde\phi:=\frac{4}{c_0}\phi$. If we express the generating functional of the free theory in terms of the rescaled dimer field, 
\beq
{\tilde W}_{\mbox{free}}[\,\tilde\phi\,]\,:=\,W_{\mbox{free}}\left[\frac{c_0}{4}\tilde\phi\right]\,,
\eeq
then the equation of motion \eqref{mean2a} reads
\beq
\frac{\delta \tilde W_{\mbox{free}}[\tilde\phi]}{\delta \tilde\phi^*}\,=\,\upvarphi +{\cal O}\Big(c_0\Big)\,.
\eeq
Thus the source $\upvarphi$ and the rescaled dimer $\tilde\phi$ form a Legendre conjugate pair in the limit $c_0\to 0$\,. In addition, 
if we express the generating functional of the mean field theory in terms of the rescaled dimer field, 
\beq
{\tilde W}_{\mbox{mean}}[\,\tilde\phi\,;c_0\,]\,:=\,W_{\mbox{mean}}\left[\frac{c_0}{4}\tilde\phi\,;c_0\right]\,,
\eeq
then the relation \eqref{exp} can be written as
\beq \label{exp3}
\tilde W_{\mbox{mean}}\Big[\tilde\phi\,;c_0\Big]
\,=\,-\int dt \,d {\bf x}\,\big(\,\upvarphi\,\tilde\phi^*+\upvarphi^*\tilde\phi\,\big)
\,+\, \frac4{c_0}\int dt \,d {\bf x}\,|\upvarphi|^2\,+\,\tilde W_{\mbox{free}}[\tilde\phi(\upvarphi)]
\,+\,{\cal O}\Big(c_0\Big)\,.
\eeq
Therefore,
\beq\label{Legendrerel2}
\lim\limits_{c_0\to 0}\Big\{\,\tilde W_{\mbox{mean}}\left[\,\tilde\phi(\upvarphi)\,\,;c_0\right]\,-\,\frac4{c_0}\int dt \,d {\bf x}\,|\upvarphi\,|^2\Big\}
\,\,=\,\,\tilde\Gamma_{\mbox{free}}[\upvarphi]\,,
\eeq
with
\beq \label{Legendre1}
\tilde\Gamma_{\mbox{free}}[\upvarphi]\,:=\, \tilde W_{\mbox{free}}[\tilde\phi]-\int dt \,d {\bf x}\,
\big(\,\upvarphi\,\tilde\phi^*+\upvarphi^*\tilde\phi\,\big)\,, \qquad \frac{\delta \tilde W_{\mbox{free}}[\phi]}{\delta \tilde\phi^*}\,=\,\upvarphi.
\eeq
Thus, we just demonstrated that, up to a divergent contact term,
the unitary Fermi gas in $d>2$ is related to the ideal Fermi gas via a Legendre transformation
in the large-$N$ limit or, equivalently, in the mean field approximation.

We remark that the intimate relation between the free and unitary fermions in the large $N$ limit gives rise to a simple relation between the scaling dimensions of the dimer field at the two fixed points
\beq \label{scal}
\Delta_{\upphi}^{\mbox{free}}+\Delta_{\upphi}^{\mbox{int}}=d+2.
\eeq
Since in the free theory $\Delta_{\upphi}^{\mbox{free}}=2\Delta_{\psi}=d$, this implies $\Delta_{\upphi}^{\mbox{int}}=2$. The non-trivial fixed point is physically admissible only for $0<d<2$ and $2<d<4$. 
Indeed, for $d>4$ one obtains $\Delta_{\upphi}^{\mbox{int}}=2<\frac{d}{2}$ which violates the unitarity bound. Moreover, in $d=2$ both fixed points merge together ($\Delta_{\upphi}^{\mbox{free}}=2=\Delta_{\upphi}^{\mbox{int}}$), and only the trivial fixed point exists. Remarkably, due to simplicity of the non-relativistic vacuum, the relation \eqref{scal} receives no $1/N$ corrections in the theory of non-relativistic fermions and thus is exact.

From the point of view of the holographic duality, the Legendre transformation corresponds to a change of the boundary condition for the bulk scalar dual to the Cooper-pair field in the same theory in the bulk \cite{Klebanov:1999tb}, in agreement with the comments in \cite{Son:2008ye}. 
More precisely, 
the highest of the two scaling dimensions ($\Delta_{\upphi}^{\mbox{free}}=d$ and $\Delta_{\upphi}^{\mbox{int}}=2$) is denoted $\Delta_+$ and corresponds to an infrared (IR) stable fixed point on the boundary side and to a standard (Dirichlet-like) boundary condition on the bulk side, while the lowest dimension, $\Delta_-$, corresponds to an ultraviolet (UV) stable fixed point and to an exotic (Neumann-like) boundary condition.

We conclude that, 
in the large-$N$ limit, the dimer effective theory of the ideal and the unitary Fermi gases for $0<d<4$ are related via a Legendre transformation and should thus share the same set of conserved currents and symmetries.\footnote{For the interacting system, however, most of these symmetries are expected to be broken by $1/N$ corrections.} For this reason, although we are primarily interested in the unitary Fermi gas in the large N limit, it is sufficient from now on to focus on the theory of the ideal Fermi gas.

\section{Higher symmetries of the Schr\"odinger equation}\label{Symmetry}

\subsection{The Schr\"odinger group of kinematical symmetries}\label{kinematicalsym}

In mathematical terms, the Galilei principle of relativity is encoded in the Galilei group.
For this reason the structure of this group plays an important role in non-relativistic physics \cite{LevyLeblondbook}. In $d$ spatial dimensions the
group acts on the spatial coordinates ${\bf x}$ and time $t$ as
\beq \label{intro1} (t,{\bf x})\to g(t,
{\bf x})=(t+\beta, \mathscr{R} {\bf x}+{\bf v} t+ {\bf a}),
\eeq
where $\beta \in \mathds{R}$; ${\bf v}, {\bf a}\in
\mathds{R}^{d}$ and $\mathscr{R}$ is a rotation matrix in $d$ spatial dimensions. 
In quantum mechanics, the Galilei group acts by projective representations on the Hilbert space of solutions to the Schr\"odinger equation when the potential is space and time translation invariant.\footnote{Of course,
for a single particle such a potential must be constant.} In other words, in such case any solution is transformed to a solution of the form
\begin{equation} \label{intro2bis}
\psi(t,{\bf x})\to 
\gamma\big(g(t,{\bf x})\big) \,\psi\big(g^{-1}(t,{\bf x})\big)\,,
\end{equation}
where $\gamma$ is a phase factor compatible with the group multiplication laws \cite{Weinberg}.
For example, a scalar wave function $\psi$ describing a single particle of mass $m$ transforms under a pure Galilei boost $g_{{\bf v}}$ as
\begin{equation} \label{intro2}
\psi(t,{\bf x})\to \exp \left[ -\frac{im}{2}({\bf v}^2 t- 2\, {\bf v}\cdot {\bf x}) \right]
\psi\big(g^{-1}_{{\bf v}}
(t,{\bf x})\big).
\end{equation}
The presence of the mass-dependent phase factor in the transformation law implies a superselection rule forbidding
the superposition of states of different masses, known as the \textit{Bargmann superselection rule} \cite{Bargmann}.
This rule constrains the dynamics and states that
every term in the Lagrangian of a non-relativistic Galilei-invariant theory must conserve the total mass. 
For this reason, the mass plays the role of a conserved charge in non-relativistic physics.

By enlarging the Galilei group through a central extension, known as the mass operator (or alternatively the particle number operator), we can make the representations unitary \cite{LevyLeblondbook,Weinberg}.
The centrally extended Galilean group is sometimes referred to as the \textit{Bargmann group} \cite{Duval:1984cj}.
Its Lie algebra consists of the following generators: the mass $\hat{M}$; one time
translation $\hat{P}_t$\,; $d$ spatial translations $\hat{P}_{i}$\,; $\frac{d(d-1)}{2}$ spatial rotations $\hat{M}_{ij}$ and $d$ Galilean boosts
$\hat{K}_{i}$\,. The non-trivial commutators are
\begin{equation} \label{NRCFT1}
  \begin{split}
    & [\hat{M}_{ij},\hat{M}_{kl}]=i(\delta_{ik} \hat{M}_{jl}-\delta_{jk}\hat{M}_{il}-\delta_{il} \hat{M}_{jk}+\delta_{jl} \hat{M}_{ik})\,, \\
    & [\hat{M}_{ij},\hat{K}_{k}]=i(\delta_{ik} \hat{K}_{j}-\delta_{jk} \hat{K}_{i})\,, \qquad [\hat{M}_{ij},\hat{P}_{k}]=i(\delta_{ik}\hat{P}_{j}-\delta_{jk} \hat{P}_{i})\,, \\
    & [\hat{P}_i,\hat{K}_j]=-i\delta_{ij}\hat{M}, \qquad [\hat{P}_t,\hat{K}_{j}]=-i\hat{P}_{j}\,.
  \end{split}
\end{equation}
Notice that the commutation relations between the translation and Galilean boost generators are the canonical commutation relations of the Heisenberg algebra $\mathfrak{h}_d$ in $d$ space dimensions (see Appendix \ref{Weylquantization} for the definition),
where the Galilean boost generators play the role of the position operators while the role of the reduced Planck constant is played by the mass.

It is remarkable that the group of space-time symmetries of the free Schr\"odinger equation with vanishing chemical potential
\begin{equation}\label{frSchrequ}
i\, \partial_t\psi(t,\textbf{x}) = -\frac{\Delta}{2m}\,\psi(t,\textbf{x})
\end{equation}
is larger than the Bargmann group if one relaxes the restriction of unit module on the factor appearing in the transformation law.
Following Niederer \cite{Niederer}, we call \textit{kinematical symmetry of the Schr\"odinger equation} any transformation of the form \eqref{intro2bis}, where $\gamma$ is a complex factor compatible with the group structure, that maps solutions to solutions.\footnote{Mathematicians would call such transformations a ``multiplier'' representation of the symmetry group.}

First, remember that the mass is just a charge and so it has scaling dimension zero. Thus, the non-interacting system has no parameter with non-vanishing scaling dimension, which implies an additional scale symmetry.
In non-relativistic physics, this symmetry scales the time
and spatial coordinates differently
\beq \label{NRCFT3a} (t,{\bf x}) \to \left(\frac{t}{\alpha^2}\,,
\frac{\bf x}{\alpha}\right), \qquad \alpha \in \mathds{R}.  \eeq
This corresponds to the dynamical critical exponent $z=2$, which determines the relative scaling of time and space coordinates.

Second, Niederer found in \cite{Niederer} that, in addition to the scale symmetry, a discrete inversion transformation $\Sigma$ which acts on space-time as
\beq \label{NRCFT3b}
(t,{\bf x}) \to \Sigma(t,{\bf x})=\left( -\frac{1}{t}\,, \frac{\bf x}{t}\right)
\eeq
is also a symmetry of the free Schr\"odinger equation.
By conjugating a time translation $g_{\beta}$ via the inversion $\Sigma$, 
\beq \label{NRCFT3ba} \ (t,{\bf x}) \to (\Sigma^{-1}g_{\beta}\Sigma)(t,{\bf x})=\left(\frac{t}{1+\beta t}\,,
\frac{{\bf x}}{1+\beta t}\right)
\eeq
a new symmetry of the free Schr\"odinger equation is found \cite{Niederer,Hagen}. This transformation is known as \textit{expansion} and is a non-relativistic analogue of the special conformal transformations. Note that a Galilean boost $g_{{\bf v}}$ is conjugate to a spatial translation $g_{{\bf a}}$ via the inversion $\Sigma$.

The extension of the Bargmann group by scale transformations and expansions is known as the \textit{Schr\"odinger group} in $d$ spatial dimensions, denoted by $Sch(d)$. Apparently this structure was known already to Jacobi (see the conclusion of \cite{Duval:2009vt}), but was rediscovered after the advent of quantum mechanics in \cite{Niederer,Hagen}. 
The Schr\"odinger group is the non-relativistic counterpart of the conformal group, though the former cannot be obtained as an In\"onu-Wigner contraction from the latter.
The Schr\"odinger group is simply generated by the Euclidean isometries (rotations and spatial translations), the time translations, the scale transformations and the inversion.\footnote{The Galilean boosts and the expansions come ``for free'' (more precisely, via conjugation of the space-time translations by the inversion).}
In addition to \eqref{NRCFT1}, the non-trivial commutators of the Schr\"odinger algebra $\mathfrak{sch}(d)$ in $d$ spatial dimensions are
\begin{equation} \label{NRCFT4}
  \begin{split}
    & [\hat{P}_i,\hat{D}]=i\hat{P}_i\,, \quad [\hat{P}_i,\hat{C}]=-i\hat{K}_i\,, \quad [\hat{K}_i,\hat{D}]=-i\hat{K}_i\,,  \\
    & [\hat{D},\hat{C}]=2i\hat{C}\,, \quad [\hat{D},\hat{P}_t]=-2i\hat{P}_t\,, \quad [\hat{C},\hat{P}_t]=-i\hat{D}\,.  \\
  \end{split}
\end{equation}
Together, the time translation generator $\hat{P}_t$, the scale generator $\hat{D}$ and
the expansion generator $\hat{C}$ span a subalgebra $\mathfrak{sl}(2,\mathbb{R})$ of the full Schr\"odinger algebra.
These generators commute with the generators $\hat{M}_{ij}$ of the rotation subalgebra $\mathfrak{o}(d)$.
The Schr\"odinger algebra has the structure of a semi-direct sum: $\mathfrak{sch}(d) = \mathfrak{h}_d \niplus \big(\mathfrak{o}(d) \oplus \mathfrak{sl}(2,\mathbb{R})\big)$.

Finally, the ``standard'' representation of the Schr\"odinger algebra as differential operators of order one acting on the one-particle wave function $\psi(t,{\bf x})$ is
\begin{equation} \label{NRCFT4a}
\begin{split}
 & \hat{P}_i=-i\partial_i, \qquad \hat{P}_t=i\partial_t, \qquad \hat{M}=m,  \\
 & \hat{M}_{ij}=-i(x_i\partial_j-x_j\partial_i), \\
 & \hat{K}_i=m x_i+i t \partial_i, \\
 & \hat{D}=i\left(2\,t\,\partial_t+x^i\partial_i+\frac{d}{2}\right), \\
 & \hat{C}=i\left(t^2\partial_t+t \Big(x^i \partial_i+\frac{d}{2}\Big)\,\right)+\frac{m}{2}\,x^2.
\end{split}
\end{equation}

\subsection{The Weyl algebra of higher symmetries}\label{Weylhigh}

The algebra of space-time symmetries of the free single-particle Schr\"odinger equation is actually
much larger than the Schr\"odinger algebra.
More precisely, the Weyl algebra (see Appendix \ref{Weylquantization} for the definition) is realised as an infinite-dimensional symmetry algebra of the free Schr\"odinger equation, as was pointed out in the inspiring work \cite{Valenzuela:2009gu}. Here, we further prove that the Weyl algebra is the \textit{maximal} algebra of space-time symmetries of the Schr\"odinger equation. In the present context, this result can be used as
the non-relativistic counterpart of the theorem of Eastwood \cite{Eastwood:2002su} on the maximal symmetry algebra of the massless Klein-Gordon equation (see \textit{e.g.} Section 4 of \cite{Calimanesti} for a review).
Accordingly, the Weyl algebra (and, possibly, its proper matrix-valued extension) provides a non-relativistic higher-spin algebra which is the precise analogue of Vasiliev's (possibly extended) higher-spin algebras \cite{Vasiliev:2003ev}.

\subsubsection{The maximal symmetry algebra of the Schr\"odinger equation}

In order to make precise and rigorous statements analogous to the known results on the conformal scalar field, let us start with some definitions mimicking the ones of \cite{Eastwood:2002su,Calimanesti}.
A \textit{symmetry of the Schr\"odinger equation} 
is a linear differential operator $\hat{A}(t,\hat{\mathbf{X}},\hat{P}_t,\hat{\mathbf{P}})$ obeying
to the condition 
\begin{equation}
\widehat{S}\,\hat{A}\,=\,\hat{B}\,\widehat{S}\,,
\label{symgen}
\end{equation}
for some linear differential operator $\hat{B}$,
where $\widehat{S}$ is the \textit{Schr\"odinger operator} defined by
\begin{equation}\label{Schrop}
\widehat{S}\,:=\,\hat{P}_t \,-\, \hat{H}\,,
\end{equation}
and $\hat{H}$ is a Hamiltonian of a massive non-relativistic particle taking the usual form
\begin{equation}\label{SchrHam}
\hat{H}(\hat{\mathbf{X}},\hat{\mathbf{P}})=\frac{\hat{\mathbf{P}}^2}{2m}\,+\,V(\hat{\mathbf{X}})\,.
\end{equation}
The Schr\"odinger equation reads
\begin{equation}\label{SchrEqu}
i\, \partial_t\psi(t,\textbf{x}) \approx \hat{H}\psi(t,\textbf{x})\quad\Longleftrightarrow \widehat{S}\psi(t,\textbf{x})\approx 0,
\end{equation}
where the ``weak equality'' symbol $\approx$ stands for an equality valid when the Schr\"odinger equation is satisfied.
By definition, any symmetry
$\hat{A}$ preserves the space Ker$\widehat{S}$ of solutions to the Schr\"odinger equation \eqref{SchrEqu}: it maps any solution $\psi$ to a solution $\psi^\prime=\hat{A}\psi$.
The general solution of the Schr\"odinger equation \eqref{SchrEqu} is of course 
\begin{equation}
\psi(t,\textbf{x})\, =\, \hat{U}(t)\,\psi(0,\textbf{x})\,,
\label{timevolpsi}
\end{equation}
where
\begin{equation}
\hat{U}(t)=\exp(- i t\hat{H})
\label{evolutionop}
\end{equation}
is the time evolution operator.
Obviously, the time evolution
\begin{equation}
\hat{F}(t)\,=\,\hat{U}(t)\,\hat{F}(\hat{\mathbf{X}},\hat{\mathbf{P}})\,\hat{U}^{-1}(t)\,=\,\hat{F}\big(\hat{\mathbf{X}}(t),\hat{\mathbf{P}}(t)\big)\,,
\label{timevol}
\end{equation}
of any spatial differential operator $\hat{F}(\hat{\mathbf{X}},\hat{\mathbf{P}})$ defines a symmetry of the Schr\"odinger equation in the above sense.
It is clear that $\hat{F}(t)$ maps solutions to solutions, where the initial wave functions are related by the initial operator $\hat{F}(0)=\hat{F}$. The condition \eqref{symgen} is satisfied with $\hat{A}=\hat{B}=\hat{F}(t)$ since
$i\partial_t \hat{F}(t)=[\hat{H},\hat{F}(t)]$, which follows from \eqref{timevol}. Note that \eqref{timevol} is the inversed ($t\to -t$) time evolution of $\hat{F}(\hat{\mathbf{X}},\hat{\mathbf{P}})$ in the Heisenberg picture.\footnote{Notice that in \cite{Bekaert:2011cu}, the inversed time evolution in the Heisenberg picture was written $\hat{F}(-t)$ in order to emphasise this fact. Here, we chose the simpler notation $\hat{F}(t)$ in order to avoid overloading the many formulas where such notations appear.}

A symmetry $\hat{A}$ is said to be \textit{trivial} if $\hat{A}=\hat{O}\widehat{S}$ for
some linear operator $\hat{O}$ because it maps any solution to zero. 
Such a trivial symmetry is always a symmetry of the Schr\"odinger equation,
since it obeys (\ref{symgen}) with $\hat{B}=\widehat{S}\hat{O}$.
The algebra of trivial symmetries forms a left
ideal in the algebra of linear operators endowed with the
composition $\circ$ as multiplication. Furthermore, it is also a right
ideal in the algebra spanned by all the symmetries of the Schr\"odinger equation.
Two symmetries $\hat{A}_1$ and $\hat{A}_2$ are said to be \textit{equivalent} if they differ by a trivial symmetry. The corresponding 
equivalence relation is denoted by a weak equality
\begin{equation}
\label{equivrel}
\hat{A}_1\approx \hat{A}_2\quad\Longleftrightarrow\quad \hat{A}_1= \hat{A}_2 +\hat{O}\widehat{S}\,.
\end{equation}
The \textit{maximal symmetry algebra of the Schr\"odinger equation} is the complex algebra of all inequivalent symmetries of the Schr\"odinger equation, \textit{i.e.} the algebra of all symmetries quotiented by the two-sided ideal of trivial symmetries. 
Let us show that \textit{for any time-independent Hamiltonian the maximal symmetry algebra of the single-particle Schr\"odinger equation
is isomorphic to the Weyl algebra of spatial differential operators.}\footnote{For an $n$-component wave function, the maximal symmetry algebra of the Schr\"odinger equation is isomorphic to the tensor product between the algebra of $n\times n$ square matrices and the Weyl algebra of spatial differential operators: $M_n\otimes{\cal A}_d$.}

The proof goes in three steps: Let $\hat{A}(t,\hat{\mathbf{X}},\hat{P}_t,\hat{\mathbf{P}})$ be a symmetry of the Schr\"odinger equation.
Firstly, one remarks that it is equivalent to a representative independent of the time translation generator: 
\begin{equation}
\hat{A}(t,\hat{\mathbf{X}},\hat{P}_t,\hat{\mathbf{P}})\approx \hat{A}^\prime(t,\hat{\mathbf{X}},\hat{\mathbf{P}})\,,
\end{equation}
because one may assume that the operator
$\hat{A}$ has been ordered in such a way that all the operators $\hat{P}_t$ are on the right. Thus each $\hat{P}_t$ can be traded for $\hat{H}$ since 
$\hat{P}_t \,\approx\, \hat{H}$. 
Secondly, one observes that the representative $\hat{A}^\prime$ must commute with the Schr\"odinger operator $\widehat{S}$. Indeed, the representative $\hat{A}^\prime$ is also a symmetry, so it must obey to the condition $\widehat{S}\,\hat{A}^\prime=\hat{B}^\prime\,\widehat{S}$ which is equivalent to 
\begin{equation}\label{aprime}
[\widehat{S},\hat{A}^\prime]=(\hat{B}^\prime-\hat{A}^\prime)\,\widehat{S}.
\end{equation}
As follows from the definition \eqref{Schrop} of the Schr\"odinger operator, the left-hand-side of this equation is equal to
\beq
\label{aprimeleft}
[\widehat{S},\hat{A}^\prime]=i\partial_t\hat{A}^\prime-[\hat{H},\hat{A}^\prime]
\eeq
where the time derivative acts on the explicit time dependence of the operator $\hat{A}^\prime(t,\hat{\mathbf{X}},\hat{\mathbf{P}})$.
In order to compare the left and right hand sides of Eq. \eqref{aprime}, let us assume that each side is ordered as before.
On the one hand, the left-hand-side of Eq. \eqref{aprime} is given by the expression \eqref{aprimeleft} which does not depend on $\hat{P}_t$ since both the Hamiltonian $\hat H$ and the representative $\hat{A}^\prime$ do not. On the other hand, the right-hand-side of Eq. \eqref{aprime} explicitly depends on $\hat{P}_t$ due to the presence of the Schr\"odinger operator $\widehat{S}=\hat P_t-\hat H$.
Therefore each side must vanish separately, which means that the commutator between $\hat{A}^\prime$ and $\widehat{S}$ is zero. Thirdly, this commutation relation implies that the representative $\hat{A}^\prime$ is the (inversed) time evolution of a spatial differential operator
\begin{equation}\label{aprime2}
\hat{A}^\prime(t,\hat{\mathbf{X}},\hat{\mathbf{P}})\,=\,\hat{U}(t)\,\hat{A}^\prime(0,\hat{\mathbf{X}},\hat{\mathbf{P}})\,\hat{U}^{-1}(t)\,.
\end{equation}
This becomes clear from the commutation relation \eqref{aprimeleft} 
which is the Schr\"odinger equation in the (time reversed) Heisenberg picture.
\qed

\subsubsection{The Schr\"odinger subalgebra}\label{Schrsub}

As should be expected, the reversed time evolution of the initial observables span all the inequivalent symmetries of \textit{any} Schr\"odinger equation. But how does the Schr\"odinger algebra $\mathfrak{sch}(d)$ fits into this result? And what is so special about the \textit{free} evolution?

A useful observation is that, when the particle is free ($\hat{H}=\hat{H}_{\mbox{free}}=\frac{\hat{P}^2}{2m}$) all the differential operators \eqref{NRCFT4a} are equivalent to polynomials at most of degree two in the time-evolved operators of positions and momenta.
For instance, the mass $\hat{M}=m$ is the degenerate case of degree zero.
Moreover, the time translation generator is equivalent to the quadratic Hamiltonian
$\hat{P}_t\approx\hat{H}_{\mbox{free}}=\frac{\hat{P}^2}{2m}$ and the rotation generators can be written as the angular momentum
$\hat{M}_{ij}=\hat{X}_i\hat{P}_j-\hat{X}_j\hat{P}_i$.
For the other generators, it is easier to first verify this property at time $t=0$.
The Galilean boost generators evaluated at $t=0$ are proportional to the positions, 
$\hat{K}^i\big|_{t=0}=m \hat{X}^i$ while the scale and expansion generators can be written as the quadratic polynomials, 
$\hat{D}\big|_{t=0}=-\hat{X}^i\hat{P}_i+i{d}/{2}$ and $\hat{C}\big|_{t=0}=\frac{m}{2}\,\hat{X}^2.$
All together, these differential operators at $t=0$ provide a unitary representation of the Schr\"odinger algebra on the Hilbert space of initial one-particle wave functions. Therefore, so does the (reversed) time evolutions of these observables for \textit{any} Hamiltonian. However,
the time-dependent operator $\frac{\hat{P}^2(t)}{2m}=\exp(-i\hat{H}t)\frac{\hat{P}^2}{2m}\exp(+i\hat{H}t)$ must be identified with the generator $\hat{P}_t$ in this particular realisation of the Schr\"odinger algebra, but it does not correspond to the genuine Hamiltonian $\hat H$ (except when the particle is free) and thus in general it will not generate the genuine time evolution of the wave function. In other words, the reversed time evolution of the above-mentioned generators of degree at most two are symmetries (in the sense of our definition), they satisfy to the commutation relations of the Schr\"odinger algebra, but they do not have any simple physical interpretation for a generic Hamiltonian.

In general, the transformations generated by the (reversed) time evolution of some observables are not ``kinematical'' \cite{Perroud:1977qh}, in the sense that they do not generate transformations of the form \eqref{intro2bis}.
A kinematical transformation is generated by a first-order linear differential operator (in particular, a mere change of coordinates is
generated by a vector field).
In the following, the first-order symmetries of the Schr\"odinger equation will be called \textit{kinematical symmetries}, while the higher-order symmetries will be denoted by \textit{higher symmetries} (following the usage of mathematicians). 
Note that a higher-order linear differential operator does not
generate a kinematical transformation. This explains why higher symmetries are usually not considered by physicists.
Nevertheless from the mathematical perspective, the Schr\"odinger algebra is always a subalgebra of symmetries of any one-particle Schr\"odinger equation but none of its realisation generate a kinematical representation of the Schr\"odinger group, except for the special cases of potentials determined by Niederer \cite{Niederer:1974ba}. As mentioned above, the simplest case is the free Hamiltonian, where the time evolution of the position and momentum operators is $\hat{\mathbf{X}}(t)=\hat{\mathbf{X}}\,-\,t\,\hat{\mathbf{P}}/m$
and $\hat{\mathbf{P}}(t)=\hat{\mathbf{P}}$.
In such case, the differential operators \eqref{NRCFT4a} can be rewritten in terms of the time evolved positions and momenta, 
\begin{equation} \label{NRCFTt}
\begin{split}
 & \hat{P}_t\approx \frac{\hat{P}^2(t)}{2m}=\frac{\hat{P}^2}{2m}=\hat{H}_{\mbox{free}}\,, \qquad \hat{M}=m,  \\
 & \hat{M}^{ij}=\hat{X}^i(t)\hat{P}^j(t)-\hat{X}^j(t)\hat{P}^i(t), \\
 & \hat{K}^i=m \hat{X}^i(t), \\
 & \hat{D}\approx-\hat{X}^i(t)\hat{P}_i(t)+i\frac{d}{2}, \\
 & \hat{C}\approx\frac{m}{2}\,\hat{X}^2(t).
\end{split}
\end{equation}
Furthermore, a nice observation of \cite{Valenzuela:2009gu,Appell} is that all these symmetries are equivalent to polynomials of degree two in the Galilean boost and translation generators (more precisely, $\hat{M}$ is of degree zero while by definition $\hat{P}$ and $\hat{K}$ are of degree one).
Indeed, one may replace everywhere $\hat{\mathbf{X}}(t)\rightarrow\hat{\mathbf{K}}/m$ and $\hat{\mathbf{P}}(t)\rightarrow\hat{\mathbf{P}}$ to get
\begin{equation} \label{NRCFTKP}
\begin{split}
 & \hat{P}_t\approx \frac{\hat{P}^2}{2m}\,, \\
 & \hat{M}_{ij}=\frac{\hat{K}_i\hat{P}_j-\hat{K}_j\hat{P}_i}{m}\,, \\
 & \hat{D}\approx-\frac{\hat{K}^i\hat{P}_i}{m}+i\frac{d}{2}\,, \\
 & \hat{C}\approx\frac{\hat{K}^2}{2m}\,.
\end{split}
\end{equation}
This implies that \textit{the associative algebra of polynomials in the Galilean boost and translation generators is isomorphic to the maximal symmetry algebra of the free single-particle Schr\"odinger equation.} In more mathematical terms, the realisation of the enveloping algebra ${\cal U}\big(\mathfrak{sch}(d)\big)$ of the Schr\"odinger algebra on the space of solutions to the free one-particle Schr\"odinger equation is isomorphic to the Weyl algebra ${\cal A}_d$ of spatial differential operators.

The proof is straightforward: As was already observed, the Galilean boost and translation generators play in the Schr\"odinger algebra a role equivalent to the positions and momenta in the Heisenberg algebra. Therefore, by themselves they generate algebraically the whole Weyl algebra ${\cal A}_d$ which has been shown to be isomorphic to the maximal symmetry algebra of the Schr\"odinger equation. The other generators of the Schr\"odinger algebra are functions of the Galilean boost and translation generators, so they cannot produce anything extra. \qed

\subsubsection{The maximal symmetry algebra of the Schr\"odinger action}\label{Schractsymalg}

One should scrutinise the issue of Hermiticity of the symmetries. This is important at the level of the action principle and also for the unitarity of the representations.
Let $^\dagger$ stands for the
\textit{spatial Hermitian conjugation} with respect to the spatial Hermitian form
\begin{equation} \label{spHermitform}
\langle\,\psi_1\,\mid\,\psi_2\,\rangle\,:=\,\int d\mathbf{x}\,\psi_1^*(t,\mathbf{x})\,\psi_2(t,\mathbf{x})\,,
\end{equation}
on the Hilbert space $L^2({\mathbb R}^d)$ of square-integrable functions, \textit{e.g.} $(\hat{X}^i)^\dagger= \hat{X}^i$ and $(\hat{P}_i)^\dagger= \hat{P}_i$. As usual, the scalar product \eqref{spHermitform} is time-independent for wave functions $\psi_1$ and $\psi_2$ which are solutions of the Schr\"odinger equation, as in \eqref{timevolpsi}.
The Weyl algebra of quantum observables is the real form of the complex Weyl algebra spanned by the spatial differential operators that are Hermitian. All Schr\"odinger algebra generators \eqref{NRCFTKP} at time $t=0$ are quantum observables. However, notice that the generators \eqref{NRCFT4a} containing a time derivative (\textit{i.e.}
the generators of time translations, scale transformations and expansions) are, in general, not Hermitian with respect to the spatial conjugation. Actually, the spatial conjugate of the time derivative is not well defined since one is not allowed to integrate it by part in \eqref{spHermitform}. The apparent paradox can be solved if one restricts the domain of definition of the generators to wave functions which are solutions of the Schr\"odinger operator, because then the generators are equivalent to the observables \eqref{NRCFTt}.

The spatial Hermitian conjugation can be extend to space-time differential operators. The \textit{space-time Hermitian conjugation} will be denoted by the same symbol $^\dagger$ although it is the Hermitian conjugation with respect to the space-time Hermitian form 
\begin{equation} \label{sptHermitform}
(\,\psi_1\,\mid\,\psi_2\,)\,=\,\int dt \,\langle\,\psi_1\,\mid\,\psi_2\,\rangle\,:=\,\int dt\,d\mathbf{x}\,\psi_1^*(t,\mathbf{x})\,\psi_2(t,\mathbf{x})\,,
\end{equation}
such that $t^\dagger= t$ and $(\hat{P}_t)^\dagger= \hat{P}_t$. 
However, the scale and expansion generators in the standard representation \eqref{NRCFT4a} are not Hermitian with respect to the space-time conjugation, $\hat{D}^\dagger= \hat{D}+2i$ and $\hat{C}^\dagger = \hat{C}+2it\partial_t$. Nevertheless, all the generators are equivalent to Hermitian operators (with respect to both conjugations), when the Schr\"odinger equation is satisfied, as can be seen from \eqref{NRCFTt}.

The \textit{Schr\"odinger action} for a non-relativistic massive field described by the Schr\"odinger equation \eqref{SchrEqu} can be written as the quadratic form
\begin{equation}
S[\psi]\,=\,(\,\psi\mid\widehat{S}\mid\psi\,)\,,
\label{quadract}
\end{equation}
where the Schr\"odinger operator \eqref{Schrop} is Hermitian with respect to the space-time conjugation, $\widehat{S}^\dagger=\widehat{S}$. 
The Euler-Lagrange equation extremising the quadratic action is of course the Schr\"odinger equation \eqref{SchrEqu}.
A \textit{symmetry of the Schr\"odinger action}
is an invertible linear operator $\hat{\cal U}$ preserving the
quadratic form \eqref{quadract}. In other words,
\begin{equation}\label{symmact}
\hat{\cal U}^\dagger\,\widehat{S}\,\hat{\cal U}\,=\,\widehat{S}\,.
\end{equation}
A \textit{symmetry generator of the Schr\"odinger
action} is a linear differential operator $\hat{A}$ which
is self-adjoint with respect to the quadratic form
\eqref{quadract} in the sense that $(\,\psi\mid\widehat{S}\mid\hat A\psi\,)=(\,\hat A\psi\mid\widehat{S}\mid\psi\,)$. More concretely,
\begin{equation}\label{symact}
\widehat{S}\,\hat{A}\,=\,\hat{A}^\dagger\widehat{S}\,.
\end{equation}
Any symmetry generator $\hat{A}$ defines a symmetry
$\hat{\cal U}=e^{i\hat{A}}$ of the Schr\"odinger action.
The \textit{maximal algebra of symmetries of Schr\"odinger action} is the real Lie algebra
of symmetry generators of the quadratic action endowed with $i$
times the commutator as Lie bracket, quotiented by the ideal of trivial symmetries.
One can show that \textit{the Weyl algebra of quantum observables is the maximal symmetry algebra of the Schr\"odinger action.}\footnote{For an $n$-component wave function, the maximal symmetry algebra of the Schr\"odinger action is isomorphic to the tensor product of the algebra of Hermitian $n\times n$ matrices with the Weyl algebra of quantum observables: $\mathfrak{u}(n)\otimes{\cal A}_d({\mathbb R})$.}

The proof goes as follows:
Firstly, any symmetry generator $\hat{A}$ of
the Schr\"odinger action is a symmetry of the Schr\"odinger equation with $\hat{B}=\hat{A}^\dagger$ in
the condition (\ref{symgen}), due to \eqref{symact}. 
Secondly, we have seen previously that any symmetry of the Schr\"odinger equation is equivalent to a representative which is function only of the translation and Galilean boost generators. Such a representative automatically commutes with the Schr\"odinger operator  $\widehat{S}$.
Thirdly, any symmetry of the
quadratic action that commutes with $\widehat{S}$ must be Hermitian with respect to the space-time conjugation, $\hat{A}=\hat{A}^\dagger$, as can be seen from \eqref{symact}.
Consequently, the representative must be Hermitian, \textit{i.e.} a quantum observable. \qed

From the point of view of holography, the precise identification of the maximal algebra of rigid symmetries of the (non-relativistic) CFT is of prime importance since it should correspond to the symmetry transformations preserving the vacuum of the bulk theory, \textit{e.g.} in the usual AdS/CFT the isometry group of $AdS$ is isomorphic to the conformal group
of the boundary. In the generalisation of the holography conjecture of \cite{Klebanov:2002ja,Sezgin:2002rt} to any spacetime dimension, the maximal symmetry algebra of the massless Klein-Gordon action \cite{Eastwood:2002su} is precisely isomorphic to the higher-spin algebra of Vasiliev equations \cite{Vasiliev:2003ev} which appears as the algebra preserving the AdS solution.
The maximal symmetry algebra of the Schr\"odinger action could play an analogous role in a non-relativistic version of higher-spin gravity. This expectation is rather natural given the fact that Vasiliev theory is formulated in a frame-like language (à la Cartan) with a connection one-form taking values in the relativistic higher-spin algebra which can be replaced by its non-relativistic analogue (see next section).

\section{Light-like dimensional reduction of currents}\label{Currents}

\subsection{Bargmann framework} 

To realise geometrically the Schr\"odinger symmetry, we first embed
the Schr\"odinger algebra in $d$ spatial dimensions $\mathfrak{sch}(d)$ into the
relativistic conformal algebra in $d+2$ space-time dimensions
$O(d{+}2, 2)$.
That the Schr\"odinger algebra can be embedded into the relativistic
conformal algebra can be made manifest at the level of the equations of motion.
More concretely, an old trick (the so-called ``Bargmann framework'' \cite{Duval:1984cj,Duval:2009vt,Duval:1990hj}) is the derivation of the free Schr\"odinger equation from the massless Klein-Gordon equation via a Kaluza-Klein reduction along a null direction. 

\subsubsection{Equations of motion: from Klein-Gordon to Schr\"odinger}

Consider the massless Klein-Gordon equation in $d+2$-dimensional 
Minkowski space-time,\footnote{We follow closely \cite{Son:2008ye} (see \textit{e.g.} \cite{Duval:1990hj} for more details on the method of null dimensional reduction).}
\begin{eqnarray}\label{masslessKG}
\Box\Psi(x) \equiv -\partial_0^2\Psi(x) + \sum_{i=1}^{d+1} \partial_i^2\Psi(x) = 0.
\end{eqnarray}
This equation is conformally invariant.  Defining the light-cone
coordinates,
\begin{equation}
  x^\pm = \frac{x^0 \pm x^{d+1}}{\sqrt2}\,,
\end{equation}
the Klein-Gordon equation becomes\footnote{The elements of the metric are defined by $\eta_{+-}\,=\,\eta_{-+}\,=-1; \,\, \eta_{ij}\,=\,1$ and the others are zero.}
\begin{equation}
  \left(-2\frac\partial{\partial x^-} \frac\partial{\partial x^+} 
  + \sum_{i=1}^{d}\partial_i^2\right) \Psi(x) =0.
\label{almostSchr}
\end{equation}
The global coordinates $x^\mu=(x^+,x^-,\textbf{x})$ have minuscule Greek indices which will span $d+2$ values while the spatial coordinates $x^i=(\textbf{x})$ have miniscule latin indices which will span $d$ different values.\footnote{In the sequel, the index will often be left implicit for the space-time coordinates $x^\mu\equiv x$. No ambiguity arises since the spatial coordinates are written $x^i\equiv\textbf{x}$.}
If the relativistic scalar field is assumed to be of the form 
\begin{eqnarray}
 \Psi(x) = e^{-imx^-}\psi(x^+,\textbf{x})\, ,
\label{reduc_dim}
\end{eqnarray}
one can make the identification\footnote{In the same way, we denote $\partial/\partial x^+$ by $\partial_+$.} $\partial/\partial x^-:=\partial_-=-im$.
Then the
equation (\ref{almostSchr}) has the form of the Schr\"odinger equation in free space
\begin{equation}
  \left( 2im \, \partial_+ + \sum_{i=1}^{d}\partial_i^2 \right) \Psi(x) = 0.
\end{equation}
The light-cone coordinate $x^+$ can be identified with the time $t$ ($\partial_+\,=\,\partial_t$ is the time derivative) and the operator $\sum_{i=1}^{d}\partial_i^2$ is the Laplacian operator $\Delta$ in flat space,
\begin{equation}
(2im\,\partial_t\,+\Delta)\Psi(x)=0.
\end{equation}
Thanks to the dimensional reduction \eqref{reduc_dim}, the exponential can be factorised and we obtain the equation of motion for the non-relativistic scalar field \eqref{frSchrequ}.
This equation is invariant under the Schr\"odinger group $Sch(d)$ as was explained in the previous Section.  Since the
original Klein-Gordon equation has conformal symmetry, this means
that $Sch(d)$ is a subgroup of $O(d{+}2,2)$.

\subsubsection{Symmetry algebra: from conformal to Schr\"odinger}\label{confSchro}

Let us now discuss the embedding of the Schr\"odinger algebra into the conformal algebra explicitly, following the discussion in \cite{Son:2008ye}.  The conformal algebra $\mathfrak{o}(d+2,2)$ can be defined by the following commutation relations:
\begin{equation}\label{conform}
\begin{split}
  [\tilde{M}^{\mu\nu},\, \tilde{M}^{\alpha\beta}] 
      &= i(\eta^{\mu\alpha} \tilde{M}^{\nu\beta}
  + \eta^{\nu\beta} \tilde{M}^{\mu\alpha} - \eta^{\mu\beta} \tilde{M}^{\nu\alpha}
  - \eta^{\nu\alpha} \tilde{M}^{\mu\beta} ),\\
  [\tilde{M}^{\mu\nu},\, \tilde P^\alpha] &= i (\eta^{\mu\alpha}\tilde P^\nu -
   \eta^{\nu\alpha} \tilde P^\mu),\\
  [\tilde D,\, \tilde P^\mu] &= -i \tilde P^\mu, \quad 
     [\tilde D,\, \tilde K^\mu] = i \tilde K^\mu, \\
  [\tilde P^\mu,\, \tilde K^\nu] 
     &= -2i(\eta^{\mu\nu} \tilde D + \tilde{M}^{\mu\nu}),
\end{split}
\end{equation}
where Greek indices run from $0$ to $d+1$, and all other commutators are
equal to $0$.  The tilde symbols denote relativistic generators; we reserve
hatted symbols for the non-relativistic operators.  
The conformal algebra generators can be realised  as
differential operators of order one acting on the relativistic scalar field $\Psi(x)$
\begin{equation} \label{conformrealis}
\begin{split}
 & \tilde{P}_\mu=-i\partial_\mu,\quad \tilde{M}_{\mu\nu}=-i(x_\mu\partial_\nu-x_\nu\partial_\mu), \\
 & \tilde{K}_\mu=i\left(2x_\mu \left(x^\nu\partial_\nu+\frac{d}2\right)-x^2\partial_\mu\right), \quad
  \tilde{D}=i\left(x^\mu\partial_\mu+\frac{d}2\right)\,.
\end{split}
\end{equation}

We identify the
light-cone momentum $\tilde P^+=(\tilde P^0+\tilde P^{d+1})/\sqrt2$
with the mass operator $\hat{M}$ in the non-relativistic theory (in agreement with the previous identification $\partial_-=-im$).  We
now select all operators in the conformal algebra that commute with
$\tilde P^+$, \textit{i.e.} which preserve the Kaluza-Klein ansatz \eqref{reduc_dim}.  Clearly these operators form a subalgebra, and one may check that it is the Schr\"odinger algebra $\mathfrak{sch}(d)$ \cite{Sorba}.  The identification is as follows:
\begin{equation}\label{embedding}
\begin{split}
  & \hat{M} = \tilde P^+,\quad \hat{P}_t = \tilde P^-, \quad \hat{P}^i = \tilde P^i,\quad
    \hat{M}^{ij} = \tilde{M}^{ij}, \\ 
  & \hat{K}^i = \tilde{M}^{i+}, \quad \hat{D} = \tilde D + \tilde{M}^{+-}, \quad
  \hat{C} = \frac{\tilde K^+}2\,.
\end{split}
\end{equation}
From Eq.~\eqref{conform}, one finds that the
commutators between the operators \eqref{embedding} are exactly the Schr\"odinger algebra commutators \eqref{NRCFT1} and \eqref{NRCFT4}.
Furthermore, the realisation \eqref{NRCFT4a} follows from \eqref{conformrealis} via the identification \eqref{embedding}.
The maximal symmetry algebra of the massless Klein-Gordon equation \eqref{masslessKG} is the algebra of polynomials 
in the conformal generators \eqref{conformrealis} modulo the equivalence relations following from the Klein-Gordon equation \cite{Eastwood:2002su}.\footnote{The maximal symmetry algebra of the massless Klein-Gordon action was denoted by $\mathfrak{hu}(1/\mathfrak{sp}(2)[d+2,2])$ by Vasiliev in \cite{Vasiliev:2003ev}.}
The maximal symmetry algebra of the free Schr\"odinger equation \eqref{frSchrequ} is the algebra of polynomials 
in the Schr\"odinger generators \eqref{NRCFT4a} modulo the equivalence relations following from the Schr\"odinger equation.
The embedding similar to the one described above actually holds at the level of maximal symmetry algebra, as could be expected: \textit{The maximal symmetry algebra of the free Schr\"odinger equation is isomorphic to the subalgebra of the maximal symmetry algebra of the massless Klein-Gordon equation, that commutes with a translation generator in a fixed light-like direction.}

The proof is direct: The free Schr\"odinger equation is equivalent to a system of two equations: the massless Klein-Gordon equation $\Box\Psi=0$ and the null reduction $\tilde P^+\Psi=m\Psi\,$. Therefore, the maximal symmetry algebra of the Schr\"odinger equation is isomorphic to the maximal symmetry algebra of the previous system of equations. \qed

In other words, the maximal symmetry algebra of the free Schr\"odinger equation is isomorphic to the centraliser of a given light-like translation generator inside the maximal symmetry algebra of the massless Klein-Gordon equation.
Therefore, a polynomial in the conformal generators is equivalent to a polynomial in the Schr\"odinger generators
if and only if it commutes with $\tilde P^+$. Obvious examples are the polynomial in the generators \eqref{embedding} of $\mathfrak{sch}(d)$ which do commute with $\tilde P^+$. A more interesting example of the previous property is the polynomial $\alpha=\tilde K^i\tilde P_i-2\tilde M^{+i}\tilde M_{+i}\,$, quadratic in the generators of $\mathfrak{o}(d+2,2)$. With the help of the commutation relations \eqref{conform}, one can check that $\alpha$ commutes
with $\tilde P^+$. By making use of  \eqref{NRCFT4a} and \eqref{conformrealis}, one further finds that it is equivalent to a polynomial in the generators of $\mathfrak{sch}(d)$: $\alpha\approx \hat{M}^{ij}\hat{M}_{ij}+id\hat{D}+d^2/2$.

\subsection{Generalities on the currents} 


\subsubsection{Currents: from relativistic to non-relativistic ones}

A \emph{relativistic symmetric conserved current} of rank $r\geqslant 1$ is a real contravariant
symmetric tensor field $C^{\mu_1\ldots\mu_r}(x)$ obeying to the conservation law
\begin{eqnarray} \partial_{\mu_1} C^{\mu_1\ldots\mu_r}(x)\approx 0\, , \label{conservation_current}\end{eqnarray}
where the ``weak equality'' symbol $\approx\,$ stands for ``equal on-mass-shell,''
\emph{i.e.} modulo terms proportional to the equations of motion.
A \emph{generating function of relativistic conserved currents} \cite{Bekaert:2009ud} is a real function $C(x;p)$
on space-time phase-space which is (i) a formal power series in the ``momenta'' $p_\mu$ 
\begin{equation}
    C(x;p) \,=\, \sum\limits_{r\geqslant 0}\frac{1}{r!}\,C^{\mu_1\ldots\mu_r}(x)\,p_{\mu_1}\ldots p_{\mu_r}\,,
\label{j}
\end{equation}
and which is (ii) such that
\begin{equation}
    \left(\,\frac{\partial}{\partial
    x^\mu}\,\frac{\partial}{\partial p_\mu}\right)\,C(x;p)\approx 0\,. \label{conservation_general}
\end{equation}
The terminology follows from the fact that all the coefficients of order $r\geqslant 1$
in the power expansion \eqref{j} of the generating function are symmetric tensors which are all conserved, since \eqref{conservation_current} follows from expanding Eq. (\ref{conservation_general}) in power series.
In flat space-time, the indices of the ``momenta'' $p_\mu$ can be raised with the Minkowski metric. Hence, one may define the bilocal function
\begin{eqnarray}
C(x;p)\,=\Psi_1\left(x+\frac i2 \, p\right)\,\Psi_2\left(x-\frac i2 \, p\right)\,,
\label{generating_function}
\end{eqnarray}
which is a generating function of relativistic conserved currents for any pair of functions $\Psi_1$ and $\Psi_2$ satisfying the Klein-Gordon equation, as can be checked by direct computation (\textit{c.f.} \cite{Bekaert:2009ud} for more details).

In order to look for the proper implementation of the Bargmann framework in the case of conserved currents, one should write the conservation law \eqref{conservation_current} of the relativistic conserved currents $C^{\mu_1\ldots\mu_r}(x)$ in the light-cone coordinates, 
\begin{eqnarray}
 \partial_+ C^{+\mu_1\cdots\mu_{r-1}} + \partial_-C^{-\mu_1\cdots\mu_{r-1}} + \partial_iC^{i\mu_1\cdots\mu_{r-1}} \approx 0\,.
\label{conservation_explicit}
\end{eqnarray}
If the components $C^{-\mu_1\cdots\mu_{r-1}}$ of the relativistic currents are independent of $x^-$ or even vanish, then
the relativistic conservation law \eqref{conservation_explicit} embodies a collection of non-relativistic conservation laws of the type (with $s\geqslant r$)
\begin{eqnarray}\label{relconslaw}
\partial_t C^{+i_1\cdots i_{s-1}+\cdots+-\cdots-} + \partial_{i}C^{i\,i_1\cdots i_{s-1}+\cdots+-\cdots-} \approx 0\,.
\end{eqnarray}
since $\partial_+$ is identified with $\partial_t$.
As one can see, the extra light-cone directions with respect to the spatial ones imply that a single relativistic current
actually generates a collection of (not necessarily independent) non-relativistic currents.

By analogy with the relativistic definitions, one will call the following function on space-time phase-space
\begin{eqnarray} \label{rel1}
c(t,\textbf{x}\,;p_t,\textbf{p})\,:=\,C(x^+=t,x^-=0,\textbf{x}\,;p^+=-p_t,p^-=0,\textbf{p})\,
\end{eqnarray}
the \textit{generating function of non-relativistic ``currents''} obtained from the generating function $C(x,p)$ of relativistic currents.
For the bilocal generating function \eqref{generating_function}, the expression \eqref{rel1} together with the dimensional reduction ansatz \eqref{reduc_dim} lead to the following generating function of non-relativistic symmetric ``currents''
\begin{eqnarray}\label{notationai}
c(t,\textbf{x}\,;p_t,\textbf{p})&=&\psi_1\left(t-\frac i2 \,p_t,\textbf{x}+ \frac i2 \,\textbf{p}\right)\,\psi_2\left(t+ \frac i2 \,p_t,\textbf{x}- \frac i2 \,\textbf{p}\right) 
\,.
\end{eqnarray}
The non-relativistic symmetric ``currents'' $c^{(a)\,i_1 \cdots i_b}$ can now be defined from
\begin{eqnarray} \label{notationaj}
c(t,\textbf{x}\,;p_t,\textbf{p})\,=\,\,\sum \limits_{r,s} \frac{1}{r! \,s!}\,c^{(r)\,i_1 \cdots i_s}(t,\textbf{x}) \, p_{i_1} \cdots p_{i_s}\,(p_t)^r\,.
\end{eqnarray}
The word ``current'' is a slight abuse of terminology here since these symmetric tensors $c^{(a)\,i_1 \cdots i_b}$ may not be conserved, even if the tensors $C^{\mu_1\ldots\mu_r}(x)$ are.\footnote{For this reason, to avoid confusion in the following we will call them bilinears.}
For instance, thanks to the dimensional reduction ansatz,
\beq
 \Psi_1(x) = e^{-im_1x^-}\psi_1(x^+,\textbf{x})\, ,\quad  \Psi_2(x) = e^{-im_2x^-}\psi_2(x^+,\textbf{x})\, ,
\label{reduc_dim12}
\eeq
the generating function of relativistic currents can be written as
\begin{eqnarray}
C(x;p)&=&e^{-i(m_1+m_2)x^-+\frac12(m_1-m_2)p^-}C(x^+,x^-=0,\textbf{x}\,;p^+,p^-=0,\textbf{p})\,,
\end{eqnarray}
which is independent of $x^-$ if and only if $m_1+m_2=0$.
Notably the non-relativistic ``currents'' generated by \eqref{notationai} will thus only be conserved when $m_1+m_2=0$.
The explicit expressions of these currents will be given in the next subsection for the cases which are relevant for the present paper.

The symmetric tensor $c^{(r)\,i_1 \cdots i_s}$ of rank $s$ is said to be of level $r$. As explained below in detail on some specific examples, the bilinears of non-vanishing level $r\neq 0$ generated by \eqref{notationai} are not genuinely independent. Indeed, these bilinears contain time derivatives of the field which can be traded for spatial derivatives via the equation of motion. Consequently, one might scrutinise on the generating function 
\begin{eqnarray}\label{notationai0}
c(t,\textbf{x}\,;p_t=0,\textbf{p})&=&\psi_1\left(t,\textbf{x}+ \frac i2 \,\textbf{p}\right)\,\psi_2\left(t,\textbf{x}- \frac i2 \,\textbf{p}\right) 
\,,
\end{eqnarray}
of non-relativistic ``currents'', $c^{(0)\,i_1 \cdots i_s}(t,\textbf{x})$, of vanishing level as can be seen from evaluating \eqref{notationaj} at $p_t=0$. The function \eqref{notationai0} is local in time but bilocal in space.
When $|m_1|=|m_2|$, it can be interpreted physically as a composite field, at instant $t$, made of two particles with the same mass, described respectively by $\psi_1(t,\textbf{x}_1)$ and $\psi_2(t,\textbf{x}_2)$. Accordingly, in \eqref{notationai0} the coordinate $\textbf{x}$ correspond to the center of mass position. 
For $\textbf{x}_1\neq\textbf{x}_2\neq\textbf{x}$, the two bodies have a non-vanishing relative orbital angular momentum which may be reinterpreted as the spin of the two-body composite.
More technically, this reinterpretation corresponds to the decomposition of the generating function in terms of tensor fields $c^{(0)\,i_1 \cdots i_s}(t,\textbf{x})$ of ``spin'' $s$. In fact, considering bilinears of any spin is very natural in the study of general pairing.

\subsubsection{Singlet bilinears}

By analogy with the simplest prescription of Klebanov and Polyakov in \cite{Klebanov:2002ja}, one might focus on the bilinears in the $\psi$ which are singlets of the internal symmetry group, \textit{i.e.} $U(1)\times Sp\,(2N)$ here. For the unitary Fermi gas, however, the Cooper pair is the main object of interest and it is charged under $U(1)$, so one prefers to slightly relax the previous requirement.

One option is to consider all the bilinears which are singlets of $Sp\,(2N)$. Remember that $\psi^A=\psi^{\alpha,a}$ where the indices take values as $\alpha=\,\uparrow,\downarrow$ and $a=1,\dots, N$ while the orthogonal and symplectic metrics are $\delta_{AB}=\delta_{\alpha\beta}\otimes\delta_{ab}$ and $\mathbb{J}_{AB}=\epsilon_{\alpha\beta}\otimes\delta_{ab}$.
Essentially, there are only two independent ways to construct $Sp\,(2N)$-singlets out of two multiplets $\psi_1$ and $\psi_2$ transforming in the fundamental representation of $Sp\,(2N)$: either as the Hermitian form $\psi^\dagger_1\psi_2=\psi^{*A}_1\delta_{AB}\psi^B_2$ of $U(2N)$ or as the symplectic form $\psi_1\mathbb{J}\psi_2=\psi^{A}_1\mathbb{J}_{AB}\psi^B_2$ of $Sp\,(2N,{\mathbb C})$. Only the Hermitian form is invariant under $U(1)$.

The restriction to the $Sp\,(2N)$-invariant sector appears natural for the large-$N$ extension of the Fermi gas but is questionable for the physical ($N=1$) Fermi gas with internal symmetry group $U(2)\cong U(1)\times Sp\,(2)$. Motivated by this remark and the existence of the embedding $U(2)\times O(N) \subset U(1)\times Sp\,(2N)$, one may consider instead the larger sector of flavor (\textit{i.e.} $O(N)$\,) singlet bilinears. Essentially, there is only one way to construct $O(N)$-singlets out of multiplets transforming in the fundamental representation of $O(N)$: via the scalar product. However, this provides three independent $O(N)$-singlets since the multiplets $\psi^\alpha$ are complex: either as the two (up or down) Hermitian forms
$\psi_1^{\alpha}{}^\dagger\psi_2^\alpha=\psi_1^{*\alpha,\, a}\delta_{ab}\psi_2^{\alpha,\, b}$ (no sum on the index $\alpha$) or as the symplectic form $\psi_1\mathbb{J}\psi_2=\psi_1^{\alpha,\,a}\epsilon_{\alpha\beta}\delta_{ab}\psi_2^{\beta,\,b}$. Again, only the Hermitian forms are invariant under $U(1)$. Notice that the two Hermitian forms and the symplectic form together reconstruct the Hermitian form of $U(2)$. This is in agreement with the analogue of the generalised prescription of Klebanov and Polyakov in \cite{Klebanov:2002ja} since one focuses on the bilinears in $\psi$ which are in the adjoint representation of the unitary group $U(2)$, the internal symmetry of the physical unitary Fermi gas.

In both cases, there exists two types of singlet generating functions: the corresponding bilinears are either neutral or charged with respect to the $U(1)$ group associated with mass conservation. The charged bilinears transform in massive representations (of mass $2m$) of the Schr\"odinger algebra, while the neutral bilinears carry massless representations. We refer the reader to Appendix \ref{App:Sch} for a detailed discussion devoted to the unitary irreducible representations (UIRs) of the Schr\"odinger algebra.

\subsection{Singlet bilinears of the symplectic subgroup}  

\subsubsection{Neutral bilinears}

Following the above discussion, we impose that\footnote{The auxiliary relativistic scalar field $\Psi$ that we use here is Grassmann-odd and $(\Psi_1\,\Psi_2)^\dagger\,=\,\Psi_2^\dagger\,\Psi_1^\dagger$.}
\begin{eqnarray}\Psi_1=\Psi^{\dagger} \quad ;\quad \Psi_2=\Psi\end{eqnarray} 
in \eqref{generating_function} such that $m_1=-m$ and $m_2=m$ in order to construct a real current generating function denoted by $J$: 
\begin{eqnarray} \label{generating}
 J(x;p)\,&=&\Psi^\dagger\left(x+\frac i2 \,p\right)\,\Psi\left(x-\frac i2 \,p\right)=\Psi^{A*}\left(x+\frac i2 \,p\right)\delta_{AB}\Psi^B\left(x-\frac i2 \,p\right)\nonumber\\
 &=&\left[\Psi\left(x-\frac i2 \,p\right)\right]^\dagger\,\Psi\left(x-\frac i2 \,p\right)=J^*(x;p) \,.
\end{eqnarray}
This relativistic parent obeys the law of conservation \eqref{conservation_general}.
The corresponding conserved currents, satisfying \eqref{conservation_current} and \eqref{conservation_explicit}, were introduced by Berends, Burgers and vanDam \cite{Berends:1985xx} long time ago and more recently were summarised in a
generating function in \cite{Bekaert:2009ud}. 
Using \eqref{j}, one sees that they take the explicit form:
\begin{eqnarray}
\nonumber
J_{\mu_1 \ldots  \mu_r}(x) \,
&=&\,\left(-\frac i2\right) ^{\,r} \sum\limits_{s=0}^r\, (-1)^{s} \, \dbinom{r}{s}  \, \partial_{(\mu_1} \ldots  \,\partial_{\mu_s}\Psi^{\dagger}(x) \, \partial_{\mu_{s+1}} \ldots \, \partial_{\mu_r)}\Psi(x) \\
&=& \left(-\frac i 2\right)^{\,r}\,\Psi^{\dagger}(x)\,\overleftrightarrow{\partial_{\mu_1}} \ldots  \overleftrightarrow{\partial_{\mu_r}} \Psi(x)
\label{currents}
\end{eqnarray}
where the usual notation $\overleftrightarrow{\partial}$ is defined by
\begin{equation}
\Phi\overleftrightarrow{\partial_\mu}\Psi\,:=\,\Phi({\partial_\mu}\Psi)\,-\,({\partial_\mu}\Phi)\Psi\,.
\nonumber
\end{equation}
The symmetric conserved current \eqref{currents} of rank $r$
is bilinear in the scalar field and contains exactly $r$
derivatives. The currents of odd rank are absent if the field is a real Grassmann-even scalar. 

After expressing the corresponding currents in terms of the non-relativistic field by making use of the dimensional reduction ansatz \eqref{reduc_dim},
\begin{eqnarray}
J_{\underbrace{+...+}_r i_1...i_s \underbrace{-...-}_q}(x) =(-m)^q\,\left(-\frac i2\right)^{r+s}\psi^\dagger(t,\textbf{x})\,\underbrace{\overleftrightarrow{\partial_t}...\overleftrightarrow{\partial_t}}_r \overleftrightarrow{\partial_{i_1}}...\overleftrightarrow{\partial_{i_s}}\psi(t,\textbf{x})\,,
\end{eqnarray} 
one can check that they \textit{do not depend} on $x^-$: $J_{\mu_1 \cdots\, \mu_r} (x) \,=\,J_{\mu_1 \cdots\, \mu_r} (t,\textbf{x})$.
In addition, there is a relation of recurrence $J_{-\mu_1 \cdots \,\mu_r }\,=\,-m\,J_{\mu_1 \cdots \,\mu_r }$.
From the last remark and the equation \eqref{conservation_explicit}, the conservation law of neutral currents becomes:
\begin{eqnarray}
\nonumber
-\partial_+J_{- \mu_1...\mu_{r-1}}(x)+ \partial^iJ_{i \mu_1...\mu_{r-1}}(x)\approx0 \\
\Rightarrow \,m\,\partial_+J_{\mu_1...\mu_{r-1}}(x)+ \partial^iJ_{i \mu_1...\mu_{r-1}}(x)\approx0 \,.
\end{eqnarray}

One can check even more simply all these properties in terms of the generating function.
Due to the definitions \eqref{generating} and \eqref{reduc_dim}, one obtains:
\begin{eqnarray}\label{neutrgenfct}
 J(x;p)=\,e^{-mp^-}j(t,\textbf{x};p_t,\textbf{p})
\end{eqnarray}
where the generating function of non-relativistic neutral currents is 
\begin{eqnarray}\label{jgenfct}
j(t,\textbf{x}\,;p_t,\textbf{p})&=&\psi^\dagger\left(t-\frac i2 \,p_t,\textbf{x}+ \frac i2 \,\textbf{p}\right)\,\psi\left(t+ \frac i2 \,p_t,\textbf{x}- \frac i2 \,\textbf{p}\right).  
\end{eqnarray}
since $p_t=p_-=-p^+\,$.
The conservation law is 
\begin{eqnarray}
\left(-\,\frac{\partial}{\partial x^+}\frac{\partial}{\partial p^-} + \delta^{ij}\frac{\partial}{\partial x^i} \frac{\partial}{\partial p^j}\right)J(x;p)\approx0
\end{eqnarray}
since $J(x;p)$ does not depend on $x^-$,
which becomes 
\begin{eqnarray}\label{nrconslaw}
\left(\,m\,\frac{\partial}{\partial t} + \frac{\partial}{\partial x^i} \frac{\partial}{\partial p_i}\right)j(t,\textbf{x}\,;p_t,\textbf{p})\approx0
\end{eqnarray}
when expressed in terms of the generating function of non-relativistic neutral currents via \eqref{neutrgenfct}.
The neutral non-relativistic conserved currents which are generated as in \eqref{notationaj} read
\begin{eqnarray}
 j^{(r)}_{i_1 \cdots i_s}(t,\textbf{x})=(-1)^r\left( -\frac i2 \right)^{r+s} \psi^\dagger(t,\textbf{x}) \underbrace{\overleftrightarrow{\partial_t}\cdots\overleftrightarrow{\partial_t}}_r\,\overleftrightarrow{\partial_{i_1}}\cdots\overleftrightarrow{\partial_{i_s}}\psi(t,\textbf{x})
\end{eqnarray}
and are related to the relativisitic neutral currents as follows:
\begin{eqnarray}
  J_{\underbrace{+...+}_r i_1...i_s \underbrace{-...-}_q}(x) = (-1)^{r+q}m^q\,j_{i_1\cdots i_s}^{(r)}(t,\textbf{x})\,.
\end{eqnarray}

Let us give few examples in order to make contact with the standard conserved currents of low rank. The ``current'' of rank zero is the number density $n$
\begin{eqnarray}\label{Jrankzero}
J=j^{(0)} = \psi^\dagger(t,\textbf{x}) \psi(t,\textbf{x})= n\,.
\end{eqnarray}
For rank one, the relativistic current is expressed by 
\begin{eqnarray}
 J_\mu(x) = -\frac i2 \Psi^\dagger(x)\,\overleftrightarrow{\partial_\mu}\Psi(x) 
\end{eqnarray}
and it leads to the mass density $\rho$, the energy density 
\beq
\epsilon\,=\,\frac1{2m}\,\partial_i\psi^\dagger\partial^i\psi
\eeq 
and the momentum density $j_i$ (our notations and conventions are as in \cite{Nishida,Son:2008ye}):
\begin{eqnarray}
\left\{
\begin{array}{rl}
J^+\,&=\,m\,j^{(0)}\,=\,m\,\psi^\dagger(t,\textbf{x})\,\psi(t,\textbf{x})\,=\, m\,n \,=\,\rho \\
J^-\,&=\,j^{(1)}\,=\,\frac i2 \,\psi^\dagger(t,\textbf{x})\,\overleftrightarrow{\partial_t}\,\psi(t,\textbf{x})\,\approx\, \epsilon\,-\,\frac1{4m}\Delta n \\
J_i&=j^{(0)}_{\,i}\,=\, -\frac i2 \,\psi^\dagger(t,\textbf{x})\,\overleftrightarrow{\partial_i}\,\psi(t,\textbf{x}) =  j_{\,i}\,.
\end{array}
\right.
\label{neutral_one}
\end{eqnarray}
The relevant law of conservation is the continuity equation: $\partial_t \rho+\partial_ij^{\,i}\approx0\,.$
Notice that the total energy is given by 
\beq
E\,=\,\int d{\bf x}\,\epsilon\approx \int d{\bf x}\,j^{(1)}
\eeq
modulo a boundary term.
For rank two, one obtains:
\begin{eqnarray}
J_{\mu\nu}(x) = -\frac 14\Psi^\dagger(x)\,\overleftrightarrow{\partial_\mu}\overleftrightarrow{\partial_\nu}\Psi(x) 
\end{eqnarray}
which leads to
\begin{eqnarray}
\left\{
\begin{array}{rl}
J^{++}\,&=\,m^2 j^{(0)}\,=\,m^2\,\psi^\dagger(t,\textbf{x})\,\psi(t,\textbf{x})\,=\,m^2\,n\,=\,m\,\rho\\
\label{+-}
J^{+-}\,&=\,m\,j^{(1)}\,=\,\frac i2 \,m\,\psi^\dagger(t,\textbf{x})\,\overleftrightarrow{\partial_t}\,\psi(t,\textbf{x})
\,\approx\,m\,\epsilon\,-\,\frac1{4m}\Delta n\\
J^+_i\,&=m\,j^{(0)}_{\,i}\,=\,-\frac i2 m\, \psi^\dagger(t,\textbf{x})\,\overleftrightarrow{\partial_i}\,\psi(t,\textbf{x})\,=\,m\,j_i\\
J^{--}\,&= \,j^{(2)}\,=\, - \frac 14 \psi^\dagger(t,\textbf{x}) \overleftrightarrow{\partial_t}\,\overleftrightarrow{\partial_t} \psi(t,\textbf{x})\\ 
J^{-}_i\,&=\,j^{(1)}_{\,i}\,=\,\frac 14 \psi^\dagger(t,\textbf{x})\,\overleftrightarrow{\partial_t}\,\overleftrightarrow{\partial_i}\,\psi(t,\textbf{x})
\,=\,m\,j_{\,i}^\epsilon+\frac14\partial_i\partial_t n\\
J_{ij}\,&=\,j^{(0)}_{ij}\,=\, -\,\frac 14 \psi^\dagger(t,\textbf{x})\, \overleftrightarrow{\partial_i}\,\overleftrightarrow{\partial_j}\,\psi(t,\textbf{x})
 \,=\,m\,\Pi_{ij} - \, \frac 14 (\,\partial_i \partial_j \,-\, \delta_{ij} \,\Delta)n\,\label{Pij}
\end{array}
\right.
\end{eqnarray}
where  
\beq
j_i^\epsilon=-\frac1{2m}(\partial_t\psi^\dagger\partial_i\psi+\partial_i\psi^\dagger\partial_t\psi)
\eeq is the  energy current 
and 
\begin{eqnarray}\label{PiSon}
\Pi_{ij} \, =\frac 1{2m} (\partial_i \psi^\dagger \, \partial_j\psi \,+\, \partial_j\psi^\dagger\,\partial_i\psi)\,-\,\frac 1{4m} \,\delta_{ij}\,\Delta n 
\end{eqnarray}
is the stress tensor in the conventions of \cite{Nishida,Son:2008ye}. The conserved currents $j^{(0)}_{ij}$ and $\Pi_{ij}$ are physically equivalent since they differ only by a trivially conserved current.
The supplementary laws of conservation are :
\begin{eqnarray}
\left\{
\begin{array}{l}
\partial_t \epsilon + \partial_i j^{\epsilon\,i}\approx0\,, \\
\partial_t j^i +\partial_j \Pi^{ij} \approx0 \,.
\end{array}
\right.
\end{eqnarray}

\subsubsection{Charged bilinears}

In order to construct the second type of currents which are singlets bilinears of $Sp\,(2N)$, one chooses
\begin{eqnarray}
\Psi_1\,=\,\Psi_2\,=\,\Psi
\end{eqnarray}
and the components are contracted by the symplectic matrix $\mathbb{J}/2$.
The generating function of such charged currents is denoted by $K$ and given by
\begin{eqnarray}
K (x;p) \,&=&\, \frac12\,\Psi^A \left(x+ \frac i2\, p\right)\,\mathbb{J}_{AB} \, \Psi^B\left(x-\frac i2 \,p\right)\\
\nonumber
&=&\sum\limits_{r\geqslant 0}\frac{1}{r!}\,K^{\mu_1\ldots\,\mu_r}(x)\,p_{\mu_1}\ldots p_{\mu_r}\,.
\end{eqnarray}
Notice that it is an even function in the momenta, $K(x;p)=K(x,-p)$, thus only relativistic charged currents of even rank are non-vanishing.
It leads to the relativistic charged  currents 
\begin{eqnarray}
K_{\mu_1 \ldots  \,\mu_r}(x) \,
&=&\,
\frac12\,\left(-\frac i2\right) ^{\,r} \sum\limits_{s=0}^r\, (-1)^{s} \, \dbinom{r}{s}  \, \mathbb{J}_{AB}\,\partial_{(\mu_1} \ldots  \,\partial_{\mu_s}\Psi^A(x) \, \partial_{\mu_{s+1}} \ldots \, \partial_{\mu_r)}\Psi^B(x)
\nonumber\\
&=& \frac12\,\left(-\frac i 2\right)^{\,r}\,\mathbb{J}_{AB}\,\Psi^A(x)\,\overleftrightarrow{\partial_{\mu_1}} 
\ldots  \overleftrightarrow{\partial_{\mu_r}} \Psi^B(x)\,.
\label{charged_currents_explicit}
\end{eqnarray}
Like the neutral currents, the relativistic charged currents are conserved. However, the corresponding charged non-relativistic bilinears are not conserved, because the relativistic ones depend on $x^-$. 
Indeed,
\begin{eqnarray}\label{chgdgenfct}
K (x;p) \,=\,e^{-2imx^-}\,k(t,\textbf{x}\,;p_t,\textbf{p})\,.
\end{eqnarray}
As one can see, the generating function in this case does not depend on $p^-$. 
Therefore the conservation law becomes 
\begin{eqnarray}\label{pscons}
\left(-\,\frac{\partial}{\partial x^-}\frac{\partial}{\partial p^+} + \delta^{ij}\frac{\partial}{\partial x^i} \frac{\partial}{\partial p^j}\right)K(x;p)\approx0\,.
\end{eqnarray}

The generating function of non-relativistic charged bilinears is:
\begin{eqnarray}\label{kgenfct}
k(t,\textbf{x}\,;p_t,\textbf{p})\,&=&\,\frac12\,\psi^A\left(t-\frac i2 p_t, \textbf{x}+\frac i2 \textbf{p}\right)\,\mathbb{J}_{AB} \, \psi^B\left(t+\frac i2\, p_t, \textbf{x}-\frac i2 \,\textbf{p}\right)\,. 
\end{eqnarray}
It is not conserved but nevertheless satisfies
\begin{eqnarray}\label{pseudoconslaw}
 \left(-2im \, \frac{\partial}{\partial p_t}+\frac{\partial}{\partial x^i}\frac{\partial}{\partial p_i}\right)k(t,\textbf{x}\,;p_t,\textbf{p}) \approx 0\,,
\end{eqnarray}
as follows from \eqref{chgdgenfct}-\eqref{pscons}.
The non-relativistic charged bilinears read 
\begin{eqnarray}
 k^{(r)}_{i_1 \cdots i_s}(t,\textbf{x})=\frac{(-1)^r}2\left( -\frac i2 \right)^{r+s} \mathbb{J}_{AB} \,\psi^A(t,\textbf{x}) \underbrace{\overleftrightarrow{\partial_t}\cdots\overleftrightarrow{\partial_t}}_r\,\overleftrightarrow{\partial_{i_1}}\cdots\overleftrightarrow{\partial_{i_s}}\psi^B(t,\textbf{x})
\end{eqnarray}
and are related to the relativistic charged currents as follows:
\begin{equation}
K_{\underbrace{+...+}_r i_1...i_s}(x) = (-1)^{r}e^{-2imx^-}\,k_{i_1\cdots i_s}^{(r)}(t,\textbf{x})\,.
\end{equation}
The non-relativistic charged bilinears satisfy
\begin{eqnarray} \label{pseudo}
2im\,k^{(r+1)}_{ 
i_1
\cdots
i_s
}(t,\textbf{x})\,+\,\partial^j k^{(r)}_{j 
i_1 
\cdots 
i_s
}(t,\textbf{x})\approx 0 \,.
\end{eqnarray}

For rank zero, one gets the Cooper pair \eqref{kCooper}
\beq
K(x^-=0)\,=\,k^{(0)}\,=\,\frac12\,\psi^A(t,\textbf{x})\,\mathbb{J}_{AB}\,\psi^B(t,\textbf{x})=k\,.
\eeq
For charged bilinears of rank two, one finds:
\begin{eqnarray}
\left\{
\begin{array}{rl}
K^{++}(x^-=0)\,&=\,\,0\\
K^{+-}(x^-=0)\,&=\,\,0\\
K^{+i}(x^-=0)\,&=\,\,0\\
K^{--}(x^-=0)\,&=\,k^{(2)}(t,\textbf{x})=\, - \frac 18\mathbb{J}_{AB}\, \psi^A(t,\textbf{x}) \overleftrightarrow{\partial_t}\,\overleftrightarrow{\partial_t} \psi^B(t,\textbf{x})\\ 
K_i^-(x^-=0)\,&=\,k^{(1)}_i(t,\textbf{x})= + \frac{1}{8} \,
\mathbb{J}_{AB}\,\psi^A(t,\textbf{x})\,\overleftrightarrow{\partial_t}\,\overleftrightarrow{\partial_i}\,\psi^B(t,\textbf{x})\\
K_{ij}(x^-=0)\,&=\,k^{(0)}_{ij}(t,\textbf{x})= \,-\frac{1}{8}\,\mathbb{J}_{AB}\,\psi^A(t,\textbf{x})\, \overleftrightarrow{\partial_i}\,\overleftrightarrow{\partial_j}\,\psi^B(t,\textbf{x}) \,.
\end{array}
\right.\label{rlk}
\end{eqnarray}
These bilinears are not conserved but instead obey:
\begin{eqnarray}
\left\{
\begin{array}{rl}
\partial_ik^{(1)i} &\approx 2im\,k^{(2)} \\
\partial_ik^{(0)ij} &\approx 2im\,k^{(1)j} \,.
\end{array}
\right.
\end{eqnarray}

\subsubsection{Traceless condition} \label{traceless}

Since the massless scalar fields are conformally symmetric,
one may expect to get infinitely many \emph{traceless} conserved currents,
while the Berends-Burgers-vanDam currents generated from \eqref{generating_function} are not traceless,
even on-shell: $\partial_p^2\,C(x;p)\not\approx0$\,.
From the representation point of view, it is important that the relativistic currents are traceless in order to have irreducible conformal primary fields.
The massless Klein-Gordon equations for $\Psi_1$ and $\Psi_2$ imply the conservation condition,
$(\partial_x\cdot\partial_p)\,C(x;p)\approx0$ for the bilocal generating function \eqref{generating_function}, as well as another on-shell condition:
\begin{equation}
    \left(-\partial_p^2+\frac{1}{4}\,\partial_x^2\right) C(x;p)\approx 0\,,
    \label{2nd on-shell}
\end{equation}
which relates trace of the Berends-Burgers-vanDam currents to their d'Alembertian.
For example, Eq. \eqref{2nd on-shell} at $p=0$ for the generating function of neutral currents reads
\beq\label{trlaw}
\eta_{\mu\nu}J^{\mu\nu}=2\eta_{+-}J^{+-}+\delta_{ij}J^{ij}\approx \frac14\,\Box J\,,
\eeq
which relates the trace of the rank-two current $J^{\mu\nu}$ to the d'Alembertian of the scalar $J$.
The relativistic Eq. \eqref{trlaw}
leads to the non-relativistic relation
\beq\label{trlike}
-2mj^{(1)}+\delta_{ij}j^{(0)ij}\approx \frac14\,\Box j^{(0)}\,,
\eeq
which, in turn, gives
\beq
-2\epsilon+\delta_{ij}\Pi^{ij}\approx -\frac{d}{4m}\,\Delta \,n\,,
\eeq
due to \eqref{Jrankzero}, \eqref{Pij} and \eqref{PiSon}. This implies the standard relationship between the total energy and the pressure valid both for ideal and unitary Fermi gases \cite{Nishida,Son:2008ye}:
\beq
\int d{\bf x}\,\Pi^i{}_i\,\approx\,2\,E\,,
\eeq
modulo a boundary term. 
Notice that the analogue of the relativistic Eq. \eqref{trlaw} for the charged currents leads to the non-relativistic relation
\beq\label{trlike2}
\delta_{ij}k^{(0)ij}\approx\,\frac14\,(4m\,i\partial_t+\Delta) \,k\,,
\eeq
as can be checked using \eqref{rlk}.

Due to the second on-shell condition \eqref{2nd on-shell}, one can construct
a generating function $\bar C(x;p)$ of relativistic currents that are conserved and traceless on-shell \cite{Bekaert:2010ky}:
\begin{equation}
    \partial_p^2\,\bar C(x;p)\approx0\,,\qquad (\partial_x\cdot\partial_p)\,\bar C(x;p)\approx0\,.
    \label{traceless conservation}
\end{equation}
This can be achieved by acting with a differential operator
$\mathcal{P} _{d+2}(p,\partial_{x})$ on the generating function of
currents
\beq
\bar C(x;p)=\mathcal{P}_{d+2}(p,\partial_{x})\,C(x;p)\,.
\eeq 
The conservation of both $\bar C$ and $C$ requires that $\mathcal{P}_{d+2}$
commutes with $\partial_x\cdot\partial_p$ on-shell. If we
construct $\mathcal{P}_{d+2}$ as a power series in the transversal projector
$\pi(p,\partial_{x}):=[p^2\,\partial_x^2-(p\cdot\partial_x)^2]/4$\,,
then the conservation condition is satisfied since
$\partial_x\cdot\partial_p\,\pi=\pi\,\partial_x\cdot\partial_p$\,. The
tracelessness condition, 
\m{\partial_p^2\,\mathcal{P}_{d+2}(p,\partial_{x})\,C(x;p) \approx 0}\, can be solved recursively and the 
operator $\mathcal{P}_{d+2}$ is determined by these
conditions (up to a constant factor) \cite{Bekaert:2010ky} : 
 \begin{equation}
    \mathcal{P}_{d+2}(p,\partial_{x}):= \sum_{n=0}^\infty\, \frac{1}{n!\,
    (-p\cdot\partial_{p}-\frac{d-3}{2})_{n}}\left(\frac{1}{4}\,\pi(p,\partial_x)\right)^n\,,
    \label{Poli}
\end{equation}
where $(a)_{n}=\Gamma(a+n)/\Gamma(a)$ is the Pochhammer
symbol. More concretely, if one applies this formula to the currents of spin two, it leads to the traceless current:
\begin{eqnarray}
\bar{C}_{\mu\nu}(x) = C_{\mu\nu}(x) \,+\,\frac{1}{4(d+1)}\,(\partial_\mu\partial_\nu\,-\,\eta_{\mu\nu} \Box)C(x)\,. 
\label{C_bar}
\end{eqnarray}

Due to \eqref{neutrgenfct}, one can express the action of the two operators $p\cdot\partial_p$ and $\pi$ on the neutral current generating function as
\begin{eqnarray}
(p\cdot\partial_p)J(x;p) \,&=&\, \left(p^+\,\frac{\partial}{\partial p^+}\,-\,m\,p^-\,+\,p^i\,\frac{\partial}{\partial p^i}\right) J(x;p)\, ,\\
\pi J(x;p)\,&=&\frac 14\left[\Big(p_ip^i-2p_-p_+\Big)\Delta-\left(p^+\frac{\partial}{\partial x^+}+p^i\,\frac{\partial}{\partial x^i}\right)^2\right]J(x;p)\,,
 \end{eqnarray}
since $J(x;p)$ does not depend on $x^-$.
This is helpful for writing the neutral traceless current generating function $J$ leading, after evaluating at $p^-=0$, to the non-relativistic generating function 
\begin{eqnarray} \label{bar}
\nonumber
\bar j(t,\textbf{x}\,;p_t,\textbf{p}) &=& \sum_{n=0}^\infty\, \frac{1}{n!\,4^{2n}
    (-p_t\,\frac{\partial}{\partial p_t}\,-\,p^i\,\frac{\partial}{\partial p^i} -\frac{d-3}{2})_{n}} \times \\
&&\times \left(\,(p_ip^i)\Delta\,-\,\left(-p_t\,\partial_t\,+\,p^i\,\partial_i \right)^2\right)^n j(t,\textbf{x}\,;p_t,\textbf{p})\,.
\end{eqnarray}
Notably, this function generates currents which satisfy the non-relativistic version of the traceless condition
\beq\label{trlikecond}
\left(-2m\frac{\partial}{\partial p_t}\,+\,\delta^{ij}\frac{\partial}{\partial p^i}\frac{\partial}{\partial p^j}\right)
\bar j(t,\textbf{x}\,;p_t,\textbf{p})\,\approx\,0\,.
\eeq 
For instance, for rank two we get a simple relation 
\beq 
-2m\bar j^{(1)}+\delta_{ij}\bar j^{(0)}_{ij}\approx 0 
\eeq 
to be contrasted with \eqref{trlike}.
Notice that this shows that the higher-level $r>0$ neutral currents ${\bar j}^{(r)}_{\dots}$ are proportional to traces of currents of level zero ${\bar j}^{(0)}_{\dots}$\,.

The formula analogous to \eqref{bar} for the charged bilinears is very similar
\begin{eqnarray}
 \nonumber
\bar k(t,\textbf{x}\,;p_t,\textbf{p}) &=& \sum_{n=0}^\infty\, \frac{1}{n!\,4^{2n}
    (-p_t\,\frac{\partial}{\partial p_t}\,-\,p^i\,\frac{\partial}{\partial p^i} -\frac{d-3}{2})_{n}} \times \\
&&\times \left(\,(p_ip^i)(-4im\,\partial_t+\Delta)\,-\,\left(-p_t\,\partial_t\,+\,p^i\,\partial_i \right)^2\,\right)^n k(t,\textbf{x}\,;p_t,\textbf{p})\,.
\end{eqnarray}
Notice that, since all the components $K^{+\cdots}$ vanish, the charged non-relativistic bilinears are spatially traceless:
$\delta_{ij}\bar k^{(a)ij \cdots}\approx 0$ to be contrasted with \textit{e.g.} \eqref{trlike2}. Remarkably, the generating function $\bar k$ gives rise to the non-relativistic spatially traceless tensors $\bar k^{(0)}_{i_1\dots i_r}$ which are actually non-relativistic conformal primary fields\footnote{For a definition of a non-relativistic conformal primary field see Appendix \ref{App:Sch}.} (such as the scalar Cooper-pair field) while the higher-level ones $k^{(r)}_{i_1\dots i_r}$ for $r>0$ are their descendants as can be seen from Eq. \eqref{pseudo}.

\subsection{Singlet bilinears of the orthogonal subgroup}  

Since the $Sp\,(2N)$-singlet bilinears have been investigated above in much detail and the $O(N)$-singlet bilinears are their natural extension, the presentation of the latter bilinears will be brief.

The neutral relativistic currents are now split in up and down ones, as one chooses in \eqref{generating_function}
\begin{eqnarray}
\Psi_1\,=\,(\Psi^{\alpha})^\dagger\,,\quad\Psi_2\,=\,\Psi^\alpha
\end{eqnarray}
with $\alpha=\,\uparrow\, , \downarrow$ and the $O(N)$-flavor components are contracted by the identity matrix.
The generating functions of such neutral relativistic currents are denoted by ${\cal J}^\alpha$,
\begin{eqnarray}
{\cal J}^\alpha(x;p)\,&=&\Psi^{\alpha \dagger}\left(x+\frac i2 \,p\right)\,\Psi^\alpha\left(x-\frac i2 \,p\right)=\Psi^{\alpha,a*}\left(x+\frac i2 \,p\right)\delta_{ab}\Psi^{\alpha,b}\left(x-\frac i2 \,p\right)\nonumber\\
 &=&\left[\Psi^\alpha\left(x-\frac i2 \,p\right)\right]^\dagger\,\Psi^\alpha\left(x-\frac i2 \,p\right)={\cal J}^{\alpha*}(x;p) \, ,
\end{eqnarray}
where there is no sum over the index $\alpha$.

For the charged currents, one chooses
\begin{eqnarray}
\Psi_1\,=\,-\Psi_{\downarrow}\,,\quad\Psi_2\,=\,\Psi_{\uparrow}
\end{eqnarray}
and the $O(N)$-vector components are again contracted by the identity matrix.
The generating function of such charged relativistic currents will be denoted by $\cal K$, \textit{e.g.}
\begin{eqnarray}
{\cal K}(x;p) \,&=&\, -\Psi_\downarrow^a \left(x+ \frac i2\, p\right)\,\delta_{ab} \, \Psi_\uparrow^b\left(x-\frac i2 \,p\right)=\Psi_\uparrow^a \left(x- \frac i2\, p\right)\,\delta_{ab} \, \Psi_\downarrow^b\left(x+\frac i2 \,p\right)\,.
\end{eqnarray}
Notice that the analogous generating function with up and down subscripts exchanged is not independent, more precisely it is equal to $-{\cal K}(x;-p)$.

We will not write explicitly the corresponding non-relativistic bilinears and generating functions $\texttt{j}_\alpha(t,\textbf{x}\,;p_t,\textbf{p})$ and $\texttt{k}(t,\textbf{x}\,;p_t,\textbf{p})$, since all the corresponding formulas are the straightforward analogues of the ones in the previous subsections. We just notice that the scalar bilinears $\texttt{j}_\alpha(t,\textbf{x}\,;p_t=0,\textbf{p}=\textbf{0})=n_\alpha(t,\textbf{x})$ are the density fields of the up and down fermions, while $\texttt{k}(t,\textbf{x}\,;p_t=0,\textbf{p}=\textbf{0})=k(t,\textbf{x})$ denotes the complex Cooper-pair field.
Two real fields and one complex field precisely match the entries of a $2\times2$ Hermitian matrix. For instance,
at rank and level zero
\beq\left(
\begin{array}{cc} 
-\texttt{j}^{(0)}_{\,\uparrow} & \texttt{k}^{(0)}\\
\texttt{k}^{(0)*} & \texttt{j}^{(0)}_{\,\downarrow}
\end{array} 
\right)=\left(
\begin{array}{cc} 
-\psi^*_\uparrow\cdot\psi_\uparrow & \psi_\uparrow\cdot\psi_\downarrow\\
\psi^*_\downarrow\cdot\psi^*_\uparrow & \psi^*_\downarrow\cdot\psi_\downarrow
\end{array} 
\right)=\Uppsi^\alpha\cdot\Uppsi^{*\beta}\,.
\eeq
This collection of $O(N)$-singlet bilinears of all ranks and levels appears to be very natural for our proposal of the gravity dual of the unitary Fermi gas \cite{Bekaert:2011cu}.

\section{Coupling to background fields}\label{Sources} 

The generating functional $W_{\mbox{free}}[\,h,\upvarphi\,;N]$ of connected correlators of $Sp\,(2N)$-singlet bilinears in the non-interacting Fermi gas described by the quadratic action
\beq\label{freeact33}
S_{\mbox{free}}[\,\psi\,;N]:=S[\,\psi\,;c_0=0,N]=
\int dt \,d {\bf x}\,  \psi^\dagger\left(i\partial_t+\frac{\Delta}{2m}+\mu \right)\psi\,,
\eeq
is defined by the path integral
\beq\label{W1extsp}
\exp i\,W_{\mbox{free}}[\,h\,,\upvarphi\,;N]\,=\,\int{\cal D}\psi{\cal D}\psi^\dagger\,\,\exp i \,S_{\mbox{free}}[\,\psi\,,h\,,\upvarphi\,;N]\,,
\eeq
where
\begin{eqnarray}
\label{S1extbckgd}
&&S_{\mbox{free}}[\,\psi\,,h\,,\upvarphi\,;N]:=S_{\mbox{free}}[\,\psi\,;N]\,\\
&&\qquad-\sum\limits_{r,s\geqslant0}\,\frac
1{r! \,s!}
\int dt\, d {\bf x}\,\big(j^{(r)i_1\cdots \,i_s}\,h_{i_1\cdots\, i_s}^{(r)}+k^{(r)i_1\cdots \,i_s*}\upvarphi_{i_1\cdots\, i_s}^{(r)}+k^{(r)i_1\cdots \,i_s}\upvarphi_{i_1\cdots\, i_s}^{(r)*}\big) \nonumber
\end{eqnarray}
is the free action
in the presence of $Sp\,(2N)$-invariant external tensor fields, $h_{i_1\cdots \,i_s}^{(r)}$ and
$\upvarphi_{i_1\cdots i_s}^{(r)}$, coupled respectively to the neutral and charged bilinears, $j^{(r)i_1\cdots\, i_s}$ and $k^{(r)i_1\cdots\, i_s}$.
In other words, the $Sp\,(2N)$-invariant bilinears are minimally coupled to the background fields which share the same properties, \textit{i.e.}
all $h_{i_1\cdots \,i_s}^{(r)}$ are real and $\upvarphi_{i_1\cdots i_s}^{(r)}$ are complex and vanish for odd rank $s$. 
Here and below, we will refrain from writing explicitly the similar formulas for the $O(N)$-singlet bilinears $\texttt{j}_\alpha^{(r)i_1\cdots\, i_s}$ and $\texttt{k}^{(r)i_1\cdots\, i_s}$ coupling respectively to the background fields $h_{i_1\cdots \,i_s}^{(r)\alpha}$ and
$\upvarphi_{i_1\cdots i_s}^{(r)}$ for all ranks. 
The collection of such fields will also be referred to as $h$ and $\upvarphi$ for short in order to cover the general case at once. The $Sp\,(2N)$-invariant background correspond to the particular case: $h_\uparrow=h_\downarrow$ and momentum-even $\upvarphi$ generating functions. 

The functional \eqref{S1extbckgd} is quadratic in the dynamical field $\psi$ (since the kinetic term and the bilinears are), therefore
the path integral \eqref{W1extsp} can easily be evaluated formally since it is a Gaussian integral. In order to write the generating functional of connected correlators in a compact form, one should start by writing \eqref{S1extbckgd} manifestly as a quadratic form. This can be done elegantly via the Weyl quantisation (reviewed in Appendix \ref{Weylquantization}) performed on the space-time phase-space, following the same procedure as in the relativistic case \cite{Bekaert:2009ud,Bekaert:2010ky}. In other words, the canonical commutation relations
\eqref{ccr} must be supplemented by $[\,\hat{P}_t\,,\,\hat{T}\,\,]\,=\,i $, where $\hat{T}$ denotes the operators corresponding to multiplication by the time coordinate $t$.\footnote{If not specified, the notations and definitions in this Section are the straightforward extension of the ones in Appendix \ref{Weylquantization}.}

Let us stress that all the steps performed in the subsection \ref{Legendreconj} can be adapted to apply in the presence of background tensor fields as well, because the external fields of non-vanishing rank do not play any role in these specific manipulations (only the scalar fields such as the Cooper pair and the dimer are pertinent in that case). In other words, the interacting and the non-interacting Fermi gases in the presence of background fields are still related, in the mean field approximation, by a Legendre transformation over the (properly shifted and/or rescaled) scalar charged dimer field. 

\subsection{Quadratic functional} \label{quadraticfunctional}

The free action \eqref{freeact33} in the absence of background can of course be written as a Schr\"odinger action \eqref{quadract}
\beq\label{freehphi0}
S_{\mbox{free}}[\,\psi\,;N]\,=\,(\psi\mid\hat{S}_{\mbox{free}}\mid\psi)\,=\,\delta_{AB}\,(\psi^A\mid\hat{S}_{\mbox{free}}\mid\psi^B)\,,
\eeq
where the operator
\beq\label{Hamfreediffop}
\hat{S}=\hat{P}_t-\hat{H}_{\mbox{free}}\,,
\eeq
is the Schr\"odinger operator \eqref{Schrop} for the free Hamiltonian $\hat{H}_{\mbox{free}}=\mathbf{\hat P}^2/2m$.
The crucial observation of this Section is that even the minimal coupling terms in \eqref{S1extbckgd} can be explicitly written 
as a quadratic functional via integrations by part. Let us perform this rewriting in the generic case, \textit{i.e.} let us consider the following minimal coupling
\begin{eqnarray}
\label{mincpling}
\sum\limits_{r,s\geqslant0}\,\frac
1{r! \,s!}
\int dt\, d {\bf x}\,c^{(r)}_{i_1\cdots \,i_s}(t,{\bf x})\,f^{(r)i_1\cdots\, i_s}(t,{\bf x})
\end{eqnarray}
between a collection of external symmetric tensor fields $f_{i_1\cdots\, i_s}^{(r)}$ and the non-relativistic bilinears 
\begin{eqnarray}
 c^{(r)}_{i_1 \cdots i_s}(t,\textbf{x})&=&(-1)^r\left( -\frac i2 \right)^{r+s} \,\psi_1(t,\textbf{x}) \underbrace{\overleftrightarrow{\partial_t}\cdots\overleftrightarrow{\partial_t}}_r\,\overleftrightarrow{\partial_{i_1}}\cdots\overleftrightarrow{\partial_{i_s}}\psi_2(t,\textbf{x})\nonumber\\
&=&\frac1{2^{r+s}} \,\psi_1(t,\textbf{x}) \underbrace{\overleftrightarrow{{\hat P}_t}\cdots\overleftrightarrow{{\hat P}_t}}_r\,\overleftrightarrow{\mathbf{\hat P}_{i_1}}\cdots\overleftrightarrow{\mathbf{\hat P}_{i_s}}\psi_2(t,\textbf{x})
\end{eqnarray}
defined by \eqref{notationai}-\eqref{notationaj}.
The main idea is to integrate by parts all momentum operators acting on $\psi_1$ inside \eqref{mincpling}, in order to have all operators acting on $\psi_2$. One may convince oneself that taking into account the ordering and the change of signs will result in the equality
\begin{eqnarray}
\label{mincpling2}
&&\int dt\, d {\bf x}\,c^{(r)}_{i_1\cdots \,i_s}(t,{\bf x})\,f^{(r)i_1\cdots\, i_s}(t,{\bf x})\\
&&\qquad=
\frac1{2^{r+s}}
\int dt\, d {\bf x}\,\,\psi_1(t,\textbf{x})\,\{\,\cdots\,\{f^{(r)i_1\ldots i_s}(\hat{\mathbf{T}},\hat{\mathbf{X}})\,,\,\hat{P}_t\}\,,\,\cdots\,,\, \hat{P}_t\}\,,\,\hat{\mathbf{P}}_{i_1}\}\,,\,\cdots\,,\, \hat{\mathbf{P}}_{i_s}\}
\,\psi_2(t,\textbf{x})
\nonumber
\end{eqnarray}
where $\{\,\,\,,\,\,\}$ denotes the anticommutator and implicitly $r$ operators $\hat{P}_t$ appear in the above formula.
Therefore, the minimal coupling \eqref{mincpling} can be rewritten in a compact form as the quadratic functional
\begin{eqnarray}
\label{mincpling2a}
\sum\limits_{r,s\geqslant0}\,\frac
1{r! \,s!}
\int dt\, d {\bf x}\,c^{(r)}_{i_1\cdots \,i_s}(t,{\bf x})\,f^{(r)i_1\cdots\, i_s}(t,{\bf x})\,=\,(\psi_1^*\mid\hat{F}\mid\psi_2)
\end{eqnarray}
where the curly bra-ket notation for the space-time Hermitian form has been introduced in \eqref{sptHermitform}
and the space-time differential operator $\hat{F}$ is given by
\beq\label{WsymF}
\hat{F}(\hat{T},\hat{\mathbf{X}};\hat{P}_t,\hat{\mathbf{P}})
\,=\,
\sum\limits_{r,s\geqslant0}\,\frac1{r! \,s!\,2^{r+s}}
\{\,\cdots\,\{f^{(r)i_1\ldots i_s}(\hat{\mathbf{T}},\hat{\mathbf{X}})\,,\,\hat{P}_t\}\,,\,\cdots\,,\, \hat{P}_t\}\,,\,\hat{\mathbf{P}}_{i_1}\}\,,\,\cdots\,,\, \hat{\mathbf{P}}_{i_s}\}\,.
\eeq
As explained in Appendix \ref{Weylquantization}, this means that the generating function
\beq\label{genfctsymb}
f(t,\textbf{x}\,;p_t,\textbf{p})\,=\,\,\sum \limits_{r, s} \frac{1}{r! \,s!}\,f^{(r)\,i_1 \cdots i_s}(t,\textbf{x}) \, p_{i_1} \cdots p_{i_s}\,(p_t)^r\,
\eeq
of symmetric tensor fields is the Weyl symbol of the operator \eqref{WsymF}.

Therefore, one finds that
the free action in the presence of $Sp\,(2N)$-invariant background fields, \textit{i.e.} \eqref{S1extbckgd}, can be written manifestly as a quadratic form
\beq\label{freehphi}
S_{\mbox{free}}[\,\psi\,,h\,,\upvarphi\,;N]\,=\,\delta_{AB}\,(\psi^A\mid\hat{S}\mid\psi^B)\,+\,\frac12\,\mathbb{J}_{AB}\,\Big[\,(\psi^A\mid\hat{\upvarphi}\mid\psi^{B*})\,-\,(\psi^{A*}\mid\hat{\upvarphi}^\dagger\mid\psi^B)\,\Big]\,,
\eeq
where the operator $\hat{S}$ is the Schr\"odinger operator \eqref{Schrop}
\beq\label{Schdiffop}
\hat{S}=\hat{P}_t-\hat{H}=\hat{S}_{\mbox{free}}-\hat{H}_{\mbox{int}}\,,
\eeq
defined in terms of the Hamiltonian
\beq\label{Hamintdiffop}
\hat{H}=\hat{H}_{\mbox{free}}+\hat{H}_{\mbox{int}}\,.
\eeq
The operators $\hat{H}_{\mbox{int}}$ and $\hat{\upvarphi}$
are the images under the Weyl map of the generating functions of the background fields $h(t,\textbf{x}\,;p_t,\textbf{p})$ and $\upvarphi(t,\textbf{x}\,;p_t,\textbf{p})$ respectively.

More generally, 
the free action in the presence of $O(N)$-invariant background fields can be written as follows:
\begin{eqnarray}\label{freehphiON}
S_{\mbox{free}}[\,\psi\,,h\,,\upvarphi\,;N]&=&\,(\psi_\uparrow\mid\hat{S}_\uparrow\mid\psi_{\uparrow})+(\psi_{\downarrow}\mid\hat{S}_\downarrow\mid\psi_\downarrow)\nonumber\\
&&+\,(\psi_\uparrow\mid\hat{\upvarphi}\mid\psi_{\downarrow}^*)+(\psi_{\downarrow}^*\mid\hat{\upvarphi}^\dagger\mid\psi_\uparrow)\,,
\end{eqnarray}
where the flavor indices have been left implicit and the two (up and down) Schr\"odinger operators $\hat{S}_\alpha$ are built from the corresponding interaction Hamiltonians $h_\alpha(t,\textbf{x}\,;p_t,\textbf{p})$.

Let us elaborate on some physical interpretations of this rewriting by concentrating first on the simplest case where there is no coupling to the charged fields ($\upvarphi=0$). 
As can be seen from \eqref{freehphi}, the free action in the presence of only $U(1)\times Sp\,(2N)$-invariant background fields can be rewritten as
a Schr\"odinger action \eqref{quadract} where the Hamiltonian is of the form \eqref{Hamintdiffop}, \textit{i.e.} the usual potential term $V(t,\textbf{x})$ is replaced by a general function on space-time phase-space $h(t,\textbf{x}\,;p_t,\textbf{p})$. In particular, a scalar background field $h(t,\textbf{x})$ coupling to the particle density $n(t,\textbf{x})$ can obviously be interpreted as a position- and time-dependent external potential term in a standard Schr\"odinger action.

In the more general case where the charged sources are present, another suggestive way of interpreting \eqref{freehphi}-\eqref{freehphiON} is by casting it in the Nambu-Gor'kov form.
In order to write \eqref{freehphi} in terms of the Nambu-Gor'kov field \eqref{Nambu}, it is necessary to perform  integrations by part in the term $(\psi_{\downarrow}\mid\hat{S}_\downarrow\mid\psi_\downarrow)$ of \eqref{freehphiON}. This can be formalised by introducing the operation $^\tau$ defined by $\hat{F}^\tau(\hat{T},\hat{\mathbf{X}};\hat{P}_t,\hat{\mathbf{P}}):=\hat{F}(\hat{T},\hat{\mathbf{X}};-\hat{P}_t,-\hat{\mathbf{P}})$ such that
\beq
\label{bypart}
(\,\psi_1\mid\hat{F}\mid\psi_2\,)\,=\,-\,(\,\psi_2^*\mid\hat{F}^\tau\mid\psi^*_1\,)\,.
\eeq
Notice that the minus sign in \eqref{bypart} arises because the fundamental fields are Grassmann odd
and the complex conjugation appears in accordance to the definition of the space-time Hermitian form \eqref{sptHermitform}.\footnote{  Mathematically, the operation $^\tau$ is a linear antiautomorphism of the Weyl algebra.
The operation $^\tau$ must be contrasted with the Hermitian conjugation $^\dagger$ which is an antilinear antiautomorphism obeying to
$ (\psi_1\mid\hat{F}\mid\psi_2)\,=\,(\psi_2\mid\hat{F}^\dagger\mid\psi_1)^*$\,.
}
The fact that the neutral (charged) $Sp\,(2N)$-invariant generating function is a real (respectively, momentum-even) function translates into the fact that the operator $H_{\mbox{int}}$ (resp. $\hat{\upvarphi}$) is Hermitian: $\hat{H}^\dagger_{\mbox{int}}=\hat{H}_{\mbox{int}}$ (resp. $\tau$-symmetric: $\hat{\upvarphi}^\tau=\hat{\upvarphi}$). 
The latter properties together with \eqref{bypart} imply the following relations
\begin{eqnarray}
\delta_{AB}\,(\psi^A\mid\hat{S}\mid\psi^B)
&=&\delta_{ab}\,(\psi_\uparrow^a\mid\hat{S}\mid\psi_\uparrow^b)\,-\,\delta_{ab}\,(\psi_\downarrow^{a*}\mid\hat{S}^\tau\mid\psi_\downarrow^{b*})\,,
\label{rell1}\\
\mathbb{J}_{AB}\,(\psi^A\mid\hat{\upvarphi}\mid\psi^{B*})
&=&2\,\delta_{ab}\,(\,\psi_\uparrow^a\mid\hat{\upvarphi}\mid\psi_\downarrow^{b*})\,,
\label{rell3}\\
\mathbb{J}_{AB}\,(\psi^{A*}\mid\hat{\upvarphi}^\dagger\mid\psi^B)
&=&-\,2\,\delta_{ab}\,(\,\psi_\downarrow^{a*}\mid\hat{\upvarphi}^\dagger\mid\psi_\uparrow^b)\,.
\label{rell2}
\end{eqnarray}
The relations \eqref{rell3}-\eqref{rell2} show that \eqref{freehphi} is indeed a particular case of \eqref{freehphiON} (remember that for the $Sp\,(2N)$-invariant background $\hat{S}_{\uparrow}=\hat{S}_{\downarrow}=\hat{S}$).
More generally, the properties of the $O(N)$-invariant generating functions translate into  $H^\dagger_\alpha=H_\alpha$.
The relation \eqref{rell1}
allows to rewrite the quadratic functional \eqref{freehphiON} in the compact form of a Schr\"odinger action in terms of the Nambu-Gor'kov field \eqref{Nambu}
\beq\label{freephi2}
S_{\mbox{free}}[\,\Uppsi\,,h\,,\upvarphi\,;N]
\,=\,
(\,\Uppsi\mid\hat{\cal S}\mid\Uppsi\,)
\,=\,\int dt\, d {\bf x}\, \Uppsi^\dagger\,\left(
\begin{array}{cc} 
\hat{S}_\uparrow & \hat{\upvarphi}\\
\hat{\upvarphi}^\dagger & -\hat{S}_\downarrow^\tau
\end{array} 
 \right)\,\Uppsi\, ,
\eeq
where the Schr\"odinger operator is the $2\times 2$ matrix
\beq\label{2x2Schr}
\hat{\cal S}\,=\,\left(
\begin{array}{cc} 
\hat{S}_\uparrow & \hat{\upvarphi}\\
\hat{\upvarphi}^\dagger & -\hat{S}_\downarrow^\tau
\end{array} 
 \right)\,.
\eeq
This suggestive rewriting is one of the main results of this Section, because it allows many further insights.
The Schr\"odinger matrix-operator $\hat{\cal S}=\hat{\cal S}_{\mbox{free}}-\hat{\cal H}_{\mbox{int}}$
is the difference of the free Schr\"odinger $2 \times 2$ matrix-operator 
\beq
\hat{\cal S}_{\mbox{free}}\,=\,\left(
\begin{array}{cc}
 \hat P_t\,-\frac{\hat P^2}{2m} & 0\\
0& \hat P_t\,+\frac{\hat P^2}{2m}
\end{array}
 \right)
\eeq
and the interaction Hamiltonian
\beq\label{2x2Schrop}
\hat{\cal H}_{\mbox{int}}\,=\,\left(
\begin{array}{cc} 
\hat{H}_{\uparrow\,\mbox{int}} & \hat{\upvarphi}\\
\hat{\upvarphi}^\dagger & -\hat{H}_{\downarrow\,\mbox{int}}^\tau
\end{array} 
 \right)
\eeq
containing the background fields.
As one can see, the free action in the presence of general background fields can be rewritten in a form which generalises
\eqref{free2a} in the sense that, in the $2\times 2$ matrix, the free Schr\"odinger operators $i\partial_t\pm(\frac{\Delta}{2m}+\mu)$ on the diagonal are replaced by the most general ones and the field $\upvarphi$ is replaced by a general differential operator $\hat{\upvarphi}$. 
Notice that the Schr\"odinger matrix-operator \eqref{2x2Schr} is Hermitian with respect to the simultaneous combination of matrix and space-time Hermitan conjugations. For notational simplicity, this operation will also be denoted by $^\dagger$ since no ambiguity arises. This Hermiticity property of \eqref{2x2Schr} can be made manifest in terms of Pauli matrices:
\beq\label{2x2SchrPauli}
\hat{\cal S}\,=\,i\partial_t\, \sigma_0 -\hat{\cal H}\,,\qquad \hat{\cal H}=\hat{H}_0\,\sigma_0+\hat{H}_1\,\sigma_1+\hat{H}_2\,\sigma_2+\hat{H}_3\,\sigma_3\,,
\eeq
since the coefficients
\beq\label{pr1}
\hat{H}_0=\frac12(\hat{H}_\uparrow-\hat{H}_\downarrow^\tau)\,,\quad\hat{H}_1=-\frac12(\hat{\upvarphi}+\hat{\upvarphi}^\dagger)\,,\quad\hat{H}_2=-\frac{i}2(\hat{\upvarphi}-\hat{\upvarphi}^\dagger)\,,\quad\hat{H}_3=\frac12(\hat{H}_\uparrow+\hat{H}_\downarrow^\tau)\,,
\eeq 
are all space-time Hermitian operators.
It is important to stress that in the particular case of a $Sp\,(2N)$-invariant background the operators $\hat{H}_i$ ($i=1,2,3$) are
$\tau$-symmetric while $\hat{H}_0$ is a $\tau$-antisymmetric operator:
\beq\label{pr2}
\hat{H}_0^\tau=-\hat{H}_0\,,\quad\hat{H}_i^\tau=\hat{H}_i\,,\quad(i=1,2,3)\,.
\eeq
More generally, in the presence of an $O(N)$-invariant background the free action takes the form of a Schr\"odinger action
with the most general $2\times2$ Hermitian matrix-operator.
As we demonstrate in the following, these differences between $Sp\,(2N)$- and $O(N)$-invariant backgrounds play an important role in the correct identification of the gauge symmetry algebra and of a putative dual bulk spectrum.

The generating functional \eqref{W1extsp} of connected correlators of singlet bilinears in the non-interacting Fermi gas
can now be evaluated formally due to the quadratic form of \eqref{freephi2}:
\beq\label{free5}
W_{\mbox{free}}[\,h,\upvarphi\,;N]
\,=\, -i
N\,\mbox{Tr}\log\hat{\cal S}\,=:\,N\,W_{\mbox{free}}[\,h,\upvarphi\,]
\,
\eeq
where $\hat{\cal S}$ is given by \eqref{2x2Schr}.
A crude but standard (BCS theory) approximation of such a complicated object would be to evaluate it in the case where the background fields are constant in space-time and momentum coordinates (in which case only the correlators of the number-density and of the Cooper-pair are evaluated). Another possible approximation is the assumption that the background fields are weak in which case one might start a perturbative expansion in powers of the background fields along the lines of \cite{Bekaert:2010ky}. Notice that the trace in the functional \eqref{free5} corresponds to an integral over the energy and momentum flowing along the fermion loop. This functional can be obtained as a light-like dimensional reduction from its higher-dimensional relativistic counterpart by fixing, in the integral over the corresponding relativistic momentum, one of the light-like component to be equal to $m$ instead of integrating over it.

Finally, since the Schr\"odinger matrix-operator $\hat{\cal S}$ is Hermitian, it is formally diagonalisable via a generalised unitary Bogolioubov transformation $\Uppsi\mapsto \Uppsi^{\,\prime}=\hat{\cal U}^{-1}\Uppsi$ , in the sense that $\hat{\cal S}^\prime=\hat{\cal U}^\dagger\hat{\cal S}\hat{\cal U}=(i\partial_t+\hat{H}^\prime_0)\,\sigma_0+\hat{H}^\prime_3\,\sigma_3$. In general, the operators $\hat{H}^\prime_0$ and $\hat{H}^\prime_3$ depend on both background fields $h$ and $\upvarphi$.
In terms of the new quasi-particle field $\Uppsi^{\,\prime}$, the quadratic form \eqref{freephi2} can be written as a sum of two Schr\"odinger actions:
\beq\label{freehphi6}
S_{\mbox{free}}[\,\psi^\prime\,,h\,,\upvarphi\,;N]\,=\,
\sum\limits_{\alpha=\uparrow,\,\downarrow}(\Uppsi^{\alpha\prime}\mid\hat{S}^\prime_\alpha\mid\Uppsi^{\alpha\prime})\,.
\eeq
Physically, this means that the free action in the presence of background fields describes (up and down) quasi-particles governed respectively by two Hamiltonian operators depending on both background fields $h$ and $\upvarphi$.
Again, this is nothing but a natural generalisation of the BCS theory.

\subsection{Gauge and rigid symmetries}

This subsection is devoted to the analysis of the gauge 
symmetries of the free classical action (in the presence of background fields) and of the corresponding effective action. Due to the simple expression of these actions (respectively, ``quadratic form'' and ``trace-log''), their symmetries are manifest. These symmetries are important because, as usual, the gauge invariance of the effective action encodes the Ward-Takahashi identities (here, on the connected correlators of bilinears).
The algebraic structure and physical interpretation of these symmetries will be addressed in more details in the next subsection.

Note that any quadratic functional such as \eqref{freephi2} is formally invariant if a transformation,
\beq\label{transfos1}
\Uppsi\longrightarrow \hat{\cal U}^{-1}\,\Uppsi\,,
\eeq
of the field $\Uppsi$ in the fundamental representation of \emph{invertible} matrix-operators $\hat{\cal U}^{-1}$ is compensated by a suitable transformation,
\beq\label{transfos2}
\hat{\cal S}\longrightarrow \hat{\cal U}^{\dagger}
\hat{\cal S}\,\hat{\cal U}\,,
\eeq
of the Hermitian Schr\"odinger matrix-operator \eqref{2x2Schr}. These finite transformations of $\hat{\cal S}$ correspond to gauge transformations of the background fields, as will be shown explicitly below. Physically, this means that the group of invertible $2\times 2$ matrix-operators can be interpreted as the group of gauge symmetries of the free classical action $S_{\mbox{free}}[\,\psi\,,h\,,\upvarphi\,;N]$
in the presence of a general $O(N)$-invariant background.
The corresponding infinitesimal transformations span the Lie algebra 
of $2\times 2$ matrix-operators. This Lie algebra of infinitesimal gauge symmetries is nothing else but the complex algebra $M_2({\mathbb C})\otimes {\cal A}_{d+1}({\mathbb C})$, \textit{i.e.} the tensor product of the algebra $M_2$ of $2\times 2$ matrices and the Weyl algebra ${\cal A}_{d+1}$ of space-time operators (both algebras are over $\mathbb C$).

On the other hand, any trace functional such as \eqref{free5} is formally invariant under the subgroup of \emph{unitary} matrix-operators ($\hat{\cal U}^{\dagger}=\hat{\cal U}^{-1}$), because the Schr\"odinger matrix-operator $\hat{\cal S}$ transforms in the adjoint representation
\beq\label{transfos3}
\hat{\cal S}\longrightarrow \hat{\cal U}^{-1}
\hat{\cal S}\,\hat{\cal U}\,,
\eeq
of this subgroup.
The generating functional $W_{\mbox{free}}[\,h,\upvarphi\,;N]$
of connected correlators arises from integrating out the fundamental fields $\Uppsi$. More precisely, it arises from one-loop diagrams for the fermions and it can be interpreted as the background effective action of the free theory.
Physically, the symmetries \eqref{transfos3} of the $O(N)$-invariant background effective action $W_{\mbox{free}}[\,h,\upvarphi\,;N]$ can be interpreted as the subset of gauge symmetries of the classical action which remain manifestly preserved at quantum level. The other transformations are in general anomalous because the trace in \eqref{free5} is only invariant under the adjoint transformation \eqref{transfos3}, hence not always under \eqref{transfos2}.\footnote{However, since the trace in \eqref{freephi2} implicitly requires a regularisation in order to be well defined, notice that its finite or its logarithmically divergent parts may admit more symmetries than the full regularised effective action (\textit{c.f.} \cite{Bekaert:2010ky} for more comments in the relativistic case).}
As one can see, formally the group of unitary matrix-operators may always be preserved at quantum level in the present construction. The corresponding algebra of infinitesimal transformations is the real Lie algebra of Hermitian $2\times2$ matrix-operators. As was explicitly shown in Eq. \eqref{2x2SchrPauli}, this real algebra is spanned by the linear combinations of sigma matrices with coefficients in the real Weyl algebra, hence it isomorphic to $\mathfrak{u}(2)\otimes {\cal A}_{d+1}({\mathbb R})$, \textit{i.e.} the tensor product 
of the algebra $\mathfrak{u}(2)$ of Hermitian $2\times2$ matrices and the Weyl algebra of Hermitian operators (both algebras are over $\mathbb R$).\footnote{In more abstract terms, the algebra $M_2({\mathbb C})\otimes {\cal A}_{d+1}({\mathbb C})$ is ${\mathbb Z}_2$-graded with respect to the eigenvalues $\pm 1$ of the Hermitian conjugation $^\dagger$. A real form of this complex algebra is the subalgebra of Hermitian $2\times 2$ matrix-operators (elements of eigenvalue $+1$).}

In order to describe the gauge symmetries \eqref{transfos2} more explicitly, let us consider infinitesimal transformations near the identity: $\hat{\cal U}=\hat{1}+i\hat{A}$ where the 
infinitesimal generator $\hat{A}$ is a general $2\times 2$ matrix-operator expressed in the form
\beq \label{generator}
\hat{A}\,=\,\left(
\begin{array}{cc}
\hat a_{\uparrow} & \hat b\\
\hat c & -\hat a_{\downarrow}^\tau
\end{array}
 \right)\,.
\eeq
The space-time operators $\hat a_{\uparrow}$, $\hat a_{\downarrow}$, $\hat b$ and $\hat c$ are infinitesimal gauge parameters.
The infinitesimal version of \eqref{transfos2} now reads
\beq
\delta \hat{\cal S}\,=\, i\,(\,\hat{\cal S}\,\hat{A}\,-\,\hat{A}^{\dagger}\,\hat{\cal S})\,.
\eeq
Since the free Schr\"odinger matrix-operator $\hat{\cal S}_{\mbox{free}}$ is kept fixed in the variation of the total Schr\"odinger matrix-operator $\hat{\cal S}=\hat{\cal S}_{\mbox{free}}-\hat{\cal H}_{\mbox{int}}$, one obtains
$\delta \,\hat{\cal S}\,=\,-\,\delta \hat{\cal H}_{\mbox{int}}\,=\,i\,(\,\hat{\cal S}\,\hat{A}\,-\,\hat{A}^{\dagger}\,\hat{\cal S}\,)$ which decomposes as
\beq\label{gtransfo}
\delta \hat{\cal H}_{\mbox{int}}\,=\,i\,(\,\hat{A}^{\dagger}\,\hat{S}_{\mbox{free}}\,-\,\hat{S}_{\mbox{free}}\,\hat{A} \,) \,+\, 
i\,(\,\hat{\cal H}_{\mbox{int}}\,\hat{A}\,-\,\hat{A}^{\dagger}\,\hat{\cal H}_{\mbox{int}} \,)
\,.
\eeq
Although the term of degree one in $\hat{\cal H}_{\mbox{int}}$ in Eq. \eqref{gtransfo} is crucial for having exact symmetries of the action, for the sake of simplicity in the following subsection we will concentrate on the term of degree zero in order to discuss the interpretation of the gauge symmetries.

In terms of the corresponding Weyl symbols, the transformation \eqref{gtransfo} reads
\begin{eqnarray}\label{gtransfinfsymb}
\delta {\cal H}(t,\textbf{x}\,;p_t,\textbf{p})
&=&i\,\Big(\,{\cal A}^*(t,\textbf{x}\,;p_t,\textbf{p})\,\star\,{\cal S}_{\mbox{free}}\,-\,{\cal S}_{\mbox{free}}\,\star\,{\cal A}(t,\textbf{x}\,;p_t,\textbf{p}) \,\Big)\\
&&\,+\, 
i\,\Big(\,{\cal H}(t,\textbf{x}\,;p_t,\textbf{p})\,\star\,{\cal A}(t,\textbf{x}\,;p_t,\textbf{p})\,-\,{\cal A}^*(t,\textbf{x}\,;p_t,\textbf{p})\,\star\,{\cal H}(t,\textbf{x}\,;p_t,\textbf{p})\,\Big)\,,\nonumber
\end{eqnarray}
where
\beq\label{calH}
{\cal H}(t,\textbf{x}\,;p_t,\textbf{p})\,=\,\left(
\begin{array}{cc}
 h_{\uparrow}(t,\textbf{x}\,;p_t,\textbf{p}) & \upvarphi(t,\textbf{x}\,;p_t,\textbf{p})\\
 \upvarphi(t,\textbf{x}\,;p_t,\textbf{p}) & - h_{\downarrow}(t,\textbf{x}\,;-p_t,-\textbf{p})
\end{array}
 \right)
\eeq
is the Weyl symbol of the interaction Hamiltonian matrix-operator $\hat{\cal H}_{\mbox{int}}$,
\beq\label{calA}
{\cal A}(t,\textbf{x}\,;p_t,\textbf{p})\,=\,\left(
\begin{array}{cc}
 a_{\uparrow}(t,\textbf{x}\,;p_t,\textbf{p}) &  b(t,\textbf{x}\,;p_t,\textbf{p})\\
 c(t,\textbf{x}\,;p_t,\textbf{p}) & - a_{\downarrow}(t,\textbf{x}\,;-p_t,-\textbf{p})
\end{array}
 \right)
\eeq
is the Weyl symbol of the infinitesimal matrix-operator $\hat A$,
and
$\star$ stands for the Moyal product on the space-time phase-space (\textit{c.f.} Appendix \ref{Weylquantization}) defined by
\beq
\,\star\, =\,\,\exp\left[\,\frac{i}{2}\,\left(-\frac{\overleftarrow{\partial}}{\partial t}\,\frac{\overrightarrow{\partial}}{\partial p_t}+\frac{\overleftarrow{\partial}}{\partial p_t}\,\frac{\overrightarrow{\partial}}{\partial t}+\frac{\overleftarrow{\partial}}{\partial x^i}\,\frac{\overrightarrow{\partial}}{\partial p_i}-\frac{\overleftarrow{\partial}}{\partial p_i}\,\frac{\overrightarrow{\partial}}{\partial x^i}\right)\right]\,,
\label{sptiMoyalproduct}
\eeq
where the left and right arrows indicate on which side the corresponding derivative acts.
The above Weyl symbols \eqref{calH}-\eqref{calA} should be interpreted as generating functions of symmetric tensor fields via the corresponding analogue of the power series expansion in momenta \eqref{genfctsymb}. 
In other words, the infinitesimal gauge transformation
\eqref{gtransfinfsymb} can be written explicitly in terms of tensor fields only but the resulting expression would be rather complicated in complete generality. For the sake of simplicity, in the following subsection this will be done only to the lowest zeroth order in the background fields.

What is the relation of the gauge symmetries of the free action in the presence of background fields and the rigid symmetries
of the Schr\"odinger action investigated in Section \ref{Symmetry}?
As can be seen from the conditions \eqref{symmact} and \eqref{symact} defining, respectively, the symmetries of the Schr\"odinger action and their generators, they can be seen as gauge symmetries of the free action preserving the background fields, \textit{e.g.} $\delta\hat{\cal H}_{\mbox{int}}=0$.
In the absence of any background field ($h=\upvarphi=0\leftrightarrow \hat{\cal H}_{\mbox{int}}=0$), the classical action \eqref{S1extbckgd} reduces to the free Schr\"odinger action \eqref{freeact33}. 
Therefore the symmetries of the free Schr\"odinger action can be seen as the subalgebra of gauge symmetries that preserve the absence of background fields. 
The maximal symmetry algebra of the free Schr\"odinger action for two-component wave functions has been identified in subsection \ref{Schractsymalg} with the real Lie algebra $\mathfrak{u}(2)\otimes {\cal A}_{d}(\mathbb{R})$ of quantum observables. Physically, this means that the algebra $\mathfrak{u}(2)\otimes {\cal A}_{d+1}$ of $2\times 2$ Hermitian space-time operators can be seen as arising from gauging 
the algebra $\mathfrak{u}(2)\otimes {\cal A}_{d}$ of rigid symmetries via the Noether procedure, \textit{c.f.} the minimal coupling \eqref{S1extbckgd}. As usual in non-relativistic physics, the gauging amounts to an arbitrary dependence on the time coordinate $t$. Here, one adds an arbitrary dependence on the time momentum $\hat{P}_t=i\partial_t$ of the transformation parameters. However, only the arbitrary time dependence is genuinely non-trivial because, on-shell, any time derivative can be traded for the Laplacian.
A related subtlety is that the charged non-relativistic bilinears are not Noether currents since they are not conserved. Thus, strictly speaking, the coupling
\eqref{S1extbckgd} to external fields is not a pure minimal coupling \textit{\`a la} Noether. As will be seen in the next subsection, the pseudo ``conservation laws'' of the charged bilinears are thus not associated with genuine rigid symmetries. Their related local symmetries simply allow to get rid of the charged background fields $\upvarphi^{(r)}$ with level $r>0$, as is consistent with the fact that the bilinears $k^{(r)}$ with $r>0$ are descendants. A somewhat similar result is actually true even for the neutral background fields and currents.

As a side remark, let us notice that the restriction to the $Sp\,(2N)$-invariant background fields subsector
is a consistent truncation. However, it seems that the corresponding non-relativistic higher-spin algebra has no relativistic parent algebra. Let us describe in some details the subalgebra of symmetries related to the restriction to the $Sp\,(2N)$-invariant subsector. In order to describe this subtle subalgebra, some algebraic technology is needed.
More precisely the operation $^\tau$, defined on the algebra of space-time operators in subsection \ref{quadraticfunctional}, can be extended to a linear antiautomorphism of the algebra of matrix-operators by defining
\beq\label{pr5}
\sigma_0^\tau=\sigma_0\,,\quad\sigma_i^\tau=-\sigma_i\,,\quad(i=1,2,3)\,.
\eeq
The algebras of $2\times 2$ matrices and of space-time operators are ${\mathbb Z}_2$-graded with respect to the eigenvalues $\pm 1$ of $^\tau$ and decompose as: $\mathfrak{u}(2)\cong
\mathfrak{u}(1)\oplus\mathfrak{sp}(2)$ (since $\sigma_0$ is of eigenvalue $+1$ and the Pauli matrices $\sigma_i$ are of eigenvalues $-1$) and ${\cal A}_{d+1}={\cal A}_{d+1}^{\mbox{even}}\oplus {\cal A}_{d+1}^{\mbox{odd}}$ (where even/odd refer to the momentum parity).
The eigenvalue $-1$ of this antiautomorphism correspond to the property \eqref{pr2}.
The corresponding real subalgebra of $2\times 2$ matrix-operators
is isomorphic to $\big(\mathfrak{u}(1)\otimes{\cal A}_{d+1}^{\mbox{odd}}\big)\oplus\big(\mathfrak{sp}(2)\otimes {\cal A}_{d+1}^{\mbox{even}}\big)$.
As one can clearly see, this subalgebra for the $Sp\,(2N)$-invariant subsector is much more complicated than the corresponding algebra of infinitesimal gauge transformations, $\mathfrak{u}(2)\otimes{\cal A}_{d+1}$, for the $O(N)$-invariant sector.
Moreover, the operation $^\tau$ seems to have no counterpart in the relativistic construction of Vasiliev \cite{Vasiliev:2003ev}. This provides a strong motivation for focusing on the flavor-invariant (\textit{i. e.} $O(N)$-invariant) bilinears when looking for a bulk dual.

\subsection{Gauge symmetries to lowest order}\label{loword}

Since, as any operator, the infinitesimal gauge parameter $\hat A$ in Eq. \eqref{generator} is the sum of a Hermitian and an anti-Hermitian operator, it is enough to consider these two cases of gauge parameters separately.

If the operator-matrix $\hat{A}$ is Hermitian, it becomes
\beq
\hat{A}\,=\,\left(
\begin{array}{cc}
\hat a_{\uparrow} & \hat b\\
\hat b^{\dagger} & -\hat a_{\downarrow}^\tau
\end{array}
 \right)\,=\,\hat{A}^\dagger
\eeq
where the operators $\hat a_{\uparrow}$ and $\hat a_{\downarrow}$ are Hermitian.
Then we obtain that \eqref{gtransfo} can be written as
\beq
 \left(
\begin{array}{cc} 
\delta\hat{H}_{\uparrow\,\mbox{int}} & \delta\hat{\upvarphi}\\
\delta\hat{\upvarphi}^\dagger & -\delta\hat{H}_{\downarrow\,\mbox{int}}^\tau
\end{array} 
 \right)\,=\,-\,i\,\left(
\begin{array}{cc}
[\hat P_t\,-\frac{\hat P^2}{2m}\,,\,\hat a_{\uparrow}]  & [\hat P_t\,,\hat b]- \frac1{2m}\,\{\hat P^2\,,\hat b\}\\
\,[\hat P_t\,,\hat b^{\dagger}]\,+\frac1{2m} \{\hat P^2\,,\hat b^{\dagger}\} & -[\hat P_t\,+\frac{\hat P^2}{2m}\,,\,
\hat a_{\downarrow}^\tau]
\end{array}
 \right)\,,
\eeq
modulo the linear term in the backgrounds which will always be dropped from now on.
This transformation is equivalent to the following infinitesimal transformation:
\beq\label{gtrintud}
\delta\hat{H}_{\mbox{int}}^\alpha\,=\,-i \,\left[\hat P_t\,-\frac{\hat P^2}{2m}\,,\,\hat a^\alpha\right]
\eeq
for the (up and down) interaction Hamiltonians, and
\beq\label{gtrupvar}
\delta\hat{\upvarphi}\,=\,-i\,[\hat P_t\,,\hat b]\,+\frac{i}{2m}\,\{\hat P^2\,,\hat b\}
\eeq
for the off-diagonal term.
The transformation \eqref{gtrintud} reads in terms of the corresponding Weyl symbols
\beq\label{gtrintud2}
\delta h^\alpha(t,\textbf{x}\,;p_t,\textbf{p})\,=\,-\,i\,\left[\,p_t\,-\frac{\mathbf{p}^2}{2m}\,\stackrel{\star}{\,,}\,a^\alpha(t,\textbf{x}\,;p_t,\textbf{p})\,\right]\,=\,\left(\frac{\partial}{\partial t} \,+\,\frac1{m}\,p^i\,\frac{\partial}{\partial x^i}\right)\,a^\alpha(t,\textbf{x}\,;p_t,\textbf{p})
\eeq
where $\star$ stands for the Moyal product \eqref{sptiMoyalproduct} on the space-time phase space.
The above Weyl symbols should be interpreted as generating functions of symmetric tensor fields via the corresponding analogue of the power series expansion in momenta \eqref{genfctsymb}. This leads to the following gauge transformations at order zero in the neutral background fields 
\beq\label{gtrintud3}
\delta h_{i_1\cdots\, i_s}^{(r)\,\alpha}=\partial_t a_{i_1\cdots\, i_s}^{(r)\,\alpha} +\frac{s}{m}\,\partial^{}_{(i_1}a_{i_2\cdots\, i_s)}^{(r)\,\alpha}
\eeq
where the round bracket stands for the symmetrisation over all indices with weight one, \textit{e.g.}
$h_{(i_1\cdots\, i_s)}=h_{i_1\cdots\, i_s}$.
These gauge symmetries of the neutral background fields are thus the pendant of the conservation laws of the neutral currents encoded in \eqref{nrconslaw}. These symmetries indeed leave invariant the minimal coupling terms on-shell, as can be checked explicitly by integrating by parts and making use of the conservation laws.
The gauge symmetries, in the case of neutral background field such that $h_\uparrow=h_\downarrow$, generalise to higher spins the non-relativistic general-coordinate 
symmetries discussed in \cite{Son:2008ye}.\footnote{Explicitly, the dictionary between notations of \cite{Son:2008ye} and ours is: 
$A_0=-\frac1{m}h^{(0)}+\frac1{8m}(\partial_i\partial_j-\delta_{ij}\Delta)h^{(0)ij}+\frac1{4m}\Delta h^{(1)}-\frac1{4m}\partial_t\partial_i h^{(1)i}$, 
$A_i=-h^{(0)}_i$, $\Phi=-h^{(1)}$, $B_i=-m\,h^{(1)}_i$, $h_{ij}=-m\,h^{(0)}_{ij}$ and $\xi^-=-\frac1{m}\,a^{(0)}$, $\xi_i=a_i^{(0)}$, $\xi^t=-a^{(1)}$.
Employing these identifications we recover the gauge transformations of \cite{Son:2008ye} to zeroth order in the background fields. More precisely, we find a higher-spin generalization of transformations of \cite{Son:2008ye} since only transformations which originate from the relativistic spin one and two gauge transformations were considered in \cite{Son:2008ye}.}
Similarly, the infinitesimal transformations corresponding to \eqref{gtrupvar}
can also be written in terms of the Weyl symbols as 
\begin{eqnarray}\label{gtrintud2a}
\delta \upvarphi(t,\textbf{x}\,;p_t,\textbf{p})&=&-i\,\Big[ p_t\stackrel{\star}{\,,} b(t,\textbf{x}\,;p_t,\textbf{p})\Big]\,+\frac{i}{2m}\,\Big\{\mathbf{p}^2
\stackrel{\star}{\,,}b(t,\textbf{x}\,;p_t,\textbf{p})\Big\}\nonumber \\
&=&\left(\partial_t \,+\,\frac{i}{m}\,\Big(\textbf{p}^2-\frac{\Delta}4\Big)\right)\,b(t,\textbf{x}\,;p_t,\textbf{p}).
\end{eqnarray}
This leads to the following gauge transformations at order zero in the charged background fields
\beq
\delta \upvarphi_{i_1\cdots\, i_s}^{(r)}=\left(\partial_t - \frac{i}{4m}\,\Delta\right)b_{i_1\cdots\, i_s}^{(r)} +\frac{i\,s(s-1)}{m}\,\delta^{}_{(i_1i_2}b_{i_3\cdots\, i_s)}^{(r)}\,.
\eeq
These transformations actually correspond to the tracelessness-like condition for the charged currents $k$, \textit{i.e.} of the type \eqref{trlike2}. If we instead had made use of the traceless currents $\bar k$, then the above transformation would take the simpler form of a Weyl transformation $\delta \bar\upvarphi_{i_1\cdots\, i_s}^{(r)}=\frac{i\,s(s-1)}{m}\delta^{}_{(i_1i_2}\bar b_{i_3\cdots\, i_s)}^{(r)}\,.$ Such kind of higher-spin generalisations of linearised Weyl transformations appear in conformal higher-spin gravity \cite{Fradkin:1985am}.

If the matrix-operator $\hat{A}$ is anti-Hermitian, it is of the form
\beq
\hat A\,=\,i\,\left(
\begin{array}{cc}
\hat c_{\uparrow} & \hat d\\
\hat d^\dagger & -\hat c_{\downarrow}^\tau
\end{array}
 \right)\,.
\eeq
where the operators $\hat c_{\uparrow}$ and $\hat c_{\downarrow}$ are Hermitian.
Then we obtain that \eqref{gtransfo} can be written as
\beq
 \left(
\begin{array}{cc} 
\delta\hat{H}_{\uparrow\,\mbox{int}} & \delta\hat{\upvarphi}\\
\delta\hat{\upvarphi}^\dagger & -\delta\hat{H}_{\downarrow\,\mbox{int}}^\tau
\end{array} 
 \right)\,=\,
\left(
\begin{array}{cc}
\{\hat P_t\,-\frac{\hat P^2}{2m},\,\hat c_{\uparrow}\}  & \{\hat P_t,\hat d\}-\frac 1{2m}\,[\hat P^2,\hat d\,]\\
\{\hat P_t,\hat d^{\dagger}\}\,+\frac1{2m}[\hat P^2,\hat d^{\dagger}] & -\{\hat P_t\,+\frac{\hat P^2}{2m},\,\hat c^\tau_{\downarrow}\}
\end{array}
\right)\,, 
\eeq
which is equivalent to the following infinitesimal transformation:
\beq\label{gtrintud4}
\delta\hat{H}_{\mbox{int}}^\alpha=\left\{\hat P_t\,-\frac{\hat P^2}{2m}\,,\,\hat c^\alpha\right\}
\eeq
for the (up and down) interaction Hamiltonians, and
\beq\label{gtrupvar3}
\delta\hat{\upvarphi}=\{\hat P_t\,,\hat d\}-\frac 1{2m}\,[\hat P^2\,,\hat d\,]\,.
\eeq
This leads to the following gauge transformations at order zero in the background fields 
\beq\label{gtrintud5}
\delta h_{i_1\cdots\, i_s}^{(r)\,\alpha}\,=\,  2\,r\,c_{i_1\cdots\, i_s}^{(r-1)\,\alpha} +\frac1{m}\,\Big(\,
\frac14\,\Delta\,c_{i_1\cdots\, i_s}^{(r)\,\alpha} -s(s-1)\delta^{}_{(i_1i_2}c_{i_3\cdots\, i_s)}^{(r)\,\alpha}\Big)
\eeq
and
\beq\label{gtrupvar4}
\delta \upvarphi_{i_1\cdots\, i_s}^{(r)}\,=\,2\,r\,d_{i_1\cdots\, i_s}^{(r-1)} +\frac{i\,s}{m}\,\partial^{}_{(i_1}\, d_{i_2\cdots\, i_s)}^{(r)}\,.
\eeq
The first important observation to be made is that the first term in these transformations for level $r\neq 0$ is of Stuckelberg type and therefore allows to get rid (at this order in the background expansion) of all tensor fields
of non-vanishing level $r>0$. This is natural since the bilinears to which they couple are not independent: the neutral (respectively, charged) bilinears of the non-vanishing level $r>0$ are traces (respectively, descendants) of the ones with $r=0$. One should be careful that it is not clear whether this gauge choice is accessible at non-linear level. In addition, the non-vanishing levels are useful for the closure of the non-Abelian gauge algebra.
Moreover, these Stuckelberg-like transformations might be anomalous at quantum level.
In any case, the second terms in the transformations \eqref{gtrintud5}-\eqref{gtrupvar4} are more familiar:
they correspond respectively to Weyl-like (Fradkin-Tseytlin's) transformations of the neutral tensor fields
and to Maxwell-like (Fronsdal's) transformations of the charged tensor fields. They correspond respectively to the trace-like (or pseudo-conservation) conditions on the neutral (or charged) bilinears \eqref{trlikecond} (or \eqref{pseudoconslaw}\,). The gauge symmetries \eqref{gtrintud5}, in the case of neutral background field such that $h_\uparrow=h_\downarrow$, generalise to higher spins the non-relativistic Weyl
symmetries discussed in \cite{Son:2005rv}.

Let us stress that it is very useful to make use of the traceless currents $\bar k$, because the  transformations $\delta \bar\upvarphi_{i_1\cdots\, i_s}^{(r)}$ take a simpler form for the part independent of the background fields. However, the explicit form of the non-linear completion would be much more complicated, which is why we refrained from making direct use of them in this Section. Nevertheless, one should observe that the scalar charged background field at level zero, \textit{i.e.} the dimer  $\upvarphi=\bar\upvarphi^{(0)}$ coupling to the Cooper pair, transforms linearly under the symmetries. More precisely,
$\delta \bar\upvarphi^{(0)}$ is linear in the background field. This property should be useful to write the symmetry transformations of the Legendre transform $\Gamma[\,h\,,\upvarphi\,;N]$ of the background effective action $W_{\mbox{free}}[\,h\,,\upvarphi\,;N]$ with respect to the dimer.
Anyway, at leading order in $1/N$, the bulk dual of the ideal and of the unitary Fermi gases has the same symmetries. Only the $1/N$ corrections are expected to break the higher-spin symmetries \cite{Girardello:2002pp}.

\section{Conclusion and outlook}\label{Conclusion} 

Recent advances in holographic duality motivated us to investigate the symmetries and the currents of non-relativistic free fermions. Since in the large-$N$ limit the unitary and free Fermi gases 
are Legendre conjugate of each other, our studies might be useful for a better understanding of the strongly-coupled many-body problem of unitary fermions. We identified the maximal symmetry algebra of the free single-particle Schr\"odinger equation with the Weyl algebra of quantum observables. This higher-spin algebra is an infinite-dimensional extension of the well-studied Schr\"odinger algebra. Further, by applying the light-like dimensional reduction to relativistic Noether currents we constructed the infinite collection of non-relativistic ``currents'' bilinear in the elementary fermions. In addition, the formalism of Weyl quantisation allowed us to express the minimal coupling of these bilinears to background sources in a compact way. The final result is formally identical to the Nambu-Gor'kov formulation of the BCS theory except that the chemical potential and the Cooper-pair source are replaced by space-time differential operators.

One of the leitmotives behind our work is the null reduction method, advocated as ``Bargmann framework'' in \cite{Duval:1984cj,Duval:2009vt,Duval:1990hj}, which allows to obtain non-relativistic structures from given relativistic ones.
The other way around, \textit{i.e.} a null lift (or ``oxydation'') of a given non-relativistic structure to its higher-dimensional relativistic counterpart, is sometimes called an ``Eisenhart lift''. One should stress that the higher-dimensional counterpart of a consistent non-relativistic field theory may be sick as a relativistic quantum field theory \textit{per se}. For instance, the spin-statistics theorem does not apply to non-relativistic theories so it may be violated in the Eisenhart lift. Therefore, in general the relativistic higher-dimensional theory should be understood as an auxiliary tool.\footnote{In any case, \textit{a priori} the Eisenhart lift should not be trusted beyond tree level. Nevertheless, this restriction might be overcome by working with the quantum effective action since then all Feynman diagrams become trees (written in terms of full propagators and of proper vertices).} The results of the present paper demonstrate the usefulness of the Eisenhart lift for the free and the unitary Fermi gases.

The Bargmann framework might also apply to the holographic duality 
in the sense that the AdS/CFT correspondence might lead to the AdS/unitary fermions correspondence upon null reduction, along the lines of \cite{Goldberger:2008vg} and as proposed in \cite{Bekaert:2011cu}.
In these proposals, the background bulk geometry is an asymptotically AdS space-time (rather than the Schr\"odinger manifold, as proposed in \cite{Son:2008ye,Balasubramanian:2008dm}) possessing a nowhere vanishing covariantly constant null vector field.\footnote{Such space-times would be called asymptotically AdS Bargmann manifolds in the terminology of \cite{Duval:1984cj}. They can somehow be interpreted physically as gravitational waves propagating in AdS with parallel rays.} The isometry group of AdS is broken to the Schr\"odinger subgroup by the dimensional reduction itself.
A nice property of this approach is that if the dimensional reduction is performed on both sides of the correspondence, then the validity of the holographic duality between the pair of relativistic parent theories would ensure the duality between the pair of reduced non-relativistic theories, at least in the large-$N$ limit.
Notice that, in this picture, the reduced holographic duality should be between a non-relativistic conformal field theory living on the boundary of a Newton-Cartan space-time and a non-relativistic gravity theory in its interior.
Indeed, the reduction of vacuum Einstein equations along a non-vanishing covariantly-constant (or at least Killing) null vector field leads to the Newton-Cartan equations describing in a geometric fashion the non-relativistic gravity theory of Newton \cite{Duval:1984cj,Julia:1994bs}.

So, with these various results in mind, let us come back to our original question: \textit{What is an educated guess for a gravity dual of unitary and free fermions?}
On the boundary side, the Bargmann framework allowed us to understand the higher-spin symmetries of the free fermions and to obtain from the relativistic massless Grassmann-odd scalar free theory the corresponding currents and couplings to background sources. Our results closely resemble the boundary data in the AdS/O(N) correspondence mentioned in the introduction.\footnote{Interestingly, an Euclidean $Sp\,(2N)$ vector model with anticommuting scalars has recently been conjectured to be dual to Vasiliev's higher-spin gravity on de Sitter space \cite{Anninos:2011ui}.}
On the bulk side, one might thus speculate that the null reduction of a higher-spin gauge theory would be a natural candidate. 
Assuming that the Bargmann framework can be applied to both sides of the correspondence, the gravity dual of the ideal and unitary Fermi gases should be a non-relativistic higher-spin gravity theory obtained directly from Vasiliev equations upon light-like reduction.\footnote{An alternative, more along the lines of \cite{Son:2008ye,Balasubramanian:2008dm}, would be to look for a natural embedding of the Schr\"odinger manifold as a natural background for some (possibly modified) version of Vasiliev equations.} 
Looking in the catalogue of Vasiliev theories in any dimension \cite{Vasiliev:2003ev}, one can see that the 
flavor-singlet bilinear sector of the large-$N$ extension of the unitary fermions in $d$ space dimension should be dual to the null-reduction of classical Vasiliev theory on $AdS_{d+3}$ with $\mathfrak{u}(2)$-valued tensor gauge fields of all integer ranks.\footnote{The corresponding higher-spin algebra was denoted by $\mathfrak{hu}(2/\mathfrak{sp}(2)[d+2,2]\,)$ in \cite{Vasiliev:2003ev}. It is isomorphic to the product between $\mathfrak{u}(2)$ and the higher-spin algebra $\mathfrak{hu}(1/\mathfrak{sp}(2)[d+2,2]\,)$.} Therefore, one is led to speculate that the bulk dual of the ``physical'' (\textit{i.e.} $N=1$, $d=3$) unitary UV-stable Fermi gas might be \emph{the null dimensional reduction of the $\mathfrak{u}(2)$ higher-spin gauge theory on $AdS_6$ with the exotic ($\Delta_-=2$) boundary condition for the complex scalar field dual to the Cooper-pair field} \cite{Bekaert:2011cu}.

These speculations are supported by our results on the large-$N$ extension of the ideal and the unitary Fermi gases, so
let us summarise them with emphasis on their relevance for the above proposal:
In Section \ref{Legendre}, it was demonstrated that, in the large-$N$ limit, the generating functionals of the unitary Fermi gas and of the ideal Fermi gas are related by a Legendre transformation.
Therefore the corresponding Fermi gases can be dual to the same bulk theory for two distinct choices of boundary conditions, as in the the conjecture \cite{Klebanov:2002ja} (and its generalisation to higher dimensions). The corresponding scaling dimensions of the Cooper-pair field was found to be precisely in agreement with the mass-square $m^2=-2\,d$ of the $AdS_{d+3}$ scalar field in Vasiliev higher-spin multiplet \cite{Vasiliev:2003ev}. The holographic degeneracy is admissible in the range $0<d<4$ in agreement with the field theory prediction.
In Section \ref{Symmetry} the maximal symmetry algebra of the free Schr\"odinger action was identified and in Section \ref{Currents} it was shown that it originates from the maximal symmetry algebra of the free massless Klein-Gordon action via light-like dimensional reduction.
Since the identification of the proper higher-spin algebras is a crucial step in the construction of higher-spin gravities of Vasiliev, the embedding of the non-relativistic higher-spin algebra into its relativistic parent (as the centraliser of a given light-like momentum) provides a strong evidence for the consistency of the dimensional reduction of Vasiliev equations.
More precisely, we believe that the techniques of the light-like dimensional reduction for Einstein gravity in the frame formalism, developed in \cite{Julia:1994bs}, must have a natural higher-spin extension since Vasiliev gravity is based on a frame-like formalism à la Cartan where, in the fiber, the AdS isometry algebra for usual gravity is replaced by the higher-spin algebra.
For the relativistic conjecture \cite{Klebanov:2002ja,Sezgin:2002rt}, the validity of the holographic dictionary at the kinematical level (\textit{i.e.} two-point functions) between bilinear boundary currents and bulk gauge fields in any dimension and for any integer spin is actually a corollary of the Flato-Fronsdal theorem and its generalisation \cite{Flato:1978qz}.
The above embedding of the non-relativistic higher-spin algebra into its relativistic parent combined with the Flato-Fronsdal theorem automatically validates the holographic dictionary proposed above between $O(N)$-singlet bilinears in the non-relativistic fields on the boundary, constructed in Section \ref{Currents}, and $\mathfrak{u}(2)$-valued symmetric tensor gauge fields of all integer spins in the bulk.
In Section \ref{Sources}, the generating functional of connected correlators of $O(N)$-singlet bilinears for the non-interacting Fermi gas was computed explicitly together with the non-relativistic conformal higher-spin Ward identities.
According to the Gubser-Klebanov-Polyakov-Witten prescription, the generating functional should be equal to the on-shell bulk higher-spin action with prescribed boundary conditions while the Ward identities should be dual to the asymptotic remnant of bulk higher-spin gauge transformations.
In the large-$N$ limit, these properties would follow directly from the light-like dimensional reduction if the parent relativistic duality \cite{Klebanov:2002ja,Sezgin:2002rt} is valid.

In order to test these ideas explicitly in the bulk, various issues need to be investigated: Firstly, one should clarify how concretely the higher-spin unitary representations of the Schr\"odinger group also describe free higher-spin fields in the bulk. Secondly, the non-relativistic analogues of the Flato-Fronsdal theorem \cite{Flato:1978qz} and of the Vasiliev equations \cite{Vasiliev:2003ev} should be spelled out.
These interesting open problems may prove to be challenging exercises to perform explicitly but one should stress that they are ensured to be well posed problems because their answers have to follow from their known relativistic counterparts via the light-like dimensional reduction, since the latter is well defined and consistent.
Both at the kinematical and dynamical level, this consistency is ensured by our embedding of the non-relativistic higher-spin algebra into its relativistic parent as the centraliser of a given light-like momentum.

Endowed with these results, one could try to perform non-trivial tests of the conjecture, presumably along the lines of the encouraging results of Giombi and Yin in $AdS_4$ \cite{Giombi:2009wh}. So far most tests of the higher-spin AdS/CFT correspondence have been restricted to bulk dimensions  $D\leqslant 4$, because Vasiliev theory is technically simpler in these dimensions (due to the use of twistors, see \textit{e.g.} \cite{Vasiliev:1995dn} for a review). For this reason, technically it might be easier to check whether the null reduction of $\mathfrak{u}(2)$ Vasiliev theory around $AdS_4$ with the standard ($\Delta_+=2$) boundary condition is dual to the $d=1$ scale-invariant ``unitary'' IR-stable two-component Fermi gas. Remarkably, the latter is well-understood as it corresponds to an infinite repulsion between ``up'' and ``down'' fermions and thus is equivalent to the non-interacting one-component Fermi gas with the same density (see \emph{e.g.} \cite{Barth} and references therein).

A possible angle of attack toward a derivation of the holographic duality would be to parallel the strategy of Douglas, Mazzucato and Razamat \cite{Douglas:2010rc}. More precisely, one might consider the exact renormalisation group equation for the regularised generating functional describing free fermions in the presence of a higher-spin background. The corresponding higher-spin sources flow under the renormalisation group and one may look for a suggestive rewriting of their scale evolution as a radial evolution of higher-spin bulk fields.

The relative simplicity of the non-relativistic higher-spin algebra and of the null reduction method supports the optimistic view that the holographic dual of unitary fermions is an accessible goal worth investigating.

\vspace{3mm}

\noindent\textbf{Note added:}\footnote{We are grateful to M.A. Vasiliev for calling these points to our attention and for his useful explanations.}
After the present work was completed and submitted to arXiv, we were informed that
it has some overlap with results obtained in the context of the $Sp\,(2d,\mathbb R)$-covariant unfolded equations initiated in \cite{Vasiliev:2001zy}. In particular, the isomorphism between the maximal symmetry algebra of the free Schr\"odinger equation and the Weyl algebra of spatial differential operators follows as a corollary\footnote{See \textit{e.g.} the subsection 2.1 of \cite{Gelfond:2008ur} for a review of the general argument and its application to the case relevant here.} from the general results on global symmetries of unfolded equations upon the identification of the spatial coordinates with the twistor variables of \cite{Vasiliev:2001zy} and of the time\footnote{Notice that the latter identification of a ``time'' coordinate among the  $\mathfrak{sp}(2d,\mathbb R)$ matrix coordinates was motivated in \cite{Vasiliev:2001dc} (see \textit{e.g.} the subsection 2.2 of \cite{Gelfond:2008ur} for a concise review of this point).} coordinate with the trace of the matrix coordinates of \cite{Vasiliev:2001zy}. Moreover, the structure \eqref{notationai0} of the generating function 
of non-relativistic bilinear currents of vanishing level is a particular instance of the ``generalised stress tensor'' of \cite{Gelfond:2008ur}. Bilinear current generating functions constructed in terms
of two different solutions with opposite signs of the Planck constant, identified with the mass here, were presented before in \cite{Gelfond:2003vh}.

\section*{Acknowledgments}

We thank C. Duval, S. Golkar, M. Hassaine, P. Horvathy, D.T. Son, J. Unterberger and M. Valenzuela for discussions on related issues.
X.B. and S.M. are grateful to the Institut f\"ur Theoretische Physik (Heidelberg) and to the Laboratoire de Math\'ematiques et Physique Th\'eorique (Tours) for respective hospitality. X.B. also acknowledges the Instituto de Matem\'atica y Fisica (Talca) for kind hospitality while the paper was near completion.


\appendix

\section{Weyl quantisation}\label{Weylquantization}

The Weyl-Wigner-Gr\"onewold-Moyal formalism \cite{WW} offers a classical-like formulation of quantum
mechanics
using phase space functions as observables and the Wigner function as an
analogue of the Liouville density function.

Classical mechanics is
based on the commutative algebra of classical observables,
\textit{i.e.} real functions $f(\mathbf{x},\mathbf{p})$ on the phase space
$T^*{\mathbb R}^d\cong{\mathbb R}^{d}\times{\mathbb R}^{d*}$,
endowed with the canonical Poisson bracket
\beq\label{cPb}
\{f,g\}_{\textbf{P.B.}}\,=\,\frac{\partial f}{\partial x^i}\,\frac{\partial
g}{\partial p_{i}}\, -\,\frac{\partial f}{\partial
p_{i}}\,\frac{\partial g}{\partial x^{i}}\,.
\eeq
Quantum mechanics is
based on the non-commutative associative algebra of quantum observables,
\textit{i.e.} Hermitian operators $\hat{F}(\hat{\mathbf{X}},\hat{\mathbf{P}})$ on the Hilbert space $L^2({\mathbb R}^d)$ of square-integrable functions.
The \textit{Weyl algebra} ${\cal A}_d$ is the associative algebra of quantum observables that are polynomials in the positions and momenta.
The \textit{Heisenberg algebra} $\mathfrak{h}_d$ is the Lie algebra of quantum observables that are polynomials of degree one in the positions and momenta, it is spanned by $\hat{X}^i$, $\hat{P}_j$ and a central element $\hbar$ obeying to the \textit{canonical commutation relations}
\beq\label{ccr}
[\,\hat{X}^i\,,\,\hat{P}_j\,\,]\,=\,i\hbar\,\delta^i_j\,.
\eeq
In more abstract terms, the Weyl algebra ${\cal A}_d$ is the universal enveloping algebra ${\cal U}(\mathfrak{h}_d)$ of the Heisenberg algebra.
The Schur lemma implies that the real eigenvalue (which we denote by the same symbol $\hbar$) of the central element labels 
the UIRs of the Heisenberg algebra.   
The \textit{theorem of Stone and von Neumann} asserts that, up to equivalence, there is a unique UIR of the Heisenberg algebra $\mathfrak{h}_d$ for each real value of $\hbar\neq 0$. Moreover, the corresponding representation of ${\cal A}_d$ is faithful, which legitimates the equivalence between the abstract definitions and the concrete realisations of the Heisenberg and Weyl algebras.\footnote{For $\hbar=0$, the UIRs of $\mathfrak{h}_d$ reduce to the one-dimensional UIRs of the commutative algebra ${\mathbb R}^{d}\times{\mathbb R}^{d*}$ labeled by the eigenvalues $\mathbf{x}$ and $\mathbf{p}$ of the operators $\hat{\mathbf{X}}$ and $\hat{\mathbf{P}}$. Obviously, when $\hbar=0$ the algebra ${\cal A}_d$ is realised as the commutative algebra of polynomials $f(\mathbf{x},\mathbf{p})$ on phase space.}

The \textit{Weyl map} ${\cal W}:f(\mathbf{x},\mathbf{p})\mapsto \hat{F}(\hat{\mathbf{X}},\hat{\mathbf{P}})$
associates to any function $f$ a Weyl(\textit{i.e.}
symmetric)-ordered operator $\hat{F}$ defined by
\begin{equation}
\hat{F}\,:=\,\frac{1}{(2\pi\hbar)^{d}}\int d\,\mathbf{k}\,d\mathbf{v}\,\,{\cal
F}(\mathbf{k},\mathbf{v})\,e^{\frac{i}{\hbar}\,(\,k_i\,\hat{X}^i\,-\,v^i\,\hat{P}_i)}\,,
\label{Weylmap}
\end{equation}
where ${\cal F}$ is the Fourier transform\footnote{The Weyl map is
well defined for a much larger class than square integrable
functions, including for instance the polynomial functions,
Fourier transforms of which are distributions.} of $f$ over the
\textit{whole} phase space (in other words, over position
\textit{and} momentum spaces)
\beq
{\cal F}(\mathbf{k},\mathbf{v})\,:=\,\frac{1}{(2\pi\hbar)^{d}}\int
d\mathbf{x}\,d\mathbf{p}\,\,f(\mathbf{x},\mathbf{p})\,e^{-\frac{i}{\hbar}\,(\,k_i\,x^i\,-\,v^i\,p_i)}\,.
\eeq
The function $f(\mathbf{x},\mathbf{p})$ is called the \textit{Weyl symbol} of the
operator $\hat{F}(\hat{\mathbf{X}},\hat{\mathbf{P}})$ which need not be in the symmetric-ordered form.
A nice property of the Weyl map (\ref{Weylmap}) is that it relates
the complex conjugation $^*$ of symbols to the Hermitian
conjugation $^\dagger$ of operators, ${\cal
W}:f^*(\mathbf{x},\mathbf{p})\mapsto \hat{F}^\dagger(\hat{\mathbf{X}},\hat{\mathbf{P}})$. Consequently, the
image of a real function (a classical observable) is a Hermitian
operator (a quantum observable). The inverse ${\cal
W}^{-1}:\hat{F}(\hat{\mathbf{X}},\hat{\mathbf{P}})\mapsto f(\mathbf{x},\mathbf{p})$ of the Weyl map is called
the \textit{Wigner map}.

The canonical commutation relations \eqref{ccr} between the position and momentum
operators and the Baker-Campbell-Hausdorff formula imply two very useful equalities:
\begin{eqnarray}
\,e^{\frac{i}{\hbar}\,(\,k_i\,\hat{X}^i\,-\,v^i\,\hat{P}_i)}
&=&e^{-\frac{i}{2\hbar}\,v^i\,\hat{P}_i}\,e^{\frac{i}{\hbar}\,
k_i\,\hat{X}^i}\,e^{-\frac{i}{2\hbar}\,v^i\,\hat{P}_i}\label{usefind}\\
&=&e^{-\frac{i}{2\hbar}\,v^i\,\{\,\hat{P}_i,\,\,\,\}}\,e^{\frac{i}{\hbar}\,k_i\,\hat{X}^i}
\label{usefind2}
\end{eqnarray}
where $\{\,\,\,,\,\,\}$ denotes the anticommutator.

On the one hand, combining (\ref{Weylmap}) with (\ref{usefind2}) implies that one
way to explicitly perform the Weyl map is via some
``anticommutator ordering'' for half of the variables with respect
to their conjugates.
For instance, the image of a Weyl symbol which is a formal power series in the momenta,
\begin{eqnarray}
f(\mathbf{x},\mathbf{p}) &=& \sum\limits_{r\geqslant 0}\frac{1}{r!}\,f^{i_1\ldots i_r}(\mathbf{x})\,p_{i_1}\ldots p_{i_r}\nonumber\\
&=&f(\mathbf{x})\,+\,f^i(\mathbf{x})\,p_i\,+\,\frac12\,f^{ij}(\mathbf{x})\,p_ip_j\,+\,{\cal O}(p^3)\,,
\end{eqnarray}
can be written as
\begin{eqnarray}
\hat{F}(\hat{\mathbf{X}},\hat{\mathbf{P}}) &=& \sum\limits_{r\geqslant 0}\frac{1}{r!\,2^r}\,\{\,\cdots\,\{f^{i_1\ldots i_r}(\hat{\mathbf{X}})\,,\,\hat{\mathbf{P}}_{i_1}\}\,,\,\cdots\,,\, \hat{\mathbf{P}}_{i_r}\}\nonumber\\
&=&\hat{F}(\hat{\mathbf{X}})\,+\,\frac12\,\big(\,\hat{F}^i(\hat{\mathbf{X}})\,\hat{\mathbf{P}}_i\,+\,\hat{\mathbf{P}}_i\,\hat{F}^i(\hat{\mathbf{X}})\,\,\big)\nonumber\\
&&\,+\,\frac14\,\big(\,\hat{F}^{ij}(\hat{\mathbf{X}})\,\hat{\mathbf{P}}_i\hat{\mathbf{P}}_j\,+\,
2\,\hat{\mathbf{P}}_i\,\hat{F}^{ij}(\hat{\mathbf{X}})\,\hat{\mathbf{P}}_j\,+\,\hat{\mathbf{P}}_i\hat{\mathbf{P}}_j\,\hat{F}^{ij}(\hat{\mathbf{X}})\,\,\big)\,+\,\ldots
\end{eqnarray}

On the other hand, Eq. (\ref{usefind}) implies that one
way to explicitly perform the Wigner map is via a Fourier transformation of the ``point shifted'' integral kernel of the operator. The \textit{integral kernel} of an operator $\hat{F}$ is the
matrix element $\langle\,\mathbf{x}\mid\hat{F}\mid \mathbf{x}^\prime\rangle$
appearing in the position representation of the state $\hat{F}\mid
\psi\,\rangle$ as follows
\beq
\langle\, \mathbf{x}\mid\hat{F}\mid \psi\,\rangle\,=\,\int d\mathbf{x}^\prime
\,\, \langle\, \mathbf{x}\mid\hat{F}\mid
\mathbf{x}^\prime\,\rangle\,\,\psi(\mathbf{x}^\prime)\,,
\eeq
where the wave function in position space is
$\psi(\mathbf{x}^\prime):=\langle\, \mathbf{x}^\prime\mid\psi\,\rangle$ and the
completeness relation $\int d\mathbf{x}^\prime\mid
\mathbf{x}^\prime\,\rangle\,\langle\, \mathbf{x}^\prime\mid\,=\widehat{\,1}$ has
been inserted. The definition (\ref{Weylmap}) and the
relation (\ref{usefind}) enable to write the integral kernel
of an operator in terms of its Weyl symbol,
\begin{equation}
\langle\,\mathbf{x}\mid\hat{F}\mid
\mathbf{x}^\prime\,\rangle\,=\,\int\frac{d\mathbf{p}}{(2\pi\hbar)^{d}}\  f\,\Big(\,\frac{\mathbf{x}+\mathbf{x}^\prime}{2}\,,\,\mathbf{p}\,\Big)\,\,e^{\frac{i}{\hbar}\,(\,x^i-x^{\prime\,i})\,p_i}\,.
\label{matrixelement}
\end{equation}
Conversely, this provides an explicit form of the Wigner map
\begin{equation}
f(\mathbf{x},\mathbf{p})\,=\,\int d\mathbf{q}\,\,\langle\,\mathbf{x}-\mathbf{q}/2\mid\hat{F}\mid
\mathbf{x}+\mathbf{q}/2\,\rangle\,\,e^{\frac{i}{\hbar}\, q^i\,p_i}\,,
\label{Wignermap}
\end{equation}
as follows from the expression (\ref{matrixelement}). This shows
that indeed the Weyl and Wigner maps are bijections between the
vector spaces of classical and quantum observables. 


The \textit{Moyal product} $\star$ is the pull-back of the
composition product in the algebra of quantum observables with
respect to the Weyl map $\cal W\,$, such that
\begin{equation}
{\cal W}\big[f(\mathbf{x},\mathbf{p})\,\star\, g(\mathbf{x},\mathbf{p})\big]\,=\,\hat{F}(\hat{\mathbf{X}},\hat{\mathbf{P}})\,\,\hat{G}(\hat{\mathbf{X}},\hat{\mathbf{P}})\,.
\label{morphism}
\end{equation}
The Wigner map (\ref{Wignermap}) allows to check that the
following explicit expression of the Moyal product satisfies the
definition (\ref{morphism}),
\begin{eqnarray}
f\,\star\, g&=&f\,\,\exp\left[\,\frac{i\,\hbar}{2}\,\left(\frac{\overleftarrow{\partial}}{\partial x^i}\,\frac{\overrightarrow{\partial}}{\partial p_i}-\frac{\overleftarrow{\partial}}{\partial p_i}\,\frac{\overrightarrow{\partial}}{\partial x^i}\right)\right]\,g\nonumber\\
&=&f\,
g+\frac{i\,\hbar}{2}\left\{f\,,\,g\right\}_{\textbf{P.B.}}+{\cal
O}(\hbar^2), \label{Moyalproduct}
\end{eqnarray}
where the arrows indicate on which factor the derivatives should
act. 

Let $\hat{H}$ be a Hamiltonian operator with the corresponding Weyl symbol $h(\mathbf{x},\mathbf{p})\,$.
In the Heisenberg picture, the time evolution of a quantum
observable $\hat{F}$ (which does not depend explicitly on time) is governed
by the differential equation
\begin{equation}
\frac{d\hat{F}}{dt}\,=\,\frac{1}{i\,\hbar}\,[\hat{F},\hat{H}]
\end{equation}
or equivalently in terms of symbols
\begin{equation}
\frac{df}{dt}\,=\,\frac{1}{i\,\hbar}\,[\,f\,\overset{\star}{,}\,h\,]
\label{timevolution}
\end{equation}
where $[\,\,\,\overset{\star}{,}\,\,\,]$ denotes the
\textit{Moyal commutator} defined by
\begin{eqnarray}
&&\left[\,f\,\overset{\star}{,}\, g\,\right]:=f\,\star\, g\,-\,g\,\star\, f\nonumber\\
&&\quad=\,2\,i\,\,f\,\,\sin\left[\,\frac{\hbar}{2}\,\left(\frac{\overleftarrow{\partial}}{\partial x^i}\,\frac{\overrightarrow{\partial}}{\partial p_{i}}-\frac{\overleftarrow{\partial}}{\partial p_{i}}\,\frac{\overrightarrow{\partial}}{\partial x^{i}}\right)\right]\,g\nonumber\\
&&\quad=\,i\,\hbar\,\left\{\,f\,,\,g\,\right\}_{\textbf{P.B.}}\,+\,{\cal
O}(\hbar^2)\,, \label{Moyalcommutator}
\end{eqnarray}
as can be seen from (\ref{Moyalproduct}). The \textit{Moyal bracket} is related to the Moyal commutator by
$$\left\{\,f\,,\,g\,\right\}_{\textbf{M.B.}}:=\frac{1}{i\,\hbar}\,[\,f\,\overset{\star}{,}\,g\,]
=\left\{\,f\,,\,g\,\right\}_{\textbf{P.B.}}+{\cal O}(\hbar).\,$$
Note that the Moyal bracket $\left\{\,\,\,,\,\,\right\}_{\textbf{M.B.}}$ is a deformation of the Poisson
bracket $\left\{\,\,\,,\,\,\right\}_{\textbf{P.B.}}$, and one can see that the equation (\ref{timevolution}) is a perturbation of the Hamiltonian
flow. If either $f(\mathbf{x},\mathbf{p})$ or $g(\mathbf{x},\mathbf{p})$ is a polynomial of degree
two, then their Moyal bracket reduces to their Poisson bracket. So
when the Hamiltonian is quadratic (free) the quantum evolution of
a Weyl symbol is identical to its classical evolution.

\section{Representations of the Schr\"odinger algebra} \label{App:Sch} 

Besides the free Schr\"odinger theory, there are known examples of interacting
theories which preserve the Schr\"odinger symmetry at quantum level. Nishida and Son called them ``non-relativistic
conformal field theories'' (NRCFT) and made an important step towards a systematic understanding of this class of theories \cite{Nishida, Nishida2}.\footnote{See also earlier important works of Henkel and Unterberger \cite{Henkel} on this subject.} In this Appendix, we review their basic results and investigate the structure of the unitary irreducible representations (UIR) of the Schr\"odinger algebra.

 In close analogy with relativistic conformal field theories, it is useful to introduce primary
 operators\footnote{or quasiprimary in the language of \cite{Henkel}} in NRCFT \cite{Nishida}. A local \textit{primary 
 operator} $\mathscr{\hat O}(t,\mathbf{x})$ has a well defined ``spin'' $s_{\mathscr{\hat O}}$, scaling dimension $\Delta_{\mathscr{\hat O}}$ and mass number
 $M_{\mathscr{\hat O}}$. In other words, it carries an irreducible representation of the rotation algebra $\mathfrak{o}(d)$ and it is an eigenvector of
 the scaling and mass operators.\footnote{For $d>3$, the irreducible representations of the rotation algebra $\mathfrak{o}(d)$ are characterised by Young diagrams  rather than a single half-integer. By ``spin'', one should understand the collection of labels characterising uniquely the representation.} For a scalar primary $\mathscr{\hat O}$ with $s_{\mathscr{\hat O}}=0$ (to which we restrict our attention here for the sake of simplicity), this means
 \beq \label{NRCFT6} [\hat{D},\mathscr{\hat O}]=-i\Delta_{\mathscr{\hat O}} \mathscr{\hat O}, \quad
 [\hat{M},\mathscr{\hat O}]=M_{\mathscr{\hat O}}\mathscr{\hat O},
 \eeq
 where $\mathscr{\hat O}\equiv\mathscr{\hat O}(t=0,\mathbf{x}=\mathbf{0})$. By definition, a primary
 operator $\mathscr{\hat O}$ must also commute with $\hat{K}_i$ and $\hat{C}$
 \beq \label{NRCFT7} [\hat{K}_{i},\mathscr{\hat O}]=0, \qquad
 [\hat{C},\mathscr{\hat O}]=0. 
 \eeq
 
 Most importantly, from the primary operator $\mathscr{\hat O}$ one can build a representation\footnote{more precisely, a ``Verma module'' in mathematical jargon} of the Schr\"odinger algebra. Specifically, the primary operator is the lowest weight operator as it has the lowest scaling dimension in the representation. The \textit{descendants} are constructed by taking spatial and temporal derivatives of the primary operator $\mathscr{\hat O}$. Using the Schr\"odinger algebra it is possible to show that the generators $\hat{P}_i$ and $\hat{H}$ form a pair of canonical creation operators which increase the scaling dimension by one and two units respectively. 
 
The commutation relation
\beq \label{ap1}  [\hat{P}_i, \hat{K}_j]=-i\delta_{ij} \hat{M} \eeq
suggests that $-i\hat{K}_j$ plays the role of a canonical annihilation operator as it decreases the scaling dimension by one unit. Actually, this is only true for the massive representations (with $M_{\mathscr{\hat O}}\ne 0$). The descendants are thus higher weight operators in a massive representation. The massless case is special since $[\hat{P}_i, \hat{K}_j]=0$, and thus all Galilean boost generators $\hat{K}_j$ commute with all ``descendants'' generated by $\hat{P}_i$. Notably, there are operators in the massless representation which are both descendants and primaries. This implies that the structures of the massive and massless representations are very different and they will be discussed separately in the following. 

In a similar fashion, the commutation relation
\beq \label{ap2} [\hat{H}, \hat{C}]=i\hat{D} \eeq
hints that $i\hat{C}$ plays the role of an annihilation operator as it always decreases the scaling dimension by two units.\footnote{In order to obtain the canonical commutation relation, the operators $\hat{H}$ and $i\hat{C}$ must be properly renormalized (see \cite{Werner} for details).} Indeed, due to the unitarity bound ($\Delta_{\mathscr{\hat O}}>\frac{d}2>0$) the right-hand-side of Eq. \eqref{ap2} is never zero. Thus, for the pair $\hat{H}$ and $\hat{C}$ there is no analogous subtlety which we encountered for the pair $\hat{P}_i$ and $\hat{K}_j$ in the particular case of $M_{\mathscr{\hat O}}=0$.

\begin{figure}[t]
\begin{center}
\includegraphics[scale=0.8]{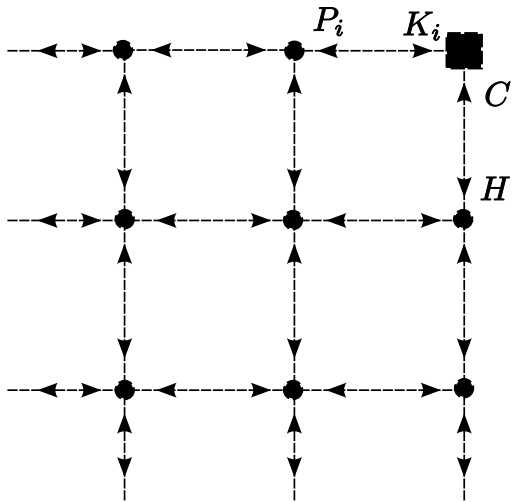}
\end{center}
\vskip -0.7cm \caption{Massive unitary irreducible representation of the Schr\"odinger algebra: The primary is represented by a full square, while descendants are depicted as solid circles.}
\label{figrep}
\end{figure}

After this general discussion we are ready to construct explicitly a massive UIR of the Schr\"odinger algebra on the basis of a primary $\mathscr{\hat O}$. In general, the representation is characterised by the scaling dimension $\Delta_{\mathscr{\hat O}}$, spin $s_{\mathscr{\hat O}}$ and mass number $M_{\mathscr{\hat O}}\ne 0$. Its structure is schematically illustrated in Fig. \ref{figrep} which makes the irreducibility of the representation manifest. We must mention that Fig. \ref{figrep} is in fact oversimplified since $\hat{P}_i$ and $\hat{K}_i$ do not commute with $\hat{H}$ and $\hat{C}$ and thus some arrows corresponding to the action of $\hat{K}_i$ and $C$ on descendants are not shown explicitly.


The operator/state correspondence of \cite{Nishida, Nishida2}\footnote{See also \cite{Tan, Werner} for the earlier quantum-mechanical formulation of this correspondence.} provides a very interesting alternative viewpoint on the massive representations. According to this correspondence the operators (the primary and descendants) of a NRCFT are mapped onto energy states of the same system placed in an external harmonic potential (with some frequency $\omega$). In particular, the primary operator corresponds to the ground state of the system of mass $M_{\mathscr{\hat O}}$ (\textit{i.e.} with particle number $N_{\mathscr{\hat O}}=\frac{M_{\mathscr{\hat O}}}{m}$) with the internal angular momentum $s_{\mathscr{\hat O}}$. The ground state in the trap reads $\ket{\psi_{\mathscr{\hat O}}}=e^{-\hat{H}/\omega}\mathscr{\hat O}\ket{0}$ and has the energy related to the scaling dimension of the primary via $E=\omega \Delta_{\mathscr{\hat O}}$. In this picture, descendants of the NRCFT simply correspond to the excited states. 
Specifically, the towers generated by $\hat{P}_i$ (see horizontal lines in Fig. \ref{figrep}) are mapped into excitations of the center-of-mass motion in the harmonic trap. Indeed the oscillator energy spectrum is equidistant with the spacing $\omega$ which matches precisely with the NRCFT result mentioned above. Explicitly, the center-of-mass excitations of the trapped system are constructed by acting repeatedly with the creation operators $\hat{Q}_i^{\dagger}=\frac{1}{\sqrt{2}}\left(\frac{\hat{P}_i}{\sqrt{\omega}}+i \sqrt{\omega} \hat{K}_{i} \right)$ on the ground state $\ket{\psi_{\mathscr{\hat O}}}$. On the other hand, one can also excite the internal motion (so called breathing modes) in the harmonic potential which is mapped into the towers generated by $\hat{H}$ in the NRCFT (see vertical lines in Fig. \ref{figrep}). Due to scale invariance the energy spectrum of breathing modes is also equidistant with the spacing $2\omega$ \cite{Werner}. The proper operator that excites the breathing modes turns out to be $\hat{B}^{\dagger}=\hat{L}^{\dagger}-\frac{\hat{Q}_i^{\dagger} \hat{Q}_i^{\dagger}}{2m_{\mathscr{\hat O}}}$, where $\hat{L}^{\dagger}=\frac{1}{2}\left(\frac{\hat{H}}{\omega}-\omega \hat{C}-i\hat{D} \right)$. Note that the pairs of operators $\hat{Q}_i^{\dagger}$, $\hat{Q}_i$  and $\hat{B}^{\dagger}$, $\hat{B}$  commute with each other, since they act on different degrees of freedom.  Finally, we mention that the operator/state correspondence makes the unitarity of the massive representation manifest, because it maps the representation onto a Hilbert space of the $N_{\mathscr{\hat O}}$-particle problem in a harmonic trap.

The light-like dimensional reduction method also provides a complementary perspective on the massive representations. Indeed, the restriction of relativistic conformal primaries to some proper subset of components leads to non-relativistic conformal primaries (with the other components being descendants). To clarify this, let us remind the definition of a primary operator in a relativistic CFT: a local primary 
 operator $\mathscr{\tilde O}(x)$ has a well defined ``spin'' $s_{\mathscr{\tilde O}}$ and scaling dimension 
 $\Delta_{\mathscr{\tilde O}}$. In other words, it carries an irreducible representation of the Lorentz algebra $\mathfrak{o}(d+1,1)$ spanned by the generators $\tilde{M}^{\mu\nu}$ and it is an eigenvector of the dilatation operator $\tilde{D}$: $[\tilde{D},\mathscr{\tilde O}]=-i\Delta_{\mathscr{\tilde O}}\mathscr{\tilde O}$ where $\mathscr{\tilde O}=\mathscr{\tilde O}(x=0)$. By definition, a relativistic primary operator $\mathscr{\hat O}$ must also commute with the conformal boost generators $\tilde{K}_\mu$: $[\tilde{K}_{\mu},\mathscr{\tilde O}]=0$. Furthermore, the dimensional reduction ansatz requires to consider an eigenvector of a null translation operator: $[\tilde{P}^+,\mathscr{\tilde O}]=M_{\mathscr{\tilde O}}\mathscr{\tilde O}$. This ansatz implies that the non-relativistic operator $\mathscr{\hat O}(t,\mathbf{x}):=\mathscr{\tilde O}(x^+=t,x^-=0,\mathbf{x})$ has mass $M_{\mathscr{\hat O}}=M_{\mathscr{\tilde O}}$. Moreover, the identification \eqref{embedding} together with the fact that $\mathscr{\tilde O}$ commutes with all conformal boost generators implies that $\mathscr{\hat O}$ commutes with the expansion generator $\hat C$. Now comes a crucial additional ansatz: let us assume that $\mathscr{\tilde O}$ further commutes with the generators $\tilde{M}^{\mu-}$ which is equivalent to the fact that all the components $\mathscr{\tilde O}^{+\dots}$ vanish. As the result, the purely spatial components $\mathscr{\hat O}^{i_1 i_2\dots}(t,\mathbf{x})$ span a non-relativistic primary with spin $s_{\mathscr{\hat O}}=s_{\mathscr{\tilde O}}$ and scaling dimension 
 $\Delta_{\mathscr{\hat O}}=\Delta_{\mathscr{\tilde O}}$, while the other components $\mathscr{\hat O}^{-\dots -i_1i_2\dots}(t,\mathbf{x})$
 are descendants. This can be verified via the identification \eqref{embedding}, the previously stated commutations and the branching rules for the restriction of $\mathfrak{o}(d+1,1)$ to $\mathfrak{o}(d)$. As a corollary, this property ensures that the charged bilinears $k^{(0)i_1\cdots i_s}(t,\mathbf{x})$ (see Section \ref{Currents} for their definition) are local non-relativistic primary operators.

Another useful perspective on the massive representations of the Schr\"odinger algebra is the so-called ``standard'' realisation of the generators. Actually, for spinning massive particles, the space-time differential operators \eqref{NRCFT4a} correspond to the ``orbital'' part of the generators which must be supplemented by a ``spinning'' (or ``internal'') part spanning an irreducible representation of the subalgebra $\mathfrak{o}(d)\oplus\mathfrak{sl}(2,\mathbb{R})$. As was mentioned in the subsection \ref{kinematicalsym}, the translation  and Galilean boost generators $\hat{P}_i$ and $\hat{K}^j$ together with the mass operator span the Heisenberg subalgebra $\mathfrak{h}_d\subset\mathfrak{sch}(d)$. The theorem of Stone and von Neumann (see Appendix \ref{Weylquantization}) implies that, given the mass $m$, there is a unique UIR of the Heisenberg subalgebra. The authors of \cite{Appell} proved that any massive representation of the Schr\"odinger algebra is equivalent to the following realisation of the remaining generators
\begin{equation} \label{NRCFTKPAppell}
\begin{split}
 & \hat{P}_t= \frac{\hat{P}^2}{2m}+\hat{L}_-\,, \\
 & \hat{M}_{ij}=\frac{\hat{K}_i\hat{P}_j-\hat{K}_j\hat{P}_i}{m}+\hat{L}_{ij}\,, \\
 & \hat{D}=-\frac{\hat{K}^i\hat{P}_i}{m}+i\frac{d}{2}+\hat{L}_0\,, \\
 & \hat{C}=\frac{\hat{K}^2}{2m}+\hat{L}_+\,,
\end{split}
\end{equation}
where the operators $\hat{L}_{ij}$, $\hat{L}_\pm$ and $\hat{L}_0$ commute with all the other generators and provide a representation of $\mathfrak{o}(d)\oplus\mathfrak{sl}(2,\mathbb{R})$ with usual notations. In a sense, the latter operators correspond to the ``spinning'' or ``internal'' part of the generators while the ``orbital'' part is entirely built out of the translation and boost generators. In order to have an irreducible representation of $\mathfrak{sch}(d)$, the internal part of the representation of $\mathfrak{o}(d)\oplus\mathfrak{sl}(2,\mathbb{R})$ should be irreducible, so it is characterised by spin and scaling dimension (for lowest weight representations). Therefore, one recovers in a different way the results obtained from the non-relativistic conformal field theory techniques. 

\begin{figure}[t]
\begin{center}
\includegraphics[scale=0.8]{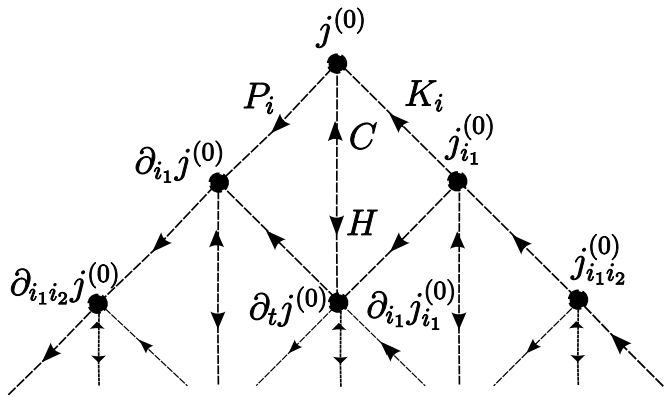}
\end{center}
\vskip -0.7cm \caption{Massless representation of the Schr\"odinger algebra: The operators are depicted as solid circles.}
\label{fig2}
\end{figure}

Let us now turn to massless representations of the Schr\"odinger algebra. As emphasized above, they have a distinct structure and are not so well understood. The representation containing \textit{e.g.} the non-relativistic currents $j^{(0)}_{i_1\dots i_n}$ (see Section \ref{Currents} for their definition) has a form of a pyramid and is illustrated in Fig. \ref{fig2}.\footnote{We are thankful to S. Golkar and D.T. Son for presenting this to us.}
The density operator $j^{(0)}=n$ is a non-relativistic primary, but not a descendant. On the other hand, the operators $\partial_{i_1} \cdots\partial_{i_n}j^{(0)}$ are both primaries and descendants. The spatial currents $j^{(0)}_{i_1 \dots i_n}$ are neither primaries nor descendants. As is clear from Fig. \ref{fig2}, this representation is not irreducible. Formally, one can generate the full representation starting from the current $j^{(0)}_{i_1 \dots i_n}$ with $n\to\infty$. The operator/state correspondence cannot be applied in a straightforward fashion to the normal-ordered neutral currents as they act trivially on the vacuum state.

In the AdS/CFT correspondence, a special role is played by the very exceptional irreducible representations of the Poincar\'e group that can be lifted to representations of the conformal group. They are called ``singletons'' and describe dynamical elementary fields living on the conformal boundary of AdS.
So an important question is: which UIRs of the Bargmann group can be extended to representations of the Schr\"odinger group? The massive (sometimes called ``physical'') representations of the Bargmann group are classified (see \textit{e.g.} \cite{LevyLeblond}) by the mass, the ``spin'' and the so-called internal energy\footnote{The internal energy is the eigenvalue of the operator $\hat{L}_-$ in \eqref{NRCFTKPAppell} and is equal to the opposite of the chemical potential $\mu$ in the free Schr\"odinger equation.} corresponding to the fact that in non-relativistic physics there is no privileged zero of the energy. One can see that \textit{all the massive representations of the Bargmann group with vanishing internal energy can be extended to representations of the Schr\"odinger group.} Indeed, conformal invariance requires that the internal energy must vanish because it is not preserved by scale transformations. Physically the internal energy may always be put to zero.\footnote{Mathematically, two massive representations of the Galilei group which only differ by the value of their internal energy are equivalent as projective representations \cite{LevyLeblond}.}
In order to complete the proof, one simply verifies that one may associate, to any representation of zero internal energy, a representation of the Schr\"odinger group (as follows from the above discussion). The only massive representations of the Schr\"odinger algebra with vanishing internal energy are those for which the UIR of the $\mathfrak{sl}(2,\mathbb{R})$ subalgebra on the internal (\textit{i.e.} spinning) degrees of freedom is trivial. Furthermore, looking at the classification of the UIRs of the Schr\"odinger group \cite{Perroud:1977qh}, one can see that \textit{the massive representations are the only non-trivial unitary irreducible representations of the Bargmann group that can be obtained as restrictions of the Schr\"odinger group.} In a sense, the analogue of the singleton representations of the Poincar\'e and conformal groups is identified with the massive representations (with zero internal energy) of the Bargmann and Schr\"odinger groups.

\end{document}